\documentclass[aps,onecolumn,preprint,superscriptaddress,nofootinbib,floats]{revtex4}
\usepackage{amsmath,amssymb,color,mathrsfs, graphicx,verbatim,epsfig, bbm, wasysym, axodraw}
\usepackage[hyperfootnotes=false]{hyperref}
\usepackage{slashed}
\allowdisplaybreaks

\def\lsim{\mathrel{\rlap{\lower4pt\hbox{\hskip1pt$\sim$}}
    \raise1pt\hbox{$<$}}}
\def\gsim{\mathrel{\rlap{\lower4pt\hbox{\hskip1pt$\sim$}}
    \raise1pt\hbox{$>$}}} 
\newcommand{\vev}[1]{ \left\langle {#1} \right\rangle }

\newcommand{\be}{\begin{eqnarray}}
\newcommand{\ee}{\end{eqnarray}}

\def\addresses#1#2{\hbox to \hsize{\@tablebox{#1}\hfil\@tablebox{#2}}}
\def\@tablebox#1{\vtop{\hsize=5in \begin{flushleft} #1 \end{flushleft}}}

\def\beq{\begin{equation}}
\def\eeq{\end{equation}}
\def\bit{\begin{itemize}}
\def\eit{\end{itemize}}
\def\beqa{\begin{eqnarray}}
\def\eeqa{\end{eqnarray}}

\def\MadGraph{{\tt MadGraph}}
\def\MadGraph5{{\tt MadGraph5}}

\def\cW{c_{W\rm hel}}
\def\sW{s_{W\rm hel}}

\newcommand{\met}{\displaystyle{\not}E_T}
\newcommand{\vecmet}{\vec{\displaystyle{\not}E}_T}

\begin{document}

\baselineskip 0.6cm

\begin{titlepage}

\thispagestyle{empty}

\begin{flushright}
PITT PACC 1315
\end{flushright}

\begin{center}

\vskip 2cm

{\Large \bf Better Hadronic Top Quark Polarimetry}

\vskip 1.0cm
{\large Brock Tweedie}
\vskip 0.4cm
{\it PITT PACC, Department of Physics and Astronomy, University of Pittsburgh, \\ Pittsburgh, PA 15260}
\vskip 2.0cm

\end{center}

\noindent   Observables sensitive to top quark polarization are important for characterizing or even discovering new physics.  The most powerful spin analyzer in top decay is the down-type fermion from the $W$, which in the case of leptonic decay allows for very clean measurements.  However, in many applications it is useful to measure the polarization of hadronically decaying top quarks.  Usually it is assumed that at most 50\% of the spin analyzing power can be recovered in this case.  This paper introduces a simple and truly optimal hadronic spin analyzer, with a power of 64\% at leading order.  The improvement is demonstrated to be robust at next-to-leading order, and in a handful of simulated measurements including the spins and spin correlations of boosted top quarks from multi-TeV $t\bar t$ resonances, the spins of semi-boosted tops from chiral stop decays, and the potentially CP-violating spin correlations induced in continuum $t\bar t$ by color dipole operators.  For the boosted studies, we explore jet substructure techniques that exhibit improved mapping between subjets and quarks.

\end{titlepage}

\setcounter{page}{1}

\section{Introduction}
\label{sec:intro}

Polarization serves as a unique tool for studying top quark production mechanisms.  As the only quark that decays before it can be depolarized by soft QCD, the top gives us direct access to its spin state through its decay angle patterns~\cite{Barger:1988jj,Kane:1991bg,Jezabek:1994qs}.  In addition to the net polarization, which can be induced by new chiral interactions or chiral particle decays, the spin correlations between top quarks in pair-production events exhibit a rich structure~\cite{Mahlon:1995zn,Stelzer:1995gc,Uwer:2004vp,Mahlon:2010gw,Baumgart:2012ay}.  The ability to view top production not just in terms of raw rate, but as a set of individual polarized processes, has been exploited repeatedly in proposals for new physics searches and categorization strategies (e.g.,~\cite{Baumgart:2012ay,Bernreuther:1993hq,Beneke:2000hk,Frederix:2007gi,Arai:2007ts,Shelton:2008nq,Degrande:2010kt,Cao:2010nw,Baumgart:2011wk,Barger:2011pu,Krohn:2011tw,Bai:2011uk,Falkowski:2011zr,Han:2012fw,Fajfer:2012si,Yang:2012ib,Gabrielli:2012pk,Falkowski:2012cu,Perelstein:2008zt,Berger:2012an,Bhattacherjee:2012ir,Belanger:2012tm,Baumgart:2013yra}), and the spin correlations in QCD $t\bar t$ production have recently been observed experimentally~\cite{Abazov:2011gi,TheATLAScollaboration:2013gja,Chatrchyan:2013wua}.  As the LHC increases in energy and luminosity, we will have the opportunity to scan both the net polarization and correlations over a broad swath of energies.  Further ahead, top quark polarization measurements will be an important aspect of future lepton and hadron accelerator programs.  Given the clear utility of these kinds of measurements, and despite their extensive previous study, the goal of this paper is to step back and ask whether they might still be systematically improved.  As we will find, there are indeed some nontrivial gains that may be achieved when measuring the polarization of top quarks that decay hadronically, gains which so far appear not to have been exploited.

In principle, the best way to estimate the spin of a top quark, relative to some prespecified quantization axis, is to focus on leptonic decays.  Because of the $V-A$ current structure of the weak interaction, the charged lepton is a ``perfect'' spin analyzer, in the sense that a 100\% polarized top quark will imprint a maximal linear bias on the lepton's decay angle distribution, as measured in the top's rest frame.  While this property nominally prefers measurements made with leptonic tops, there are two major disadvantages that complicate the accounting.  First, leptonic decay rates are small, an effect that is especially felt when we are measuring correlations and require both tops to be leptonic.  Second, leptonic decays inevitably lose some kinematic information from the neutrino.  This is again especially felt in dileptonic correlations, as reconstruction of the individual top rest frames and the precise production kinematics becomes difficult and ambiguous.  Production of tops in new physics processes with additional neutrinos or other invisible particles leads to similar kinematic complications, even for pairs of tops in the $l$+jets decay channel.  For these reasons, hadronically-decaying tops are often considered for use in polarimetry, despite the fact that the analog of the lepton, namely the down-type quark, effectively loses its identity upon reconstruction as a jet or subjet.  Hadronic tops are also obviously ``messier'' due to parton showering and hadronization, but hadronic top kinematic reconstructions at the LHC are by now routine in both threshold and boosted production regimes (e.g.,~\cite{Chatrchyan:2013xza,ATLAS:2012aj,Chatrchyan:2013lca,Aad:2012raa}).

While hadronic measurements can never be as good in principle as idealized leptonic ones, there are standard ways to salvage some of the spin sensitivity.  The simplest option is to pick the $b$-quark direction as the spin analyzer, or equivalently the direction of the hadronic $W$-boson.  This yields a spin sensitivity about 40\% as large as what was achievable with leptons.  However, the identity of the down-type quark is not actually completely lost.  $W$ bosons in top decay are produced on average with negative helicity, favoring the down-type quark to be emitted closer to the $b$-quark.  The light-quark closer to the $b$-quark is also the softer of the two $W$ decay products when viewed in the top rest frame.  Therefore, by boosting into the top rest frame and picking the light-quark jet that is better-aligned with the $b$-jet or is less energetic, the chance of guessing correctly is biased in our favor.  Using such a procedure bumps up the spin analyzing power to 50\% of the lepton's.  This has at least tacitly been considered the most sensitive available choice for hadronic top decays.

Though at first glance, it may seem that the best that we can do is to pick the jet with the highest chance of having come from the down-type quark, here we show that we can in fact do better by using simple weighted sums of the two light-quark jets' unit vectors.  When we chose these weights to be equal to the individual probabilities of coming from the down-quark, we obtain an optimal hadronic polarimeter with analyzing power of 64\% at leading order, or approximately the $W$ boson's velocity relative to the speed of light in the top rest frame.

As we will see, it is impossible to build a more powerful hadronic top spin analyzer direction at quark-level.  However, the question then arises whether this observation can be translated into gains in the performance of realistic measurements at jet- or subjet-level.  We take the opportunity to address this question under a number of different conditions, first using simple simulations of individual top decays at leading and next-to-leading order (NLO), and then moving on to complete LHC event simulations.  We pay particular attention to boosted top production, as the viable scale of new physics continues to be pushed up in many scenarios.  In doing so, we develop modified jet substructure algorithms that provide improved reconstruction of the 3-body top decay kinematics, relative to some of the common options.

In the next section of this paper, we discuss polarimetry with hadronic top quarks in full generality at parton-level, demonstrating the above claim of optimality, and exploring other aspects such as likelihood-based polarimeters and strategies when no $b$-tagging information is available.  In Section~\ref{sec:NLO} we study the stability of the hadronic polarimeters against QCD radiative corrections and the viability of the shower approximation used in the remainder of the paper.  In Section~\ref{sec:measurement}, we verify that the benefit of our optimal construction holds up in complete events with showering and jet reconstruction, taking as examples heavy $t\bar t$ resonances, chiral stop decays, and continuum $t\bar t$ production in the presence of potentially CP-violating color dipole operators.  The first two of these studies use a substructure procedure derived from the HEPTopTagger~\cite{Plehn:2010st}, with several novel modifications geared toward improving the mapping between subjets and quarks.  Section~\ref{sec:conclusions} contains our conclusions.  An appendix (\ref{sec:appendix}) includes a more in-depth discussion of the benefits of our modifications to the HEPTopTagger, as well as polarization measurements with a modified JHU top-tagger~\cite{Kaplan:2008ie}.

\section{Hadronic Polarimetry Variables}
\label{sec:variables}

\subsection{The Optimal Hadronic Spin Analyzer}  \label{sec:optimal}

\begin{figure}[tp]
\begin{center} \begin{picture}(400,125)(0,0)
\SetColor{Black} \SetWidth{1}
\LongArrow(105, 50)(  0, 50)   \Text(-10, 50)[]{$b$}
\LongArrow(105, 50)(135, 85)   \Text(144, 97)[]{$\bar d$}
\LongArrow(105, 50)( 75, 15)   \Text( 70,  5)[]{$u$}
\DashLine(105,50)(155, 50){5}
\LongArrow(305, 50)(240, 50)   \Text(232, 50)[]{$b$}
\LongArrow(305, 50)(360, 85)   \Text(372, 97)[]{$\bar d$}
\LongArrow(305, 50)(315, 15)   \Text(320,  5)[]{$u$}
\Text(195, 50)[]{\Large $\Rightarrow$}
\CArc(105,50)(30,5,45)   \Text(155,67)[]{$\theta_{W\rm hel}$}
\end{picture} \end{center}
\caption{The 3-quark top decay system as viewed in the $W$ rest frame (left), and boosted back into the top rest frame (right).  The $W$'s polar decay angle in its rest frame is $\theta_{W\rm hel}$, and is defined as the direction of the down-type quark with respect to $-\hat b$, as indicated in the figure.  The cosine of this angle, $c_{W\rm hel} \equiv \cos\theta_{W\rm hel}$, is in one-to-one mapping with a rigid body of quark momentum vectors in top-frame.  The Euler angles of this rigid body are the remaining physical degrees of freedom of the decay, assuming fixed masses, and are fully randomized for unpolarized tops.} 
\label{fig:cWhel}
\end{figure}
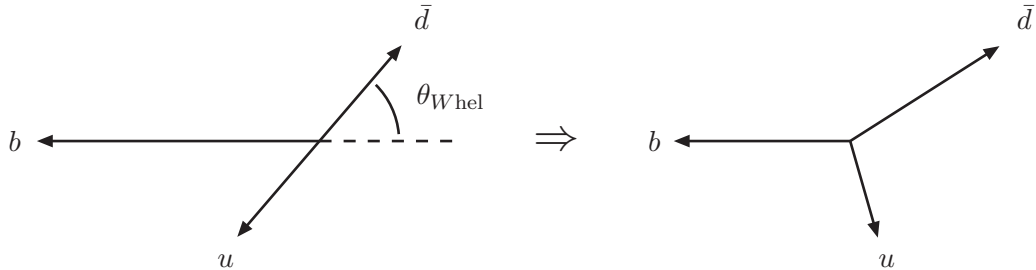

The decay angle distributions of unpolarized top quarks are fairly simple to understand.  The top quark undergoes an initial decay into $bW^+$ at a random orientation.  The $W^+$ then subsequently decays, and we will assume that this is into a $\bar d$-quark and a $u$-quark.  Here and throughout, we will not distinguish down from strange, nor up from charm, and we will default to calling the $\bar d$-quark simply the ``$d$-quark'' without an overbar.  The azimuthal orientation of the $W$ decay is also random, but the polar decay angle viewed within the $W$ rest frame, commonly called its helicity angle, exhibits a bias due to the polarizations of the $W$.  It is standard to take the ``$z$-axis'' of this decay to be the direction pointing opposite to the $b$-quark ($\hat z \equiv -\hat b$) in $W$-frame.  We will denote the cosine of this angle $\cW$, and take the convention that positive $\cW$ means that the $d$-quark is emitted in the forward hemisphere and the $u$-quark in the backward hemisphere.  The geometry is illustrated in Fig.~\ref{fig:cWhel}.  The $W$'s polarization causes $\cW$ to be distributed as
\beq
\rho\left(\cW\right) \,\equiv\,  \frac38 f_R \left(1+\cW\right)^2 + \frac34 f_0 \left(1-\cW^2\right) + \frac38 f_L \left(1-\cW\right)^2, \label{eq:Wpolar}
\eeq
where $f_R$, $f_0$, and $f_L$ are respectively the fractions of right-handed helicity, zero helicity, and left-handed helicity $W$ bosons in top-frame.  In the $V-A$ electroweak theory, $f_R$ is nearly zero, and
\beqa
f_0 & \,\simeq\, & \frac{m_t^2}{m_t^2+2m_W^2} \,\simeq\, 0.70 \nonumber \\
f_L & \,\simeq\, & \frac{2m_W^2}{m_t^2+2m_W^2} \,\simeq\, 0.30 \: ,  \label{eq:helicityFractions}
\eeqa
in the approximation $m_b = 0$ and taking $m_t = 172$~GeV.  By the approximate CP-invariance of the decay, anti-tops have a nearly identical distribution.

The introduction of top quark polarization can in principle lead to much richer patterns in the multidimensional space of decay angles.  However, the $V-A$ interaction is again highly constraining:  all polarization-sensitivity is encoded in the direction of the $d$-quark in top-frame.\footnote{We are of course assuming here that the decays are Standard Model-like.  The absence of $V+A$ structure has already been verified to the several percent level at the LHC~\cite{CMS:2013pfa,Chatrchyan:2013jna,ATLAS:2013tla}.  Weak electric/magnetic moment operators are also constrained, and will come under much greater scrutiny in the future.}  More precisely, imagine a top quark created in a generic event.  The top may be produced with net polarization due to a chiral interaction (such as single-top production), and its spin may have correlations with other parts of the event (such as with the spin of the anti-top in QCD $t\bar t$ production).  If we fix everything about the final-state spins and kinematics of the rest of the event, and trace out over the two possible top spin states, all of these effects collapse into a single vector which we can dot into the $d$-quark direction ($\hat d$) to determine the top's differential decay rate.  Call this vector $\vec P$.  It is the average top quark polarization as measured in the top's rest frame for this given set of ambient spins and kinematics.  Its magnitude varies between 0 and 1.  This polarization introduces into the top's multibody decay angle distribution an additional overall factor $1 + \vec P \cdot \hat d$.  This is the sense in which the $d$-quark is a maximal spin analyzer.  In particular, when the magnitude of $\vec P$ is 1, the $d$-quark has zero probability of being found antiparallel to it.  (The case of anti-tops is flipped, and the distribution becomes $1 - \vec P \cdot \hat d$.)  Integrating out all top decay angles except for the polar angle of $\hat d$ relative to $\vec P$, we would get the usual expression
\beq
\frac{1}{\Gamma}\frac{d\Gamma}{d\cos\theta_{d\cdot P}} \,=\, \frac{1 + P\cos\theta_{d\cdot P}}{2} \: .
\eeq

Because the $d$-quark cannot be uniquely identified, unlike in the analogous leptonic decay, this maximal spin sensitivity is inevitably lost.\footnote{Methods for measuring the charge of the progenitor quark do exist, but are not very statistically powerful for separating charge +1/3 from +2/3 (see~\cite{Krohn:2012fg}).  It might nonetheless be interesting to explore what further gains could be achieved by folding in this information.}  However, the different particles in top decay are highly kinematically correlated due to the top and $W$ mass-shell constraints, and the $W$ polar decay distribution of Eq.~\ref{eq:Wpolar}.  We therefore have the opportunity to make geometric constructions that exploit these correlations.  Generally the simplest option is to build some axis $\hat a$ from the reconstructable kinematics, and use this as a proxy for $\hat d$.  Assuming that this axis is defined independently of $\vec P$, and integrating out all decay angles except for the polar angle of this axis relative to $\vec P$, rotational invariance forces  
\beq
\frac{1}{\Gamma}\frac{d\Gamma}{d\cos\theta_{a\cdot P}} \,=\, \frac{1 + \kappa_a P\cos\theta_{a\cdot P}}{2} \: .  \label{eq:cosThetaGeneral}
\eeq
The parameter $\kappa_a$ is called the analyzing power.  It is a number between $-1$ and 1, and equals the average of $\hat a \cdot \hat d$.  (Again, for anti-tops, take $\kappa_a \to -\kappa_a$.)

A very common axis choice for hadronic top decays is $\hat b$, the direction of the $b$-quark, or equivalently the direction of the hadronic $W$ ($\hat W = - \hat b$).  This is, of course, because the $b$-quark {\it can} be unambiguously identified via $b$-tagging or kinematics.  One way to understand the $b$'s sensitivity to top spin is to pretend that the $W$ is stable, and consider decays into the two dominant $W$ spin states while preserving overall angular momentum.  The resulting analyzing power is $\kappa_b \simeq f_L-f_0 \simeq -0.40$ (and $\kappa_W = -\kappa_b$).

Another common choice is to pick the softer of the two light-quarks in top frame, or equivalently the quark that is better-aligned with the $b$-quark.  The chance that this choice picks out the $d$-quark can be determined from Eq.~\ref{eq:Wpolar} to be 61\%, and the corresponding spin analyzing power comes out to $\kappa_{\rm soft} \simeq 0.50$.  (A complete formula for $\kappa_{\rm soft}$ can be found in~\cite{Jezabek:1994qs}.)  This is the strongest hadronic top quark spin analyzer that has so far been studied.  Its advantage relative to $\hat b$ has been exploited in $t\bar t$ spin correlation measurements at the Tevatron~\cite{CDF10211} in the $l$+jets channel, and was also shown to have superior performance for measuring azimuthal decay angle sum/difference correlations in~\cite{Baumgart:2012ay}.

We can now ask whether $\kappa_{\rm soft}$ is really the best that we can do in principle, and how difficult it might be to construct a more powerful spin analyzer.  To do this, let us consider the complete multidimensional decay angle distribution, first assuming perfect knowledge of the quark identities, and then moving on to the realistic case where the light-quark identities are lost.  For each value of the $W$ helicity angle cosine $\cW$, the system of quark directions ($\hat d$, $\hat u$, $\hat b$) defines a distinct rigid body, and the remaining three angular degrees of freedom are just this object's Euler angles.  One of these angles, the overall azimuthal orientation of the system about $\vec P$, exhibits a flat distribution due to the residual rotational invariance.  (I.e., $\vec P$ breaks $SO(3)$ down to $SO(2)$.)  We will call this $\phi_{\rm global}$.  Note that this angle can usually be physically defined from the ambient system, and is therefore not simply a ``dummy'' variable, even though the top's decay is not sensitive to it.  The remaining two angles can be parametrized in many ways, for example as $\theta_{d \cdot P}$ and the relative orientation between the top decay plane and the plane defined by $\hat d$ and $\vec P$.  More generally, we can view this pair of angles as the spherical coordinates of $\vec P$ itself relative to the 3-quark rigid body.  To define these coordinates, start with the system consisting of $\vec P$ and the quarks, apply a global rotation such that the quark vectors lock in to some fixed reference orientation, and measure the position of $\hat P$ on the unit sphere.  Collectively referring to $\vec P$'s spherical coordinates as $\Omega_P$, we get
\beq
\frac{1}{\Gamma}\frac{d^4\Gamma}{d\cW d\Omega_P \, d\phi_{\rm global}} \,=\, \rho\left(\cW\right) \frac{1 + \vec P \cdot \hat d}{8\pi^2} \: .
\eeq

When we acknowledge that the $d$-quark the $u$-quark are fated to become anonymous jets, we are forced to identify $\cW \leftrightarrow -\cW$.  The forward-emitted quark in $W$-frame will be harder in top-frame and more separated in angle from the $b$-quark.  Similarly, the backward-emitted quark will be softer and better aligned with the $b$-quark.  We can therefore strip the light-quarks' flavor labels $d$ and $u$, and replace them with ``soft'' and ``hard.''  This relationship between energies and angles will ultimately be slightly scrambled by QCD showering, but alternative labeling schemes (such as purely geometric ones) will be closely related to this one, and we do not need to make these distinctions here.  The soft-quark and hard-quark each has some probability of really being the $d$-quark:
\be
p(d \to q_{\rm soft}) & \,=\, & \frac{\rho\left(-|\cW|\right)}{\rho\left(|\cW|\right) + \rho\left(-|\cW|\right)} \nonumber \\ 
&  & \nonumber \\
p(d \to q_{\rm hard}) & \,=\, & \frac{\rho\left(|\cW|\right)}{\rho\left(|\cW|\right) + \rho\left(-|\cW|\right)} \: . \label{eq:quarkProbs}
\ee
Denoting the soft-quark direction as $\hat q_{\rm soft}$ and the hard-quark direction as $\hat q_{\rm hard}$, the full differential decay distribution becomes
\be
\frac{1}{\Gamma}\frac{d^4\Gamma}{d|\cW| d\Omega_P \, d\phi_{\rm global}} & \,=\, & \big(\rho\left(|\cW|\right) + \rho\left(-|\cW|\right) \big) \times \nonumber \\ & & \frac{1 + \vec P \cdot \big[ p(d \to q_{\rm soft}) \hat q_{\rm soft} + p(d \to q_{\rm hard}) \hat q_{\rm hard}\big]}{8\pi^2} \: .  \label{eq:fullDecayFolded}
\ee
We immediately see that all spin sensitivity is aligned with the direction in brackets, which is just a weighted average of the two quark directions.  This is therefore the optimal spin analyzer direction, and the analyzing power for a given value of $|\cW|$ is this vector's length:
\be
\vec q_{\rm opt}(|\cW|) & \,\equiv\, & p(d \to q_{\rm soft}) \hat q_{\rm soft} + p(d \to q_{\rm hard}) \hat q_{\rm hard}  \nonumber \\
\kappa_{\rm opt}(|\cW|) & \,=\, & |\vec q_{\rm opt}(|\cW|)| \nonumber \\
                         & \,\simeq\,  &  \frac{\sqrt{\sW^4 m_t^2(m_t^2-2m_W^2) + (1+\cW^2)^2 m_W^4}}{\sW^2 m_t^2 + (1+\cW^2)m_W^2}  \: ,
\ee
in the limit of vanishing $b$-quark mass, and defining $\sW \equiv \sqrt{1-\cW^2}$.

\begin{figure*}[tp!]
\begin{center}
\includegraphics[width=0.44\textwidth]{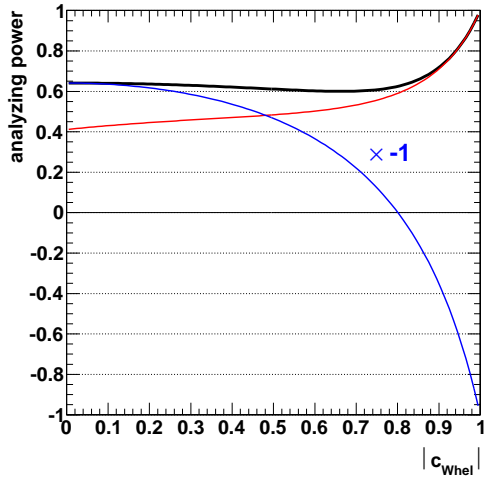}
\caption{Analyzing powers as a function of $|\cW|$:  black is $q_{\rm opt}$, red is $q_{\rm soft}$, and blue is the $b$-quark.}
\label{fig:powers}
\end{center}
\end{figure*}

\begin{table}
\begin{center}
\begin{tabular}{ l|r }
Spin Analyzer \     &  \ Power \\   \hline 
lepton/down-quark \ &    1.00  \\
neutrino/up-quark   &   -0.34  \\      
$b$-quark or $W$    &   $\mp$0.40  \\ 
soft-quark          &    0.50  \\  
optimal hadronic \  &    0.64  \\
\end{tabular}
\end{center}
\caption{Integrated leading-order analyzing powers of various top quark spin analyzers.}
\label{table:analyzers}
\end{table}

Fig.~\ref{fig:powers} shows the analyzing power of this optimal direction, and for some of the other choices, as a function of $|\cW|$.  For $|\cW| = 1$, there is essentially no ambiguity:  the $d$-quark is almost never emitted collinear to the $W$ in top-frame due to the approximate absence of right-handed $W$-polarization.  We therefore recover in that case the full spin analyzing power of the $d$-quark.  The opposite extreme is $|\cW| = 0$, in which case we have no ability to discriminate, and must simply perform an unweighted average over the two light-quark's unit vectors.  The resulting direction will be pointing along $\hat W$, but with reduced length determined by the quarks' opening angle in top-frame.  This length is just the $W$'s velocity, $\beta_W \simeq 0.64$.  In fact, the analyzing power turns out to be a fairly flat function of $|\cW|$ except near 1, as a Taylor expansion about 0 yields an accidentally small leading quadratic dependence (with coefficient roughly proportional to $2m_W^2(m_t^2-3m_W^2)/m_t^4 \simeq 0.15$).  Since $\rho(|\cW|)$ is also largest around zero, the integrated analyzing power is also quite close to $\beta_W$, smaller in ratio by less than a percent.  A list of all standard spin analyzers, now including this new one, is shown in Table~\ref{table:analyzers}.\footnote{We can also consider what we get if we simply take an {\it unweighted} sum of the two quarks' unit vectors in top-frame.  This direction has an analyzing power that roughly averages those of $\hat q_{\rm soft}$ and $\hat q_{\rm opt}$, or about 0.57.}

\subsection{Spin Analyzers Versus Likelihoods}
\label{sec:likelihood}

\begin{figure*}[tp!]
\begin{center}
\includegraphics[width=0.44\textwidth]{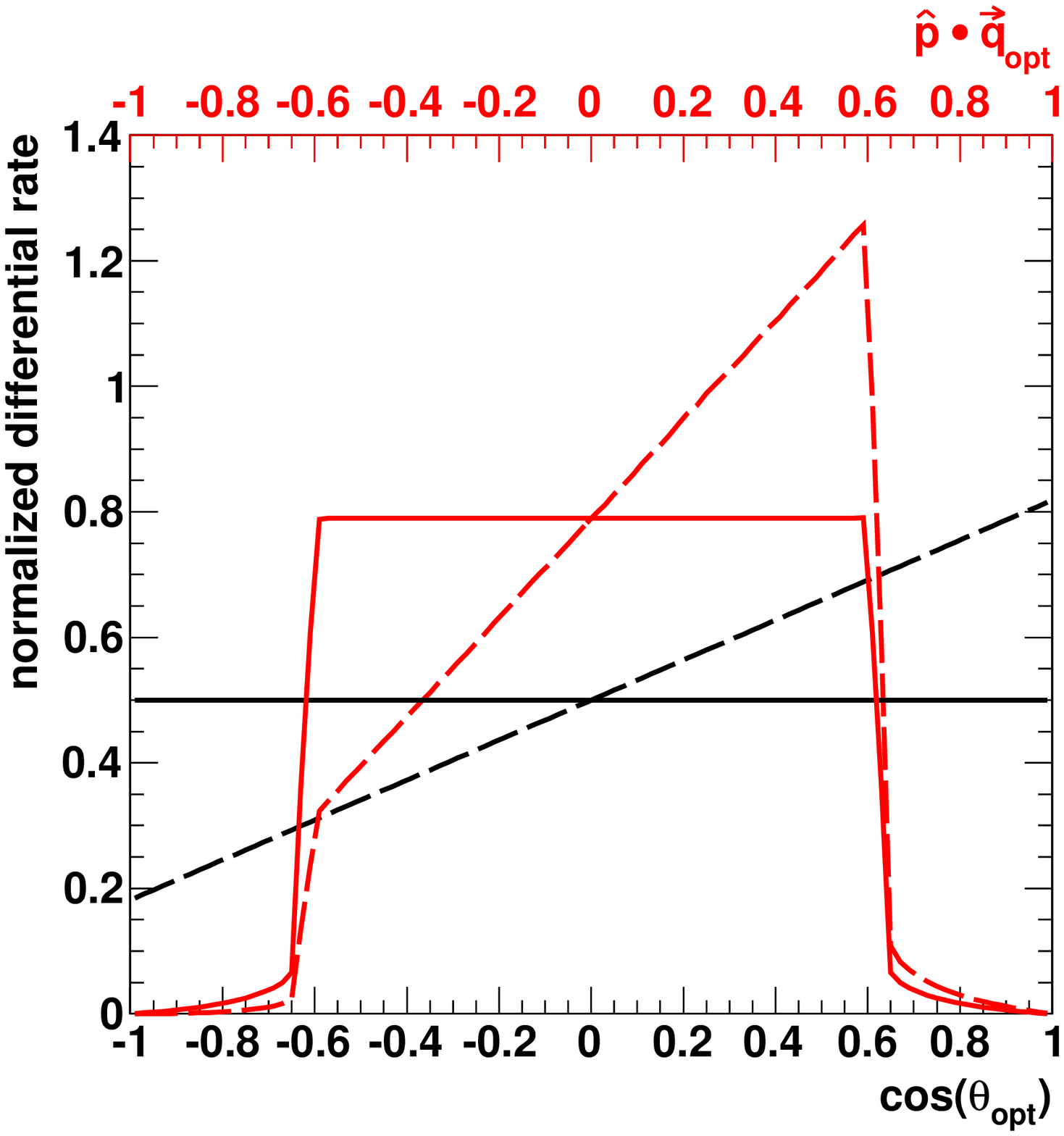}
\includegraphics[width=0.44\textwidth]{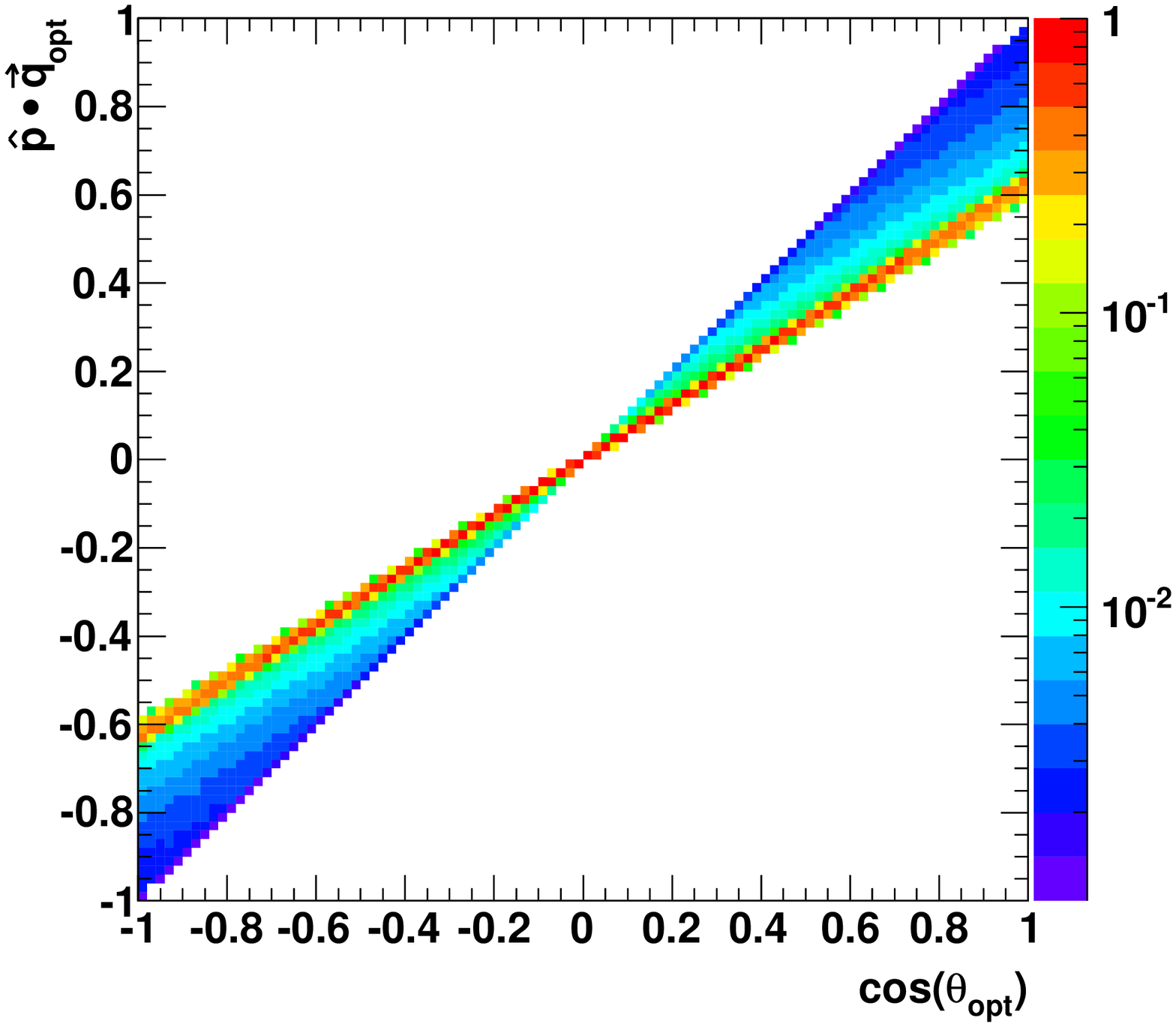}
\caption{Comparisons between the simple angular variable $\cos\theta_{\rm opt}$ and the formally more powerful $\hat P \cdot \vec q_{\rm opt}$: individual distributions for unpolarized and 100\% polarized tops (left), and the joint distribution for unpolarized tops (right).  Solid lines on the left plot are unpolarized, and dashed lines are polarized.  Black and red respectively indicate $\cos\theta_{\rm opt}$ and $\hat P \cdot \vec q_{\rm opt}$.  The density scale on the right plot is normalized against the case of perfect correlation between the two variables.}
\label{fig:likelihood}
\end{center}
\end{figure*}

While there is no way to form a better spin analyzing direction, there remains in principle a better way to utilize the information available to us over the full 4D decay phase space.  Given two physics hypotheses that yield distinct likelihood densities over an arbitrary phase space, we can foliate that space into contours of fixed likelihood-ratios.  In the present case, let us take these two hypotheses to be either unpolarized or polarized along some specific $\hat P$, with the only difference being the $\vec P \cdot \vec q_{\rm opt}$ term in Eq.~\ref{eq:fullDecayFolded}.  Our construction above almost yields likelihood-ratio contours for a given polarization $\vec P$, but not quite.  The likelihood-ratio contours can be uniquely labeled by $\hat P \cdot \vec q_{\rm opt}$, whereas the usual spin analysis of Eq.~\ref{eq:cosThetaGeneral} would form contours of $\cos\theta_{\rm opt} \equiv \hat P \cdot \hat q_{\rm opt}$.  The difference is the length of $\vec q_{\rm opt}$, again the analyzing power.  Were this analyzing power a fixed number (as is the case for $\hat d$), this difference would be immaterial, but since it is not we can ask to what extent a simple angular analysis underperforms the full likelihood analysis.  To illustrate that $\cos\theta_{\rm opt}$ and $\hat P \cdot \vec q_{\rm opt}$ are in fact distinct, we show their distributions and correlations for a simulated set of top quark decays in Fig.~\ref{fig:likelihood}.

To make some comparison between the two variables, we can consider their performance under a standard Neyman $\chi^2$ fit in the large-statistics limit.  Imagine binning over $\cos\theta_{\rm opt}$ or over the likelihood-ratio variable $\hat P \cdot \vec q_{\rm opt}$, and assigning each bin a gaussian error estimate equal to the square root of the observed bin count.  Suppose that we have unpolarized bin expectations $\mu_i$, and that polarization of strength $P$ (along the pre-specified $\hat P$) induces deviations $P\Delta\mu_i$ with $\sum_i\Delta\mu_i = 0$.  The $\Delta\mu_i$ factors encode the spin sensitivity.  A least-squares fit of $P$ on unpolarized data or data moderately affected by polarization ($|P\Delta\mu_i/\mu_i| \lsim 1$) would yield a characteristic uncertainty
\beq
\delta P \,\simeq\, \left[\sum_i \frac{\Delta\mu_i^2}{\mu_i}\right]^{-1/2} .  \label{eq:DeltaP}
\eeq
This carries over to the limit of infinitely-fine bins, and the sum in brackets can be viewed as a continuous integral over either of the two variables.
Applying this formula to $\cos\theta_{\rm opt}$ is trivial, since $\mu_i$ is flat and $\Delta\mu_i$ is a linear slope.  The analogous calculation for $\hat P \cdot \vec q_{\rm opt}$ requires slightly more care, since each value of $\kappa_{\rm opt}(\cW) \equiv |\vec q_{\rm opt}(\cW)|$ yields a different range over which the contribution is nonzero.  The results are
\be
\left[\delta P\right]_{\cos\theta} \;\;\;\;\;\; & \,\simeq\, & \sqrt{\frac{3}{N}} \frac{1}{\vev{\kappa_{\rm opt}}} \nonumber \\
\left[\delta P\right]_{\rm likelihood} & \,\simeq\, & \sqrt{\frac{3}{N}} \frac{1}{\sqrt{\vev{\kappa_{\rm opt}^2}}} \: ,
\ee
where $N = \sum_i\mu_i$ is the total sample size.

We can now clearly see in what sense the likelihood-ratio discriminator variable is more powerful: $\sqrt{\vev{\kappa_{\rm opt}^2}}$ is larger than $\vev{\kappa_{\rm opt}}$ for any distribution of $\kappa_{\rm opt}$, so the fit uncertainty $\delta P$ for the likelihood-ratio is smaller.  However, because $\kappa_{\rm opt}$ is a fairly flat function of $\cW$, and $\rho(\cW)$ is small where $\kappa_{\rm opt}$ starts to deviate, the fractional difference between $\sqrt{\vev{\kappa_{\rm opt}^2}}$ and $\vev{\kappa_{\rm opt}}$ is actually only a few parts per mil.  Therefore, at least at this idealized level, we miss very little discriminating power by using angles instead of the formally more powerful likelihood-ratios.\footnote{We can also consider the 2-bin limit, in which case $\cos\theta_{\rm opt}$ and the likelihood-ratio discriminator have identical distributions, and the effect of polarization is to simply induce an asymmetry of $P\vev{\kappa_{\rm opt}}/2$.  The uncertainty that would be returned by Eq.~\ref{eq:DeltaP} is then $\delta P \simeq (2/\sqrt{N})(1/\vev{\kappa_{\rm opt}})$, which is also what we would get by directly applying propagation-of-errors to the asymmetry formula in the moderate-asymmetry limit.  This 2-bin uncertainty is a factor of $2/\sqrt{3} \simeq 1.15$ larger than what we would have obtained by fitting the full linear shape of $\cos\theta_{\rm opt}$.  This ratio is easy to verify in toy monte carlo.}  Throughout the rest of this paper, we will default to only using $\vec q_{\rm opt}$ to define a spin analyzer direction and ignore its magnitude, with the understanding that this is nonetheless very close to the most aggressive possible approach.  We will return to using the polarization error estimator introduced in Eq.~\ref{eq:DeltaP} as we move on to comparing different observables under more realistic conditions.

Before proceeding, it is also interesting to perform a similar analysis on the other common spin analyzers, $q_{\rm soft}$ and the $b$-quark, by replacing, e.g., $\cos\theta_b$ with $\kappa_b(|\cW|)\cos\theta_b = (\hat b \cdot \vec q_{\rm opt})\cos\theta_b$.  For $q_{\rm soft}$, the improvement is again modest, close to 2\% relative.  For the $b$-quark, the improvement is significant, almost 30\% relative, yielding slightly better sensitivity even than $q_{\rm soft}$.  Most of this improvement comes from the fact that $\kappa_b$ flips sign for different values of $|\cW|$, which is corrected for by the 1D likelihood-ratio but not by the simple angular analysis.  However, capitalizing on this improved $b$-quark spin sensitivity in any case requires us to have enough information to construct $\vec q_{\rm opt}$.

\subsection{Optimizing Polarimetry Without ${\mathbf b}$-Tags}
\label{sec:untagged}

In most top quark studies, we take for granted the ability to identify $b$-jets using methods such as displaced vertices.  There is, however, an inevitable degradation as we go to higher $p_T$'s due to the collimation of tracks, and it is not currently clear to what extent this could pose a problem in studies of highly-boosted tops, such as from our heavy $t\bar t$ resonance examples in Section~\ref{subsec:resonance}.  So far, tagging $b$-subjets using displaced vertices is a relatively new endeavor, but experimental studies look promising~\cite{CMS:2013vea,ATLAS:2009elr}.  It is also worth noting that even a loose $b$-tag operating point remains highly useful, if our main interest is to discriminate one subjet out of three, rather than to separate a small $b$-enriched signal from a much larger light-flavor background.  Still, let us consider the extreme case where {\it no} tagging is available, and we are left to identify the $b$-subjet using pure kinematics.  It should be understood that this is a quite pessimistic situation, and serves as a lower bound on realistic performance.  Indeed kinematic and $b$-tagging information could likely be combined over quite a broad range of top $p_T$ scales.

Of course, in the simple 3-quark picture discussed in the previous subsections, kinematic tagging is not difficult.  If we pretend that the $W$ resonance peak is a $\delta$-function, then we can trivially pick out the two light quarks by studying the masses of all pairings.  Adding in the $W$'s Breit-Wigner lineshape does not significantly complicate the procedure, as picking the quark pair whose mass is closest to $m_W$ will still be correct the vast majority of the time.  The main context in which a more advanced procedure becomes useful is in real-life measurement, where in the highly-boosted case the quarks turn into subjets, and their 4-momenta and pairwise invariant masses become smeared out by QCD showering and instrumental effects.  The naive 4-dimensional phase space then formally becomes extended to 9-dimensional, as both the top and $W$ resonances are lifted off of their mass shells, as are the nominally massless quarks.  (The smearings can also depend on the overall $p_T$ and $\eta$ of the top-jet, adding yet two more dimensions.)  We will not attempt to tackle the full probability density over this large space, especially as many of the details are highly dependent on reconstruction algorithms and detector performance.  But we can still make progress by making a few well-motivated simplifying assumptions, and then employing the same type of strategy developed above, namely superimposing unit vectors according to relative probabilities.

It is generally safe to assume that the softest of the three subjets in top-frame is indeed from the $W$, and furnishes our $q_{\rm soft}$.\footnote{In the presence of 4-momentum smearings, we also no longer actually know if the softest jet is really $q_{\rm soft}$ versus $q_{\rm hard}$ (though the chance that it is the $b$-quark is indeed usually very small).  This potential ambiguity is mainly an issue for $|c_W| \simeq 0$, where the two quarks would appear in top-frame with nearly equal energy.  However, note that in this kinematic region, the $\hat q_{\rm opt}$ construction weights the two jets equally anyway.  Also, in attempting to use $\hat q_{\rm soft}$ as a polarimeter, we may still make a mistake and pick up $q_{\rm hard}$ instead, but at $|c_W| \simeq 0$ the two quarks have the same analyzing power. }  We then have two candidate assignments for $q_{\rm hard}$ and the $b$-quark.  The subtlety is that the $W$ decay may produce a quark with $E_{\rm hard} \simeq E_b$ in top-frame, and therefore $m(q_{\rm hard}q_{\rm soft}) \simeq m(b\,q_{\rm soft}) \simeq m_W$, in which case there is no way to unambiguously identify $q_{\rm hard}$ from the $b$ using only kinematics.  To properly account for the ambiguity, we should consider both choices simultaneously, assigning each a probability based on the two $W$ candidates' masses and an assumed joint probability distribution for the true masses $m(q_{\rm hard}q_{\rm soft})$ and $m(b\,q_{\rm soft})$.  More specifically, suppose that we order the three subjets according to their top-frame energy and label them as $j_1$, $j_2$, and $j_3$.  We identify $j_3 = q_{\rm soft}$, and then form an optimal spin analyzer given the available information, 
\be
\vec q_{\rm opt} & \,\to\, & p(W \to j_1 j_3) \times \big(\,p(d\to j_1|W\to j_1 j_3)\:\hat j_1 \,+\, p(d\to j_3|W\to j_1 j_3)\:\hat j_3\,\big) \,+ \nonumber \\
                 &         & p(W \to j_2 j_3) \times \big(\,p(d\to j_2|W\to j_2 j_3)\:\hat j_2 \,+\, p(d\to j_3|W\to j_2 j_3)\:\hat j_3\,\big) \, .
\ee
The quantities $p(W \to j_1 j_3)$ and $p(W \to j_2 j_3)$ are the relative probabilities of the $W$ decay to have produced $j_1j_3$ or $j_2j_3$ respectively, and implicitly for $j_2$ or $j_1$ to have come from the $b$-quark.  The quantities $p(d\to j_1|W\to j_1 j_3)$, etc, are the different light-quark flavor assignment probabilities as in Eq.~\ref{eq:quarkProbs}, conditioned on which choice we made for the two $W$-subjets.

To estimate $p(W \to j_1 j_3)$ and $p(W \to j_2 j_3)$, it suffices to focus on the assumed distribution of $m(q_{\rm hard}q_{\rm soft})$.  We have found that folding in more complete information by including $m(b\,q_{\rm soft})$ does not practically improve the achievable analyzing power.  This is likely due to that fact that, in a coarse-grained viewpoint, the above procedure is telling us to average the two $W$ and $b$ assignments when the candidate $W$ masses are close to each, and when they are far apart to just pick the one closer to $m_W$.  The major input here is the $W$ mass resolution model, which defines ``close'' and ``far.''  Practically any function with a prominent peak of the appropriate width suffices to model the distribution.  We take here a Breit-Wigner lineshape, with the natural width replaced by a resolution-smeared one.

The correctly-paired $W$ peak shape can vary depending on other details of the top-jet, in particular its overall $p_T$ and mass.  Different $p_T$'s can give different resolutions controlled by the detector's angular segmentation, while the center of the peak typically shifts in close correlation with the reconstructed top-jet mass.  Dealing with the former requires a detailed $p_T$-dependent resolution model, which we do not pursue, but the latter is largely corrected for by normalizing out the overall top-jet mass.  Consequently, the probabilities are computed by comparing $m(j_1j_3)/m(j_1j_2j_3)$ and $m(j_2j_3)/m(j_1j_2j_3)$ to a Breit-Wigner over the dimensionless variable $m(q_{\rm hard}q_{\rm soft})/m(b\,q_{\rm hard}q_{\rm soft})$.  The distribution is centered at $m_W/m_t \simeq 0.46$, and has a fixed width that must be determined by studying the distribution in monte carlo data with a perfect $b$-tag.

Since the performance of this method is contingent upon reconstruction details, we reserve its numerical study for Section~\ref{sec:measurement}.

\section{QCD Radiative Corrections}  \label{sec:NLO}

So far our discussion has mainly been restricted to a simple parton-level picture, as if the quarks in the leading-order decay were practically observable (if anonymous) particles.  More realistically, QCD radiative corrections are significant, forcing us to go over from a parton-level picture to a jet-level picture.  In the next section, we will study the implications in complete LHC events for several new physics scenarios.  These studies incorporate radiative corrections in an approximate way, via the leading-log, $p_T$-ordered parton shower of {\tt PYTHIA6}~\cite{Sjostrand:2006za}.  As an intermediate step, in this section we consider the corrected decays of individual tops in more detail, disconnected from any other event activity (an approach that can be formally justified in the narrow-width limit).  In particular, we would like to find out whether the leading-order construction of the optimal hadronic spin analyzer continues to offer any gains over the standard analyzers, and to what extent the parton shower accurately captures their absolute and relative performances.

To facilitate these comparisons, we have written a fast, standalone monte carlo program for polarized top decay at NLO, using the matrix elements and subtraction scheme provided in~\cite{Campbell:2012uf}.\footnote{The code has been validated on several quantities that are computed analytically in the literature, at both leading and next-to-leading order.  The NLO validations include:  corrections to the total top~\cite{Jezabek:1988iv} and $W$ decay rates (including individual dipole-regulated contributions~\cite{Catani:1996vz}), unpolarized $W$ decay kinematic distributions such as rest-frame thrust and $d\Gamma/dx_u dx_d$, the bottom quark energy spectrum~\cite{Corcella:2001hz}, the lepton and neutrino energy spectra~\cite{Czarnecki:1990pe,Czarnecki:1994pu}, corrections to the $W$ helicity fractions from the top decay (including the $\sim10^{-3}$ shift in $f_R$)~\cite{Fischer:2000kx} by fitting the leptonic polar decay distribution, corrections to the $W$ helicity angle distributions of bare quarks~\cite{Groote:2013xt}, and the corrections to the lepton and neutrino analyzing powers~\cite{Czarnecki:1990pe,Czarnecki:1994pu}.  Further cross-checks of the real emission differential decay rates have also been performed against {\tt MadGraph5}.  Interestingly, we obtain small but significant disagreements with the numerically-calculated NLO quark and jet analyzing powers of~\cite{Brandenburg:2002xr} when using their parameter choices and reconstruction logic.  For example, they predict a bare up-quark analyzing power of $-0.3167\pm0.0006$, whereas we predict $-0.2927\pm0.0007$.  For the soft-jet analyzing power defined with the Durham $k_T$ algorithm, they predict $0.4734\pm0.0007$ whereas we predict $0.4592\pm0.0007$.}  We compare this to leading-order simulations in {\tt MadGraph5}~\cite{Alwall:2011uj}, of $e^+\nu_e \to t \bar b \to (bW^+)\bar b \to (b(f\bar f')) \bar b$ at threshold, with the kinematic width of the top quark set to ``zero.''  These samples are then passed through {\tt PYTHIA6} with a few restrictions: no QED radiation (including no ISR), no hadronization, a veto on events with $g\to b\bar b$ splittings, and stable $b$-quarks.  The tops in the {\tt MadGraph5} samples are already 100\% polarized along the $z$-axis, but to make closer contact with our procedures below, we optionally randomize the orientation of the events and reweight by $1+\cos\theta_{d/l}$.  We find that results obtained with/without this additional step are statistically consistent with each other.  Both simulations set $m_t = 172$~GeV, $m_b = 4.7$~GeV, $m_W = 80.4$~GeV, and $\Gamma_W = 2.08$~GeV.  The NLO simulation uses a fixed $\alpha_s = 0.108$, whereas the shower uses an internal running $\alpha_s$.

It is instructive to first consider the corrections to semileptonic decay.  For the lepton itself, it is well-known that the the radiative corrections are extremely small, tallying to roughly $-0.001$~\cite{Czarnecki:1990pe}.\footnote{This is due to two facts.  First, an analyzing power of unity is an extremum.  In particular, it is stable at linear order to perturbations in the Born amplitudes, which means that the $O(\alpha_s)$ Born-virtual interference correction vanishes.  Second, the leading real emission diagram (using purely transverse external gluon polarizations), where the gluon is emitted off of the $b$-quark, exhibits the same maximal correlation between the top spin and lepton direction as is found in the leading-order diagram, independently of the detailed 4-body kinematics.  The only nonzero correction to $\kappa_l$ at $O(\alpha_s)$ comes from the square of the subleading real emission diagram where the gluon is attached to the top.  The correction from interference with the leading emission diagram also vanishes, again due to the extremization.}  In the parton shower approach, the lepton receives a small kinematic adjustment as the $W$ and showered bottom systems are boosted along the top decay axis to conserve 4-momentum.  The net effect on the analyzing power is nonetheless $O(0.1\%)$ or smaller, effectively in agreement with the NLO calculation.  The neutrino and the ``$b$-jet''/$W$-boson axis (built either from all recoiling quarks/gluons or from the lepton and neutrino) receive relatively much larger corrections at NLO: respectively about $+0.01$ and $-0.01$ in absolute magnitude.  The shower approximately reproduces the former upward shift, suggesting that it is indeed mainly a recoil effect.  However, by construction, the shower cannot change the momentum orientation of the radiating bottom system, and therefore predicts exactly zero shift in $b$/$W$ analyzing power.

Besides missing this small desensitization of the $b$/$W$ axis to the top polarization, the kinematics of the radiation should be correctly modeled up to $O(\alpha_s)$ by the parton shower when integrated over $W$ decay orientations, since {\tt PYTHIA6} automatically incorporates basic matrix-element matching in heavy particle decays~\cite{Norrbin:2000uu}.  Because these corrections are incoherently factorized between the $t\to Wb$ and $W \to f \bar f'$ decay steps, they lose any angular correlations between the radiation pattern of the first step and the decay orientation of the second step.  Such effects are suppressed in the soft/collinear regions of phase space that dominate the emission rate, but it is easy to imagine that their omission could lead to further percent-scale mistakes when we move on to proper jet reconstruction.  Analogous considerations apply to the parton shower initiated within the $W$ decay.

Before considering the fully hadronic decay, it is also possible at this stage to apply the optimal hadronic polarimeter construction, using the lepton and neutrino as proxies for the down- and up-quarks.  The NLO power drops slightly from the leading-order prediction, by about $0.005$.  (The parton shower exhibits an even smaller drop.)  The smallness of this correction is largely attributable to the fact that the lepton analyzing power is nearly unaffected and that the $W$ polarization state is only corrected at the percent-level~\cite{Fischer:2000kx}.  Using the NLO-corrected helicity fractions in the construction of $\hat q_{\rm opt}$, instead of the leading-order ones, has negligible impact.  It is therefore adequate to continue to use the leading-order helicity fractions given in Eq.~\ref{eq:helicityFractions} (which also justifiably neglect the bottom mass and $W$ width).

In order to study the effects of QCD corrections on the fully hadronic decay, we must introduce a jet algorithm and reconstruction cuts.  We consider three approaches:  1)~cluster into a 3-body configuration using the Durham $e^+e^-$ $k_T$ measure, as was done in in the foundational work on this topic~\cite{Brandenburg:2002xr}; 2)~cluster with the ``anti-Durham'' algorithm, the $e^+e^-$ analog of anti-$k_T$~\cite{Cacciari:2008gp}, with an angular-radius parameter of $R=0.7$ and keeping only the three most energetic jets; and 3)~a Cambridge/Aachen-based jet substructure procedure inspired by the HEPTopTagger~\cite{Plehn:2010st} (described in full detail in Section~\ref{subsec:resonance}), applied to tops that have been boosted up to 1~TeV transverse momentum.  The jet that contains the $b$-quark is tagged as the $b$-jet.  For approaches (2) and (3), we only keep events where the $b$-quark is clustered into one of the utilized jets/subjets.  We further demand that the reconstructed top mass is greater than 130~GeV and that the ratio between reconstructed $W$ and top masses lies in the window $[50,110]~{\rm GeV}/m_t$.  For the NLO simulations, we use the definition of the analyzing powers given in~\cite{Brandenburg:2002xr}, with the overall $1/\Gamma(t\to u\bar d b(g))$ normalization factor expanded to $O(\alpha_s)$, and a similar fixed-order definition for the reconstruction rate.  (The differences with respect to simple ratios are anyway sub-percent.)  For approach (3), where the induced reconstruction biases are not rotationally-symmetric in the top's rest frame, and the polar angle distributions of the various spin analyzers are no longer simple linear functions, we use forward-backward asymmetries (multiplied by two).  The differences between asymmetries in polarized and unpolarized samples serve as simple estimates of the leading-order and NLO analyzing powers.

\begin{table}
\begin{center}
\begin{tabular}{ l|ccc|ccc|ccc}
                   & \multicolumn{3}{c|}{3-Body Durham}& \multicolumn{3}{c|}{Anti-Durham $R=0.7$}& \multicolumn{3}{c}{C/A Substructure}  \\ 
                   &  \ \ \ LO \ \ \ & \ \  NLO \ \  &  shower \ & \ \ \ LO \ \ \ & \ \  NLO \ \  &  shower \ & \ \ \ LO \ \ \ & \ \  NLO \ \  &  shower \\ \hline
reco rate          &      1.000      &     1.000     &   1.000   &      0.967     &     0.923     &   0.890   &     0.905      &     0.862     &   0.836 \\ \hline
optimal hadronic \ &      0.638      &     0.574     &   0.583   &      0.630     &     0.594     &   0.610   &     0.617      &     0.578     &   0.591 \\ 
soft-jet           &      0.505      &     0.452     &   0.464   &      0.492     &     0.465     &   0.484   &     0.489      &     0.459     &   0.477 \\
$b$-jet            &      0.394      &     0.375     &   0.381   &      0.426     &     0.411     &   0.420   &     0.423      &     0.403     &   0.410 \\
\end{tabular}
\end{center}
\caption{Unpolarized event reconstruction efficiencies and analyzing powers of the different hadronic spin analyzers within simple leading-order, NLO, and parton-showered simulations of single top quarks.  The three different reconstructions labeling the upper column headings are described in the main text.  (The minus sign on the $b$-jet power is omitted.  Absolute monte carlo statistical errors on all numbers are 0.001 or smaller.)}
\label{table:NLO}
\end{table}

Table~\ref{table:NLO} contains the results of these comparisons.  Three features are notable.  First, the radiative corrections always reduce the analyzing powers, by as much as 10\% relative to their leading-order values.  Second, the ratios of the analyzing powers stay much more stable.  In particular, the optimal polarimeter is 25--30\% more powerful than the soft-jet for all simulations and all reconstructions.  Third, the parton shower always predicts slightly higher powers than what is obtained at fixed-order NLO, typically by 0.01--0.02.  While there is certainly some residual $O(\alpha_s^2)$ uncertainty on the NLO prediction, the consistently smaller corrections exhibited by the shower are suggestive, especially since it actually uses {\it larger} values of $\alpha_s$ (evaluated at the $p_T$ scales of parton branchings rather than at $m_t$).  It therefore seems quite possible that the shower is underestimating the full corrections.  However, the magnitude of that underestimate is small in an absolute sense, and the very good stability of the ratios of analyzing powers suggests that the parton shower is trustworthy for determining the relative performances of different polarimeters.

\section{Realistic Examples}
\label{sec:measurement}

There are many contexts in which a more efficient hadronic top quark polarimeter may prove useful in characterizing or searching for new physics.  Besides the fact that hadronic top decays dominate the branching fraction, events with at least one hadronic top often give us better resolution on the production kinematics, and their full kinematic reconstruction is unaffected by additional injections of $\met$ such as from neutralinos.  However, as emphasized above, realistic analyses with hadronic tops must contend with the added complications of QCD showering and hadronization.  Besides making individual light-quark identifications extremely difficult, these effects can smear out the measured decay kinematics.  This is in turn compounded by smearings intrinsic to the detectors and combinatoric ambiguities with other parts of the event.  In addition, basic kinematic cuts, crucial to ensure that the individual jets or subjets are even identifiable, can heavily resculpt the observed decay distributions.  Therefore, it behooves us to take a closer look at how our optimal hadronic polarimeter fares under such harsh conditions.

In the following subsections, we illustrate the robustness of the optimal hadronic polarimeter relative to other hadronic polarimeters within three examples of new physics affecting $t\bar t$ production in the $l$+jets channel.  The first is a set of 2.5~TeV spin-1 $t\bar t$ resonances, producing boosted tops with $p_T \sim 1$~TeV.  The couplings can be varied to exhibit purely polarized tops of either chirality, or unpolarized tops with characteristic spin correlation patterns.  The second example is chiral stop pair production, with masses near the current experimental lower limit, and producing pairs of polarized semi-boosted tops.  The third example is the introduction of chromomagnetic and/or chromoelectric dipole moment operators, which imprint themselves as (possibly CP-violating) spin correlations in the $t\bar t$ continuum.

The goal here is not to perform complete phenomenological studies, but to compare potential polarimetry performance.  Consequently, we do not include full categorizations of backgrounds, which are anyway dominantly top-like, and by default do not include pileup (though see below).  We perform one subset of studies at particle-level (after showering, hadronization, and hadron decays), and one with a simplified and somewhat pessimistic detector model, in order to try to bracket realistic performance.  The detector model is similar in spirit to {\tt Delphes}~\cite{Ovyn:2009tx,deFavereau:2013fsa}.   All non-leptonic particle energy is deposited in a $0.1\times 0.1$ granularity ``calorimeter'' in $\eta$-$\phi$ space, extending out to $|\eta| = 4.0$.  Photon energy is deposited into an ECAL, and fractionally smeared cell-by-cell as $(0.05$~GeV$^{1/2})/\sqrt{E} \oplus (0.25$~GeV$)/E \oplus 0.0055$.  Hadronic energy is deposited into an HCAL, and fractionally smeared as $(1.5$~GeV$^{1/2})/\sqrt{E} \oplus 0.05$.  (These calorimeter cell energy resolutions are taken from~\cite{Ovyn:2009tx}.)  Missing energy $x$ and $y$ components are individually smeared by $(0.7$~GeV$^{1/2})\times\sqrt{H_T}$, where $H_T$ is the sum over all visible transverse event activity.  Leptons are treated as perfectly measured.  With this detector model, possible benefits of particle/energy flow are not exploited, nor is the true segmentation of the ECAL.

At the future LHC, pileup will become a major issue, and we may wonder to what extent the hadronic observables discussed in this paper can still be faithfully reconstructed.  Some pileup removal strategy should be performed in reality, such as trimming~\cite{Krohn:2009th}, jet cleansing~\cite{Krohn:2013lba}, or one of any number of new techniques that continue to be developed.  In particular, both a recent ATLAS substructure study~\cite{Aad:2013gja} and the Snowmass 2013 study on boosted top quarks show the significant benefits of trimming individual top-jets~\cite{Calkins:2013ega}, and a recent study of boosted RPV stop substructure demonstrates a successful application of pre-trimming the entire event~\cite{Bai:2013xla}.  We have cross-checked all of the analyses below in a scenario with 140 overlayed pileup events on average,\footnote{To model the min-bias events constituting the pileup, we use {\tt PYTHIA~8.1}~\cite{Sjostrand:2007gs} tune 4C.  Poissonian fluctuations about the mean number of pileup interactions are included.} and then subtracted using a combination of (perfect) charged hadron subtraction and event-wide trimming with $R=0.2$ anti-$k_T$ jets with a fixed acceptance threshold of 25~GeV.  This simple approach by itself is adequate to largely preserve the pileup-free performance.  The lasting effects are 5--10\% losses in overall reconstruction efficiency and percent-scale weakenings of polarization sensitivity.  We take this as good evidence that, regardless of what pileup removal approaches will ultimately prove to be the most powerful, the polarization of high-$p_T$ hadronic tops should remain visible.

\subsection{Boosted Tops from Multi-TeV Resonances}
\label{subsec:resonance}

One of the simplest new phenomena involving top quarks would be a resonance in the $t\bar t$ invariant mass spectrum.  These arise in numerous models, ranging from a simple $U(1)$ extension of the gauge sector (reviewed in~\cite{Langacker:2008yv}) to theories with a partially composite electroweak sector (e.g.,~\cite{ArkaniHamed:2001nc,Agashe:2003zs}).  The Tevatron and LHC have already conducted many dedicated searches (including~\cite{Chatrchyan:2013lca,TheATLAScollaboration:2013kha,Aad:2012raa}), and current limits on several models extend up to about 2~TeV.  With the LHC poised to roughly double in energy, much higher-mass resonances will become visible.  Optimistically assuming that a resonance with large $S/B$ lies just around the corner, we study the spins and spin correlations of top pairs produced from a 2.5~TeV spin-1 resonance in the $l$+jets channel.  We implement this model by first generating SM $q\bar q \to t\bar t \to (l\nu b)(jjb)$ events at the 14~TeV LHC with {\tt MadGraph5} and {\tt PYTHIA6}~\cite{Alwall:2011uj,Sjostrand:2006za} in the invariant mass range $[2400,2600]$~GeV.  We reweight event-by-event with the 6-body matrix elements of the singly-produced resonance.  We set $\Gamma/M = 20\%$, so that most events contribute with similar weight.

We consider four variations on this model:  chiral right-handed couplings, chiral left-handed couplings, vector couplings, and axial-vector couplings.  The chiral models produce tops in essentially fixed helicity states.  The vector and axial-vector models produce tops with zero net polarizations, but with characteristic spin correlations.

Since the resonance mass is far heavier than the top mass, the tops generated in the decay are relativistic, and approaches of jet substructure are appropriate.  As a first step in global event reconstruction, and before applying any calorimeter model, we identify {\it mini-isolated} leptons in the event~\cite{Rehermann:2010vq}.  Mini-isolation works similar to normal isolation, but tallies only nearby track energy and uses a cone that shrinks with the lepton $p_T$.  Here, we take $R_{\rm iso} = \min\big((15$~GeV$)/p_T(l),0.4\big)$.  The sum of the transverse energy of all charged particles inside the cone must be dominated by the lepton:  $p_T(l)/p_T(\rm cone) > 90\%$.  (Leptons that fail this criterion are reclassified as ``hadrons.'')  The event must have one exactly mini-isolated lepton with $p_T(l) > 30$~GeV and $|\eta(l)| < 2.5$.  We then cluster the other particles or calorimeter cells in the event using the anti-$k_T$ algorithm~\cite{Cacciari:2008gp} with $R = 0.45$ in {\tt FastJet}~\cite{Cacciari:2005hq}.  At this stage, we kinematically identify the $b$-jet associated with the lepton by iterating over all jets with $p_T(j) > 50$~GeV and $|\eta(j)| < 2.5$, and keeping the hardest one that satisfies $m(bl) < 200$~GeV.  We do not insist that this jet carry a $b$-tag.

The remaining particles or calorimeter cells in the event are then reclustered into {\it fat-jets} with the Cambridge/Aachen algorithm~\cite{Dokshitzer:1997in} with $R=1.2$, keeping fat-jets with $p_T($fat-jet$) > 300$~GeV and $|\eta($fat-jet$)| < 2.5$.  The hardest identified fat-jet serves as our hadronic top-jet candidate.  There now exist many ways to process a top-jet back into a full parton-level picture of the decay (for reviews, see~\cite{Abdesselam:2010pt,Altheimer:2012mn,Altheimer:2013yza}).  We have specifically studied the behavior of the JHU top-tagger~\cite{Kaplan:2008ie} and the HEPTopTagger~\cite{Plehn:2010st}.  One of the main ways in which the two approaches differ is on the type cutoff used for defining subjets: relative $p_T$ for the former and absolute mass for the latter.  The HEPTopTagger also has $m_t$ built into its method for choosing which subjets are usable.  Both, as it turns out, can be improved, at least as far as the accuracy with which they map subjets into quarks at high top boost, and we propose using modified versions to maximize the quality of polarimetry.  Having considered novel modifications to each tagger, we present here our results with a modified HEPTopTagger.  We find that this yields $\sim$10\% better spin sensitivity than JHU due to a higher efficiency for picking up relatively soft quarks, which the JHU tagger tends to remove (at least given the settings we have chosen).  However, it should be noted that saving softer subjets for analysis could become difficult in samples contaminated by non-top backgrounds and/or pileup.  More detailed discussions of the effects of our modifications, and of the JHU tagger, can be found in Appendix~\ref{sec:appendix}.  In particular, our results with a modified JHU tagger, though somewhat more biased by the declustering criteria, exhibit very similar relative performances between the different polarization-sensitive variables considered below.

The HEPTopTagger works by recursively declustering a top-jet, shedding diffuse radiation along the way, until it resolves structures below some mass threshold.  The original algorithm, tailored to semi-boosted tops with $p_T \sim m_t$, invokes an additional filtering~\cite{Butterworth:2008iy} step to further reduce contamination.  The subjet triplet whose filtered mass is closest to $m_t$ is kept as the top candidate.  Its surviving constituents are reclustered back into three subjets, which serve as the proxies for the original quarks, and these can be fed into a set of multibody kinematic cuts to help discriminate against backgrounds.  Our observations in the highly-boosted regime under consideration here suggests that the original approach is on the one hand too aggressive at removing radiation, and on the other hand is susceptible to merging together two quarks into one subjet while creating additional spurious soft subjets.  However, this behavior can be improved with the following modified algorithm: 
\begin{enumerate}
\item  Recursively decluster the C/A fat-jet, as in the original HEPTopTagger, until we resolve structures with $m < 30$~GeV.  Do {\it not} apply a mass-drop criterion.  No radiation is thrown away.\footnote{As discussed above, an initial pileup removal step, such as charged hadron subtraction and event-wide trimming, would allow this procedure to survive in a high-pileup environment.}
\item  There must be at least three subjets to continue.  If there are more than three, consider only the hardest four in $p_T$.  Do {\it not} apply any filtering or reclustering.
\item  If a 4th-hardest subjet is present but is softer than the 3rd-hardest by a factor of more than 3, ignore it.
\item  Attempt to reconstruct the top using the hardest two subjets in combination with either the 3rd-hardest or 4th-hardest, if the latter exists and is usable according to the above criterion.  The choice that gives a mass closer to $m_t$ is used.
\item  Apply any desired multibody kinematic cuts to these three ``quarks.''  (See below.)
\end{enumerate}
For the most part, this is a simplification of the original method, though one new discrete parameter and one new continuous parameter have been introduced:  respectively, the number of subjets that we consider for the top reconstruction and the allowable relative $p_T$ threshold between the 4th-hardest and 3rd-hardest.  These exist primarily to deal with the confusions presented by FSR/ISR subjets, which can often exceed the $p_T$ of the softest quark from the top decay, and can serve as impostors by combining with the two hardest subjets to form an object with mass close to $m_t$.  

\begin{figure*}[tp!]
\begin{center}
\includegraphics[width=0.44\textwidth]{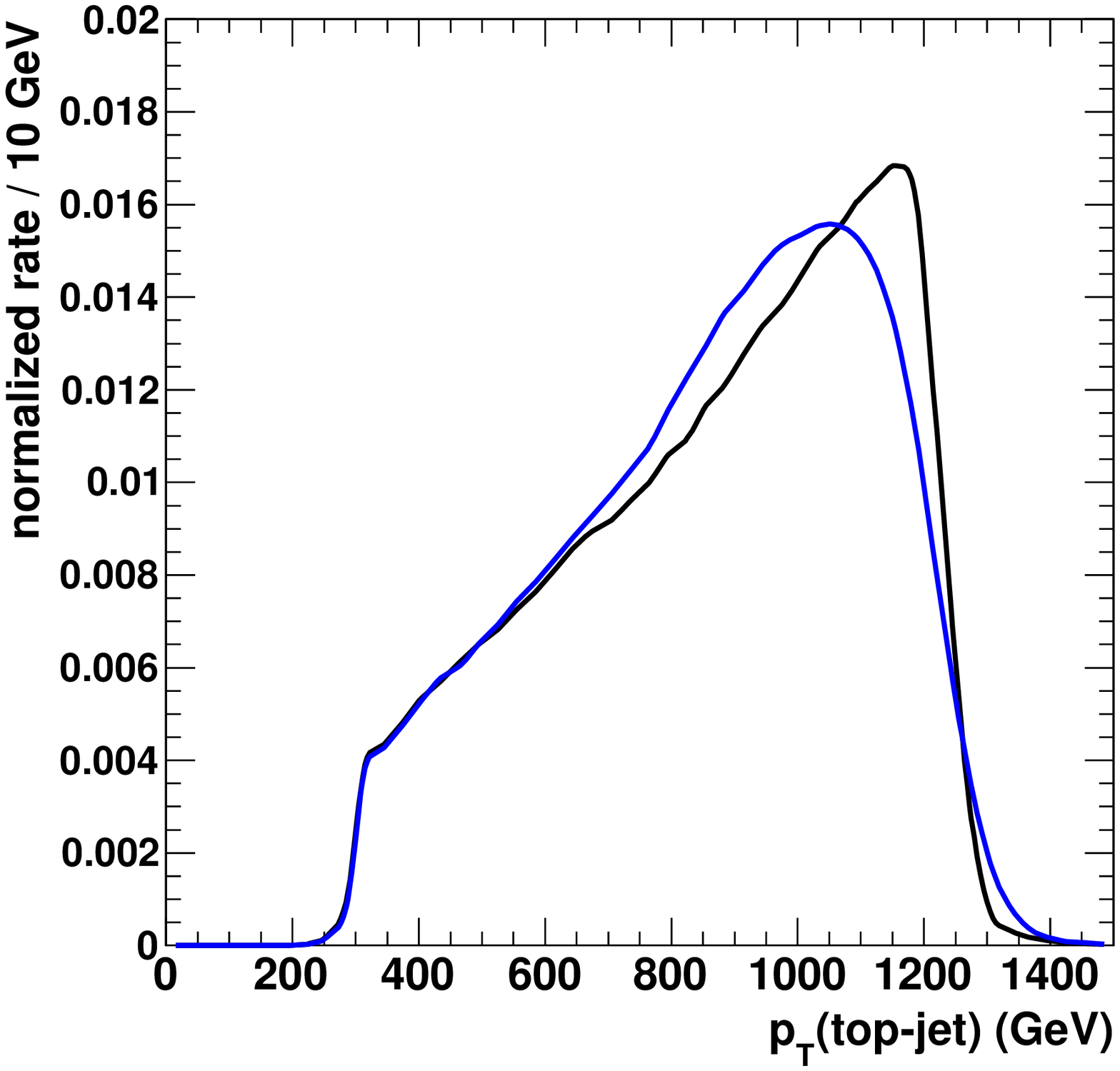}
\includegraphics[width=0.44\textwidth]{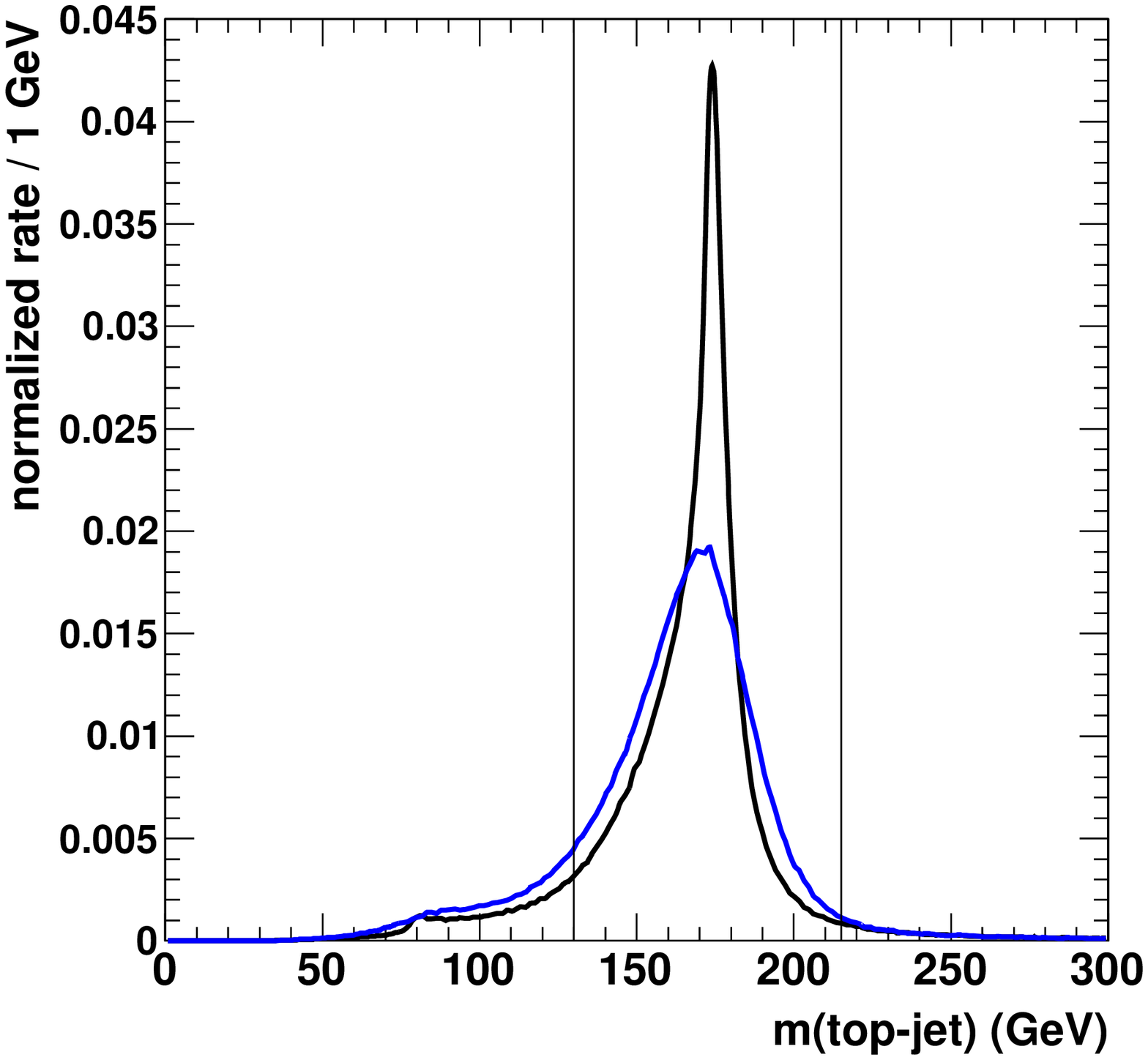}
\caption{Distributions of reconstructed top-jet $p_T$ (left) and mass (right) after declustering.  Black is particle-level, and blue is calorimeter-level.  No $b$-tags have been applied.  Mass window cuts are indicated by vertical lines.}
\label{fig:boosted_tops}
\end{center}
\end{figure*}

The vast majority of fat-jets successfully decluster into at least three subjets in this manner.  To ensure good-quality reconstruction, the final system must satisfy a top mass window constraint $m($top-jet$) = [130,215]$~GeV.  The pass rate for this cut is nonetheless substantial: 85--90\%.  The top-jet $p_T$ and mass distributions, with and without detector effects, are shown in Fig.~\ref{fig:boosted_tops}.

\begin{figure*}[tp!]
\begin{center}
\includegraphics[width=0.44\textwidth]{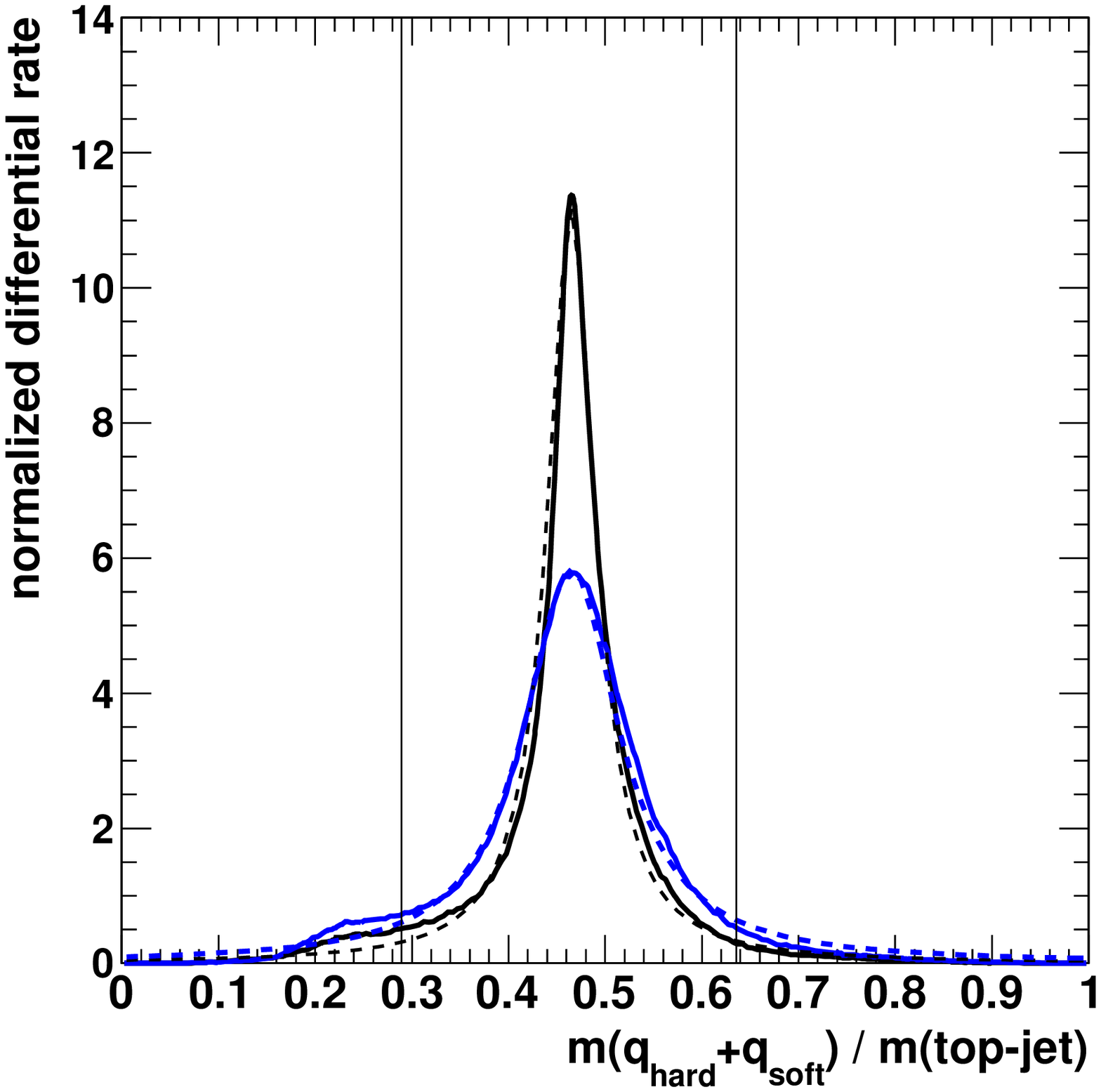}
\includegraphics[width=0.44\textwidth]{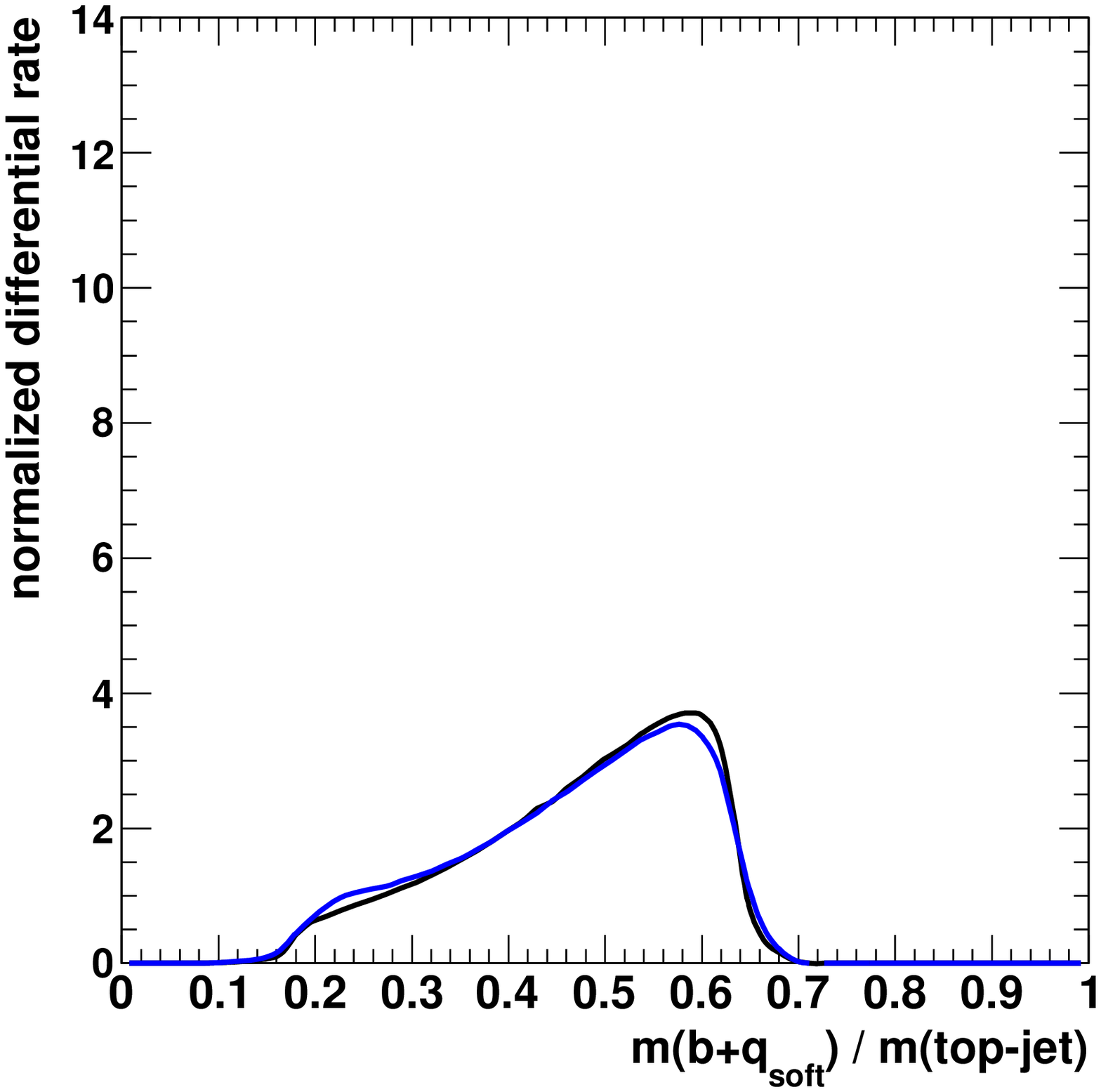}
\includegraphics[width=0.44\textwidth]{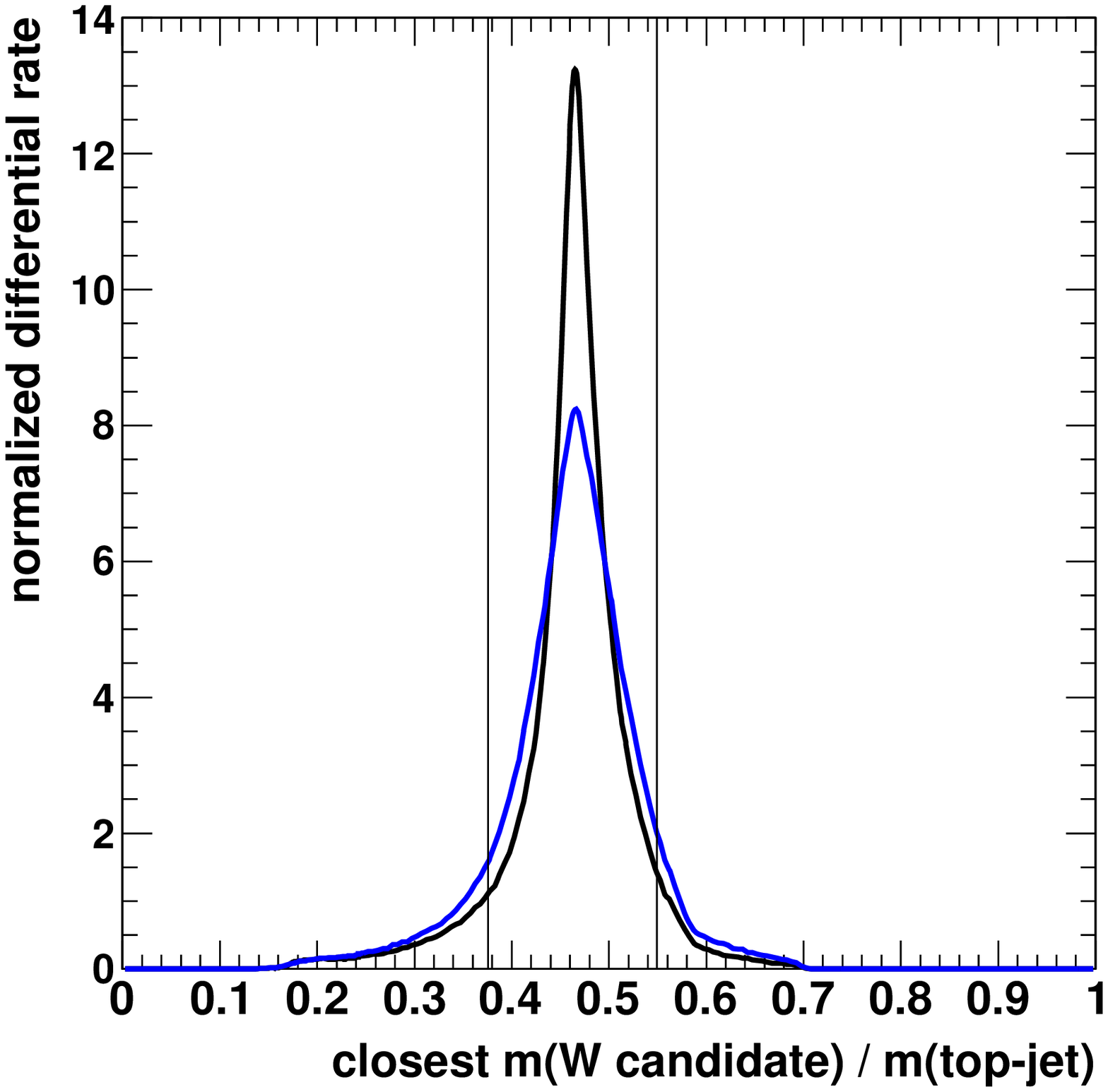}
\caption{Distributions of reconstructed $W$ boson mass relative to top-jet mass using $b$-tag subjet identification (top left), ``wrong pairing'' $b$+$q_{\rm soft}$ relative mass (top right), and subjet-pair relative mass closer to true $m_W/m_t$ without a $b$-tag (bottom).  Black is particle-level, and blue is calorimeter-level.  Dashed lines in the first plot are our Breit-Wigner parametrizations for the untagged $W$-candidate superposition method.  Mass window cuts are indicated by vertical lines.}
\label{fig:boosted_Ws}
\end{center}
\end{figure*}

We are then left with the task of identifying the $b$-quark amongst the three subjets, and making sure that the $W$ is correctly reconstructed.  We explore two extreme versions of this:  $b$-tagging either works perfectly, or we are left to identify the $b$-subjet using pure kinematics.  For the tagged analysis, we associate to each prompt $b$-flavored hadron in the event the closest subjet.  We tag any subjet with an associated $b$-hadron that is closer than the next-closest subjet.  Roughly 97\% of our top-jets contain one $b$-subjet identified in this manner.  For the untagged analysis, we assume that the softest subjet in top-frame is from the $W$, and further subdivide our approaches by either making a binary choice for the second $W$-subjet or using the superposition method outlined in Sec.~\ref{sec:untagged}.  In both of these, we normalize out the overall top-jet mass and concentrate on obtaining dimensionless masses near $m_W/m_t \simeq 0.46$.  To make a binary choice, we consider the two possible pairings that contain the softest subjet, and pick the one whose dimensionless mass comes closer to this ratio.  To instead perform a superposition of both choices, we assign relative probabilities based on an assumed Breit-Wigner profile with center at $m_W/m_t$ and width of 0.06 (0.12) for the particle-level (calorimeter-level) analysis.  In all cases, we further constrain the kinematics by imposing a cut on the reconstructed $W$/top mass ratio.  For the $b$-tagged analysis, the ratio must be in the range $[50,110]~{\rm GeV}/m_t$.  For the untagged analyses, the candidate ratio closer to $m_W/m_t$ must be in the range $[65,95]~{\rm GeV}/m_t$.  The efficiencies to pass these cuts are 80--90\%.  The cuts are only used here to reduce the effects of outlier events, though they would also serve to purify out backgrounds if relevant.  We show the associated kinematic distributions in Fig.~\ref{fig:boosted_Ws}.

\begin{figure*}[tp!]
\begin{center}
\includegraphics[width=0.44\textwidth]{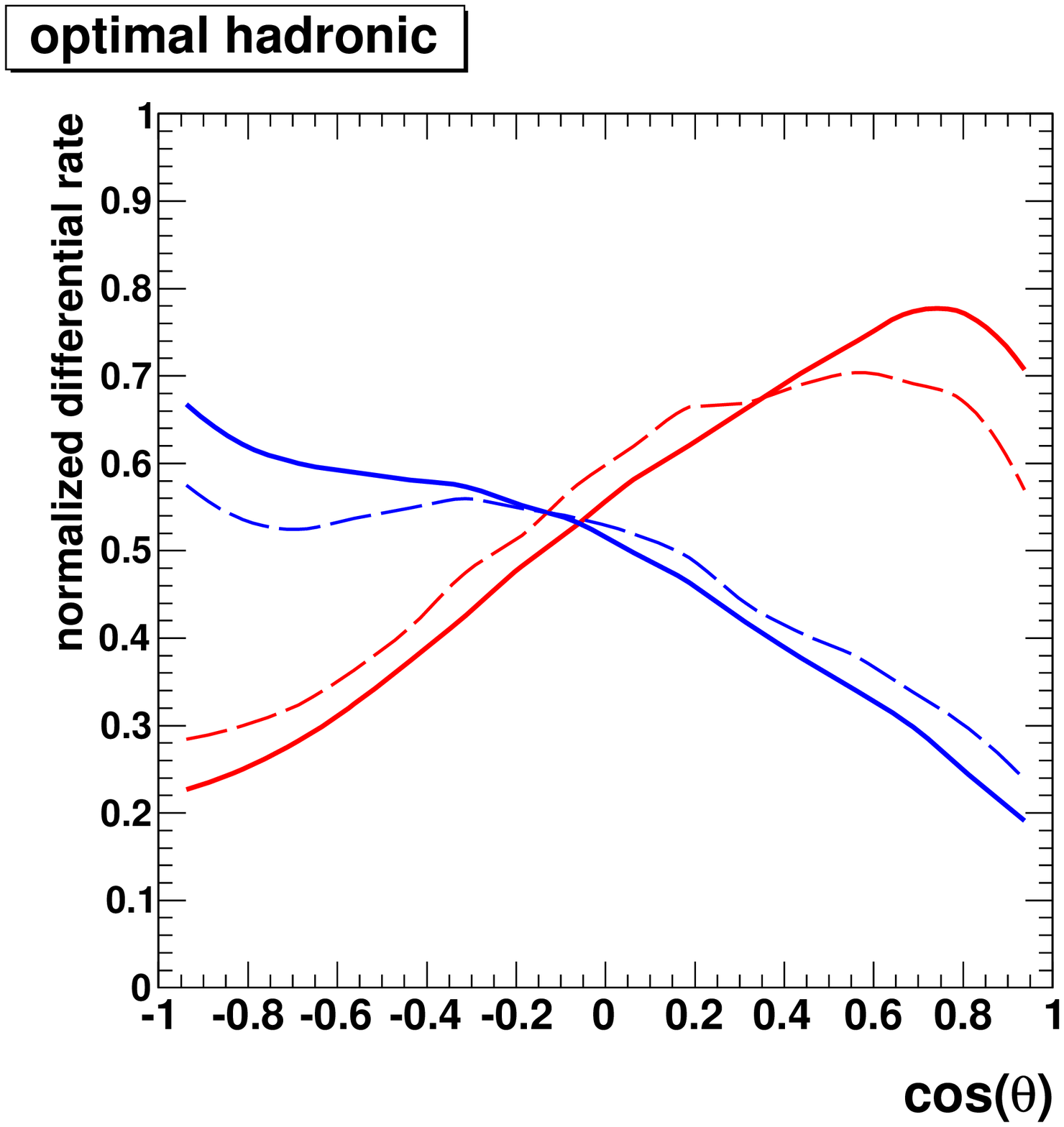}
\includegraphics[width=0.44\textwidth]{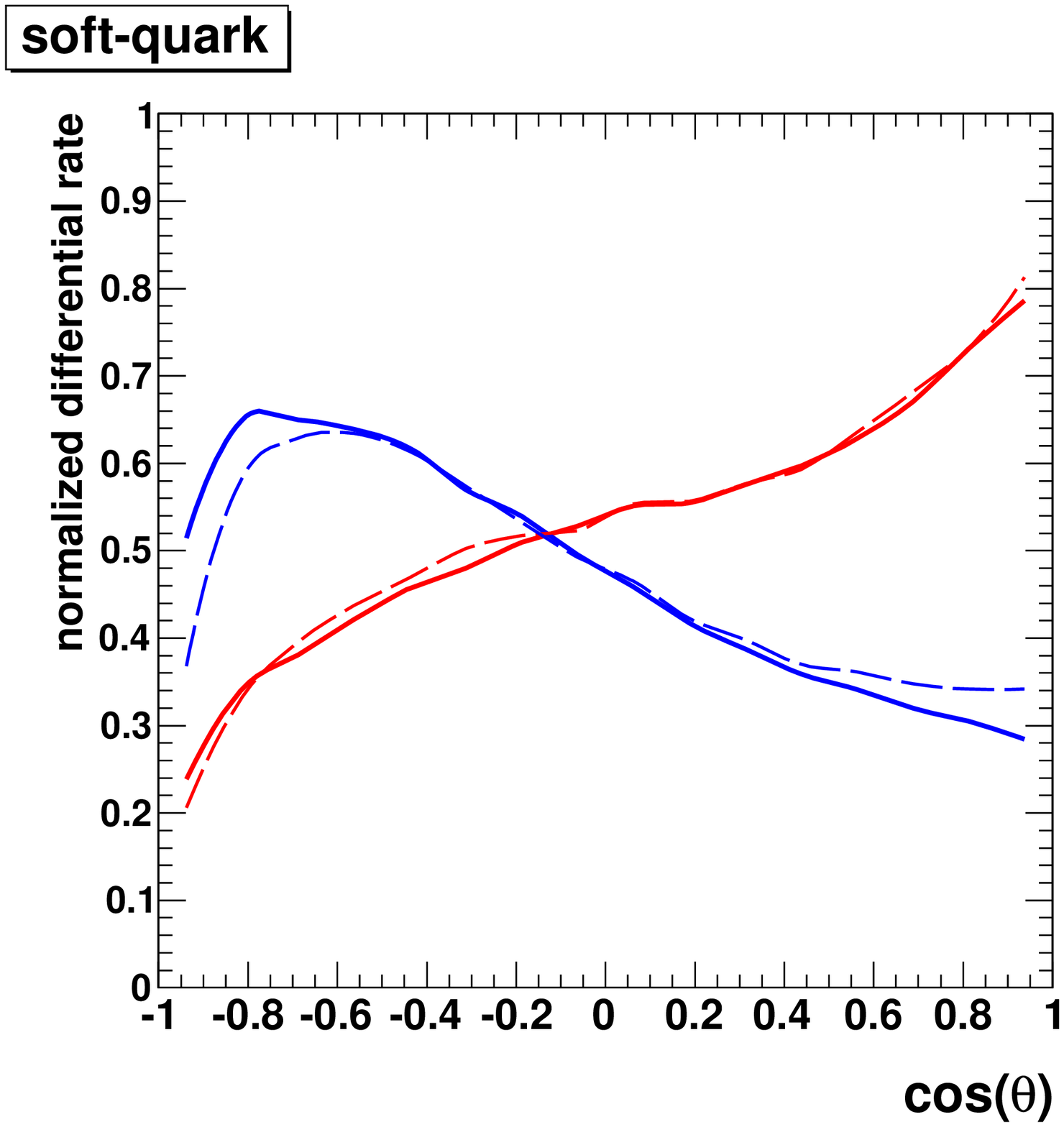}
\includegraphics[width=0.44\textwidth]{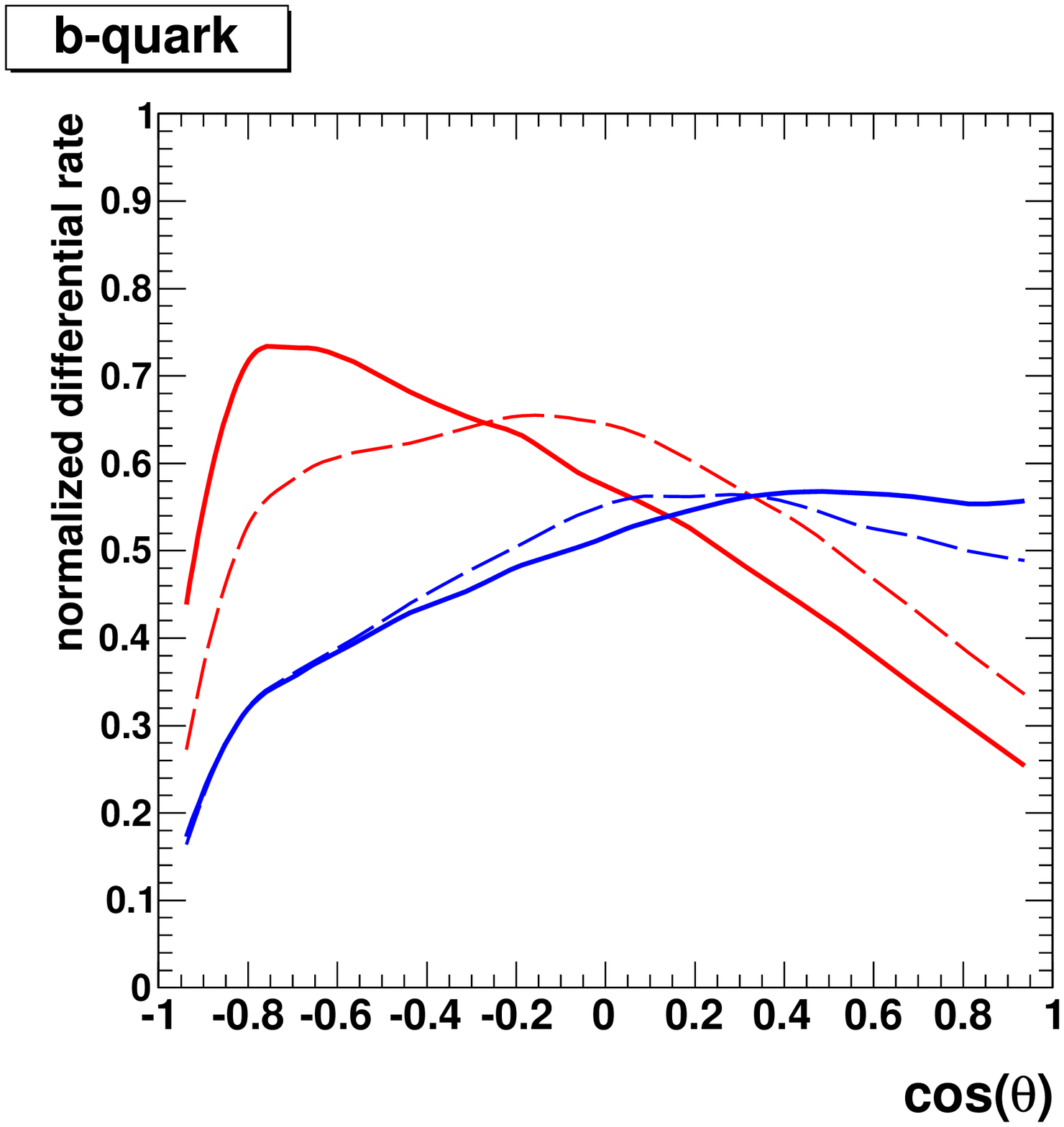}
\includegraphics[width=0.44\textwidth]{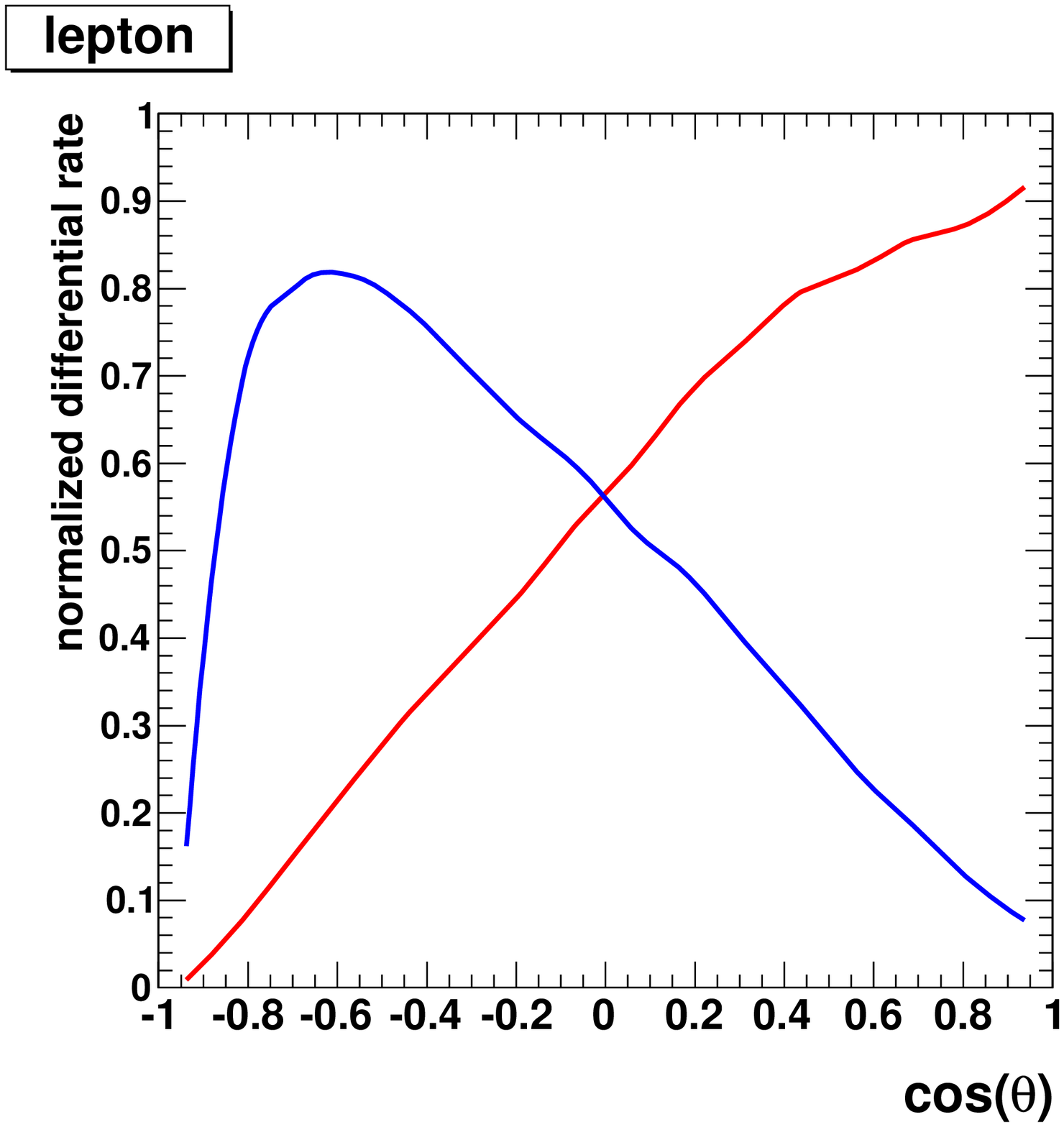}
\caption{Examples of reconstructed polar decay angle distributions for different top-jet spin analyzers:  optimal hadronic polarimeter (top left), softer light-quark (top right), $b$-quark (bottom left), and lepton from the semileptonic side of the event (bottom right).  Red indicates right-handed chirality, and blue indicates left handed chirality.  Solid is our most optimistic reconstruction:  particle-level with $b$-tags.  Dashed is our most pessimistic reconstruction:  calorimeter-level with the $W$ reconstructed kinematically using the binary choice method.  The chiralities are normalized according to their relative global reconstruction efficiencies, and such that they average to unity.}
\label{fig:boosted_cosTh}
\end{center}
\end{figure*}

\begin{figure*}[tp!]
\begin{center}
\includegraphics[width=0.44\textwidth]{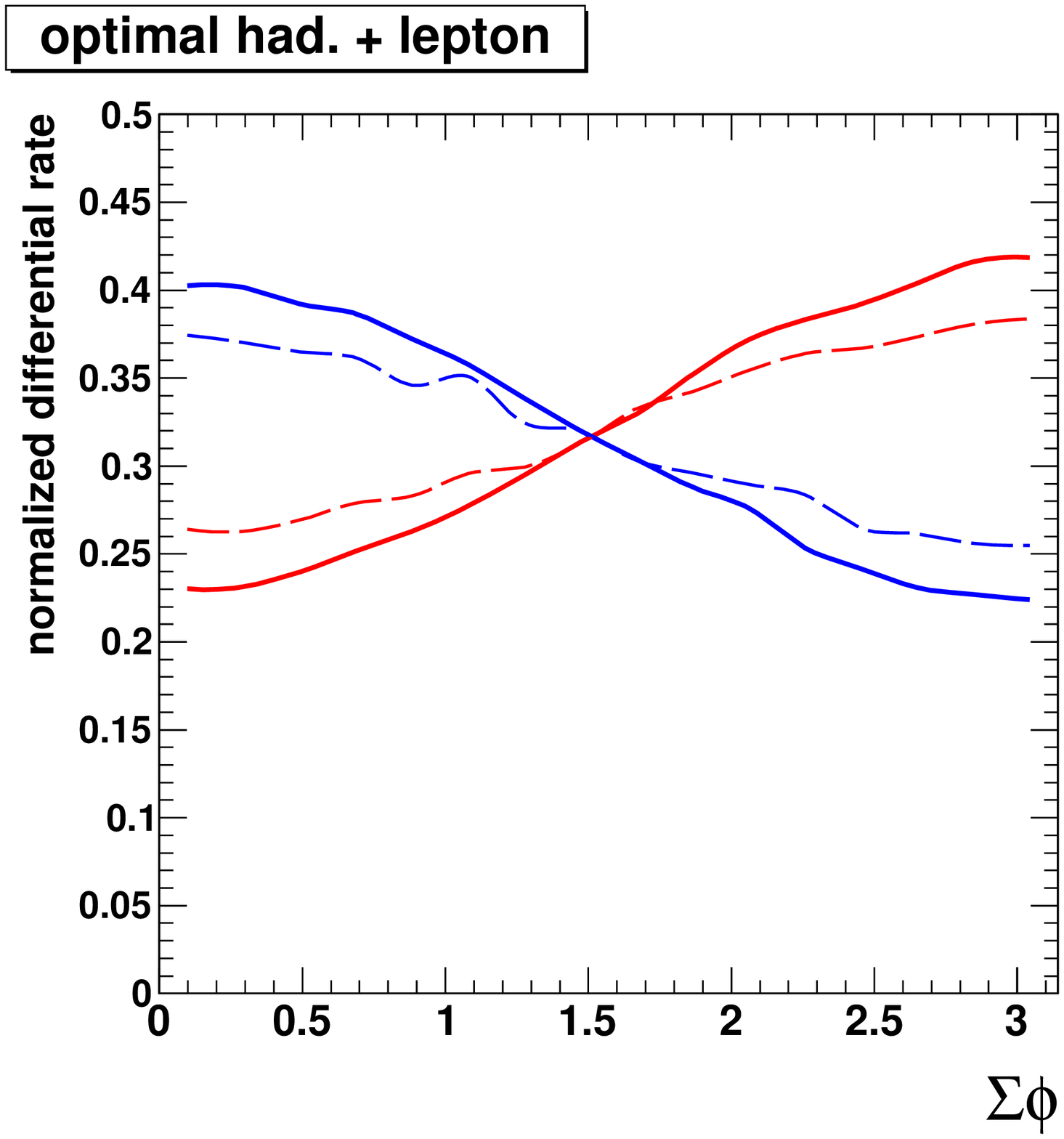}
\includegraphics[width=0.44\textwidth]{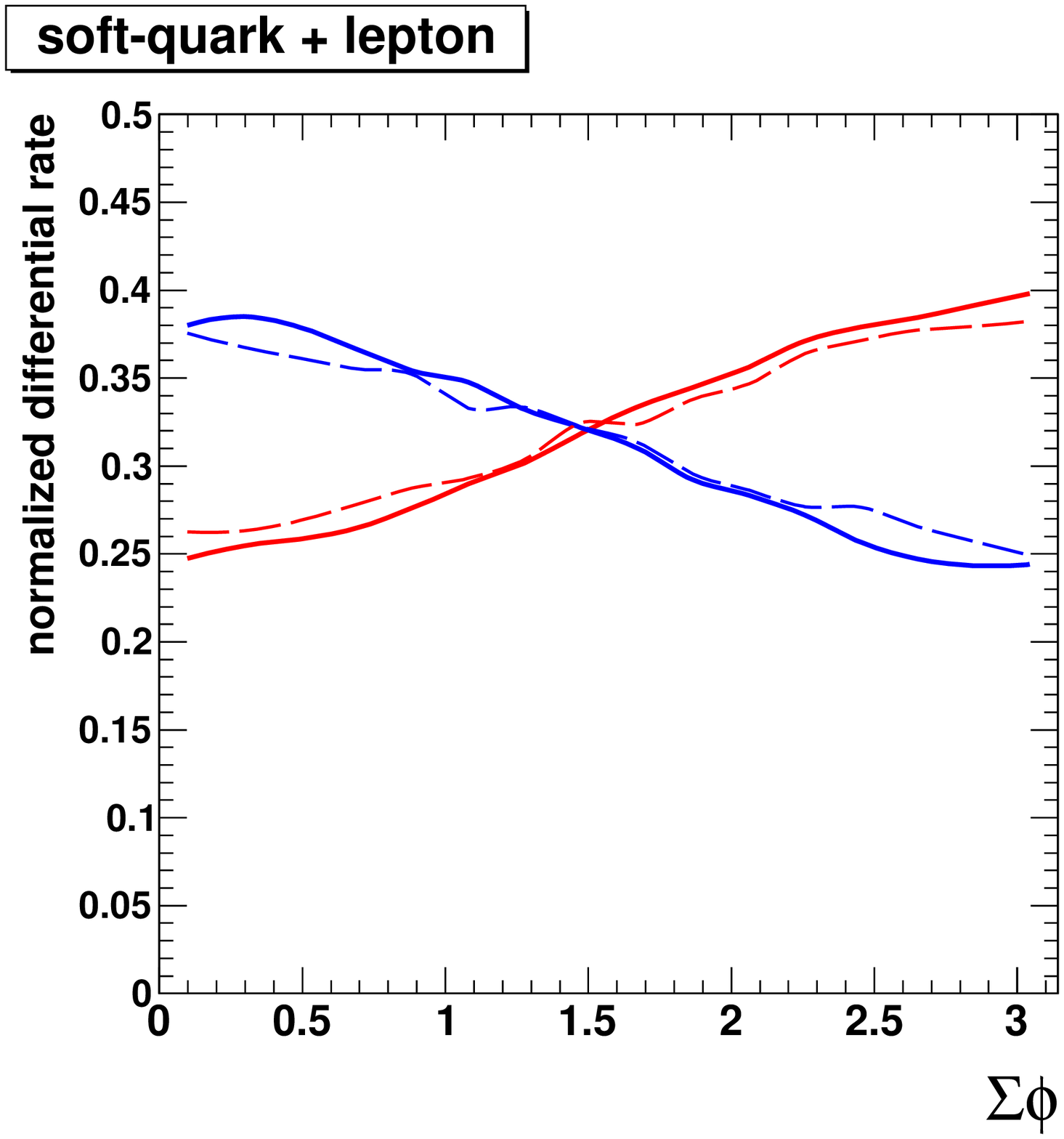}
\includegraphics[width=0.44\textwidth]{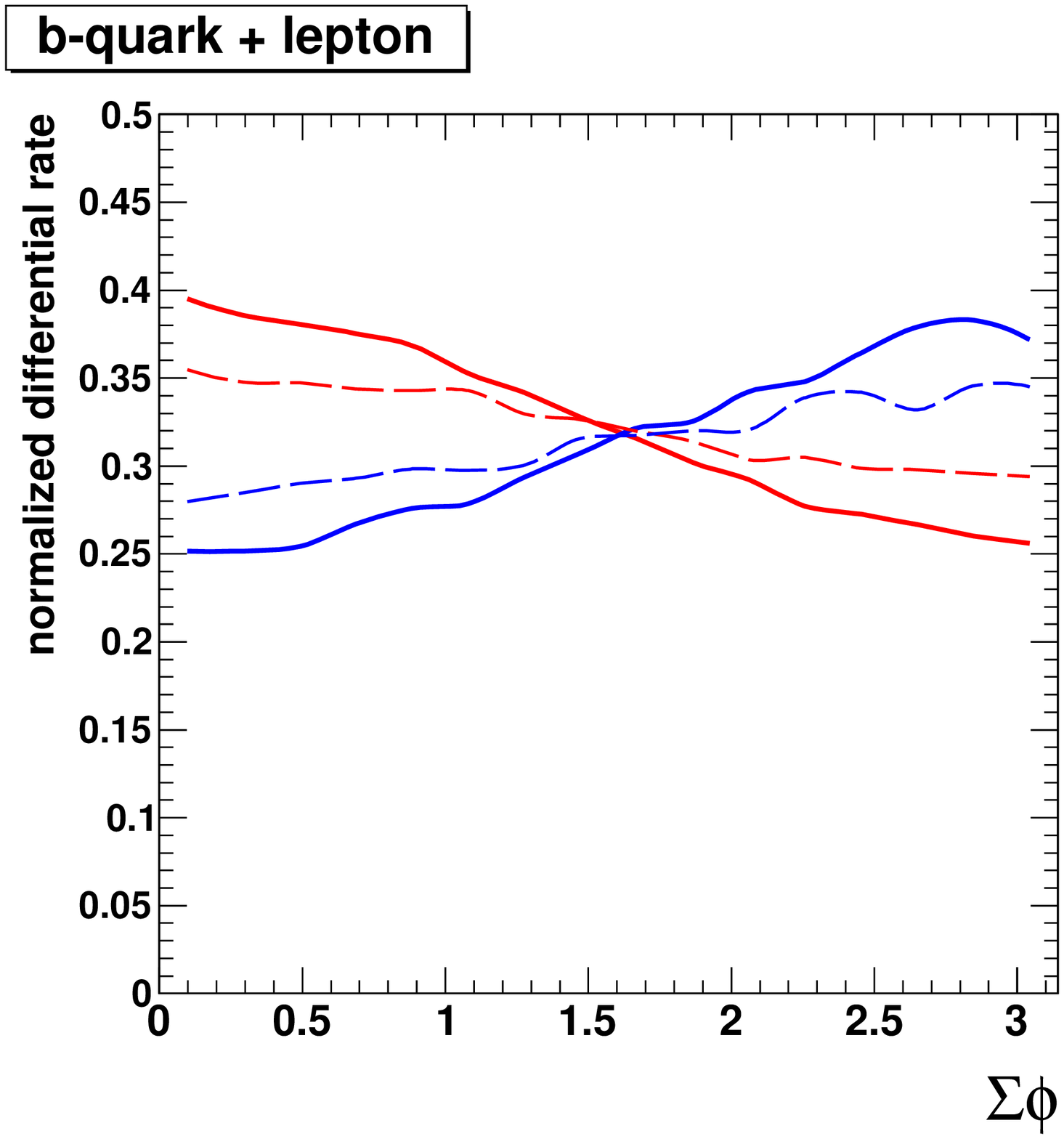}
\caption{Examples of reconstructed summed azimuthal decay angle distributions for different top-jet spin analyzers paired with the lepton from the semileptonic side of the event:  optimal hadronic polarimeter (top left), softer light-quark (top right), and $b$-quark (bottom).  Red indicates vector coupling, and blue indicates axial coupling.  Solid is our most optimistic reconstruction:  particle-level with $b$-tags.  Dashed is our most pessimistic reconstruction:  calorimeter-level with the $W$ reconstructed kinematically using the binary choice method.}
\label{fig:boosted_phiSum}
\end{center}
\end{figure*}

We now use the various sets of spin analyzer constructions to study the net helicities of tops from the chiral resonance decays and the azimuthal spin correlations of tops from the vector/axial resonance decays.  We reconstruct the global semileptonic $t\bar t$ system as usual, by solving for the neutrino $p_z$'s from $\vecmet$ and $\vec l$, and picking the solution that yields a leptonic top mass closer to $m_t$.  In the case that the solutions are complex, the magnitude of $\met$ is reduced to the point where $m_T(l,\met) = m_W$.  The $t\bar t$ system is actively boosted to rest, and then the individual tops are boosted to rest along the resonance decay axis.  To measure the helicity of the hadronic top, we use the polar decay angle of our spin analyzer with respect to this axis (orienting ``+$\hat z$'' along the hadronic top's direction of motion).  Note that the analyzing powers for antitops are the opposite of those for tops, but the helicities of the antitops from the chiral resonance are also reversed, so no charge information is required.  To measure the spin correlations, we use the azimuthal angle sum variable introduced in~\cite{Baumgart:2011wk}.  Since in this case we are not interested in observables sensitive to parity-violation, this can also be constructed without reference to the top quark charges.  (See~\cite{Baumgart:2012ay} for a proposal to measure parity-violating asymmetries with the azimuthal angle sum.)  Start by reflecting the lepton through the $t\bar t$ production plane, defined by the resonance decay axis and the beam axis.  The correlation-sensitive variable is then the unsigned azimuthal angle offset between this mirror-lepton and our hadronic spin analyzer around the resonance decay axis, which displays a modulation proportional to the difference between the resonance's vector and axial couplings to top: $g_V^2-g_A^2$.  To enhance the size of the modulation effect, which is largest at central production angles in the $t\bar t$ rest frame, we restrict this measurement to production angles whose cosines are less than 1/2.\footnote{Similar to the helicity measurements, this could also be more highly optimized by using likelihood-based observables, in this case that account for the fact that the strength of the correlation depends on the sines of the analyzer's polar decay angles and the $t\bar t$ production angle.  For the centrally-produced tops, we estimate a possible sensitivity increase of 7\%.}  Figs.~\ref{fig:boosted_cosTh} and~\ref{fig:boosted_phiSum} illustrate the impact of the resonance's coupling structure on a handful of representative distributions for various hadronic spin analyzers, as well as for the lepton.  Note that in the absence of cuts, the polar angle distributions in Fig.~\ref{fig:boosted_cosTh} would be straight lines with slopes proportional to analyzing powers.

To compare the sensitivities of the different measurements, we can apply the fit uncertainty estimator of Eq.~\ref{eq:DeltaP}.  A minor difference arises in the complete analysis, in that the reconstruction efficiencies for different top chiralities are not equal, with right-handed events being picked up about 10--15\% more often than left-handed.  (Much of this effect is due to the cuts on the leptonic side.)  Consequently, an unpolarized distribution would not look like an equal admixture of normalized right-handed and left-handed distributions.  We account for this by slightly modifying the constructions of the unpolarized distribution and the polarization-induced deviation used in Eq.~\ref{eq:DeltaP}.\footnote{In detail, suppose that we ignore the efficiency issue, simply taking the right-handed and left-handed distributions as normalized templates, and get $\mu_i$ and $\Delta\mu_i$ as before.  If we subsequently want to correct for the overall acceptance asymmetry, $A$, these should be modified to $\mu_i \to \mu_i + A\Delta\mu_i$, $\Delta\mu_i \to (1-A^2)\Delta\mu_i$.}  Numerically, the effect is small, as the uncertainty calculation effectively only feels these reconstruction biases quadratically.  Another minor point is that, for the vector versus axial cases, we are not measuring $P \propto g_Vg_A$, but, as mentioned, something proportional to $g_V^2-g_A^2$.  To keep a common ground for these different types of measurements, and also to divide out the overall statistics of the sample, we always normalize performance to what we would have obtained using perfect spin analyzers with no reconstruction biases, but with an equivalent final sample size.\footnote{While Eq.~\ref{eq:DeltaP} folds in full shape information for each polarization-sensitive observable, we note that simple 2-bin asymmetry analyses yield very similar relative performances amongst variables, if somewhat reduced absolute performances.}  Rather than displaying relative fit uncertainties, we display their inverses, so that bigger numbers (closer to one) correspond to more sensitive measurements.  The resulting normalized sensitivities can be viewed as effective analyzing powers, or effective products of analyzing powers in the case of correlations.

\begin{table}
\begin{center}
\begin{tabular}{ l|ccc|ccc }
                   & \multicolumn{3}{c|}{Particle-Level}& \multicolumn{3}{c}{Calorimeter-Level} \\ 
Spin Analyzer \    & \ \ $b$-tag \ \  & \ binary $W$ \ & \  $\sum W$ \   & \ \ $b$-tag \ \ & \ binary $W$ \ &  \  $\sum W$ \  \\   \hline 
optimal hadronic \ &      0.565       &      0.471     &      0.489      &      0.529      &      0.400     &       0.425     \\ 
soft-jet           &      0.442       &      0.430     &      0.430      &      0.411      &      0.385     &       0.385     \\ 
$b$-jet            &      0.400       &      0.272     &      0.345      &      0.390      &      0.217     &       0.319     \\ \hline
lepton             &               \multicolumn{3}{c|}{0.870}            &   \multicolumn{3}{c}{0.834} \\ 
\end{tabular}
\end{center}
\caption{Effective analyzing powers of the different spin analyzers in the boosted top chirality discrimination study, using polar decay angles for the polarization-sensitive variables.  Different columns represent different reconstruction assumptions described in the text.  The effective leptonic analyzing power, nominally unity, is shown to illustrate the degrading effects of analysis cuts.  (Absolute monte carlo statistical errors on all numbers are of order 0.002.)}
\label{table:boosted_chirality}
\end{table}

\begin{table}
\begin{center}
\begin{tabular}{ l|ccc|ccc }
                         & \multicolumn{3}{c|}{Particle-Level}& \multicolumn{3}{c}{Calorimeter-Level} \\ 
Correlation Analyzers \  & \ \ $b$-tag \ \  & \ binary $W$ \ & \  $\sum W$ \   & \ \ $b$-tag \ \ & \ binary $W$ \ &  \  $\sum W$ \  \\   \hline 
optimal had.\ + lepton   &      0.660       &      0.554     &      0.574      &      0.596      &      0.426     &       0.458     \\ 
soft-jet + lepton        &      0.513       &      0.500     &      0.500      &      0.449      &      0.420     &       0.420     \\ 
$b$-jet + lepton         &      0.468       &      0.321     &      0.401      &      0.442      &      0.215     &       0.321     \\
\end{tabular}
\end{center}
\caption{Effective products of analyzing powers of the different spin analyzers in the boosted top vector/axial discrimination study, using sums of azimuthal decay angles for the polarization-sensitive variables.  Different columns represent different reconstruction assumptions described in the text.  (Absolute monte carlo statistical errors on all numbers are of order 0.006.)}
\label{table:boosted_correlation}
\end{table}

The full set of effective analyzing powers are displayed in Table~\ref{table:boosted_chirality} for spin measurements,\footnote{We have also investigated a few other chirality discriminators not listed in the table.  The leptonic top's visible energy ratio $E(l)/\big(E(l)+E(b)\big)$~\cite{Shelton:2008nq}, appropriate to cases such as SUSY (Sec.~\ref{subsec:stop}) or dileptonic $t\bar t$ where the $\met$ is not entirely from a lone neutrino, still yields a substantial effective analyzing power of 0.69, or 79\% as powerful as a fully reconstructed leptonic top.  An untagged hadronic substructure variable was proposed in~\cite{Krohn:2011tw}.  Using the three subjets obtained with our substructure strategy, we find distributions similar to those in~\cite{Krohn:2011tw}, and compute an effective analyzing power of 0.20--0.22.  This is weaker than any of the other hadronic polarimeters studied here.  The optimal likelihood-ratio discriminator, studied in Sec.~\ref{sec:likelihood} at parton-level, still does not appear to offer any significant gain.} and in Table~\ref{table:boosted_correlation} for spin correlation measurements.  In all cases involving the hadronic top, variables utilizing the optimal hadronic polarimeter are the most powerful, with the margin depending on the assumptions going into the reconstruction.  Comparing to the next-most-powerful option, $q_{\rm soft}$, the most dramatic improvements occur when $b$-tagging information is available, amounting to 25--30\% relative.  This is comparable to the parton-level expectation of $0.64/0.50-1 = 28\%$, though both analyzers exhibit overall degradation during reconstruction.\footnote{It is instructive to look at the parton-level kinematics for events that pass our full set of reconstructions, as this gives a feeling for how much of the degradation is due to phase space cuts versus wrongly-assigned or misreconstructed partons.  Within the $b$-tagged particle-level sample, our effective $q_{\rm opt}$ and $q_{\rm soft}$ analyzing powers listed in Table~\ref{table:boosted_chirality} are 8--9\% smaller than their parton-level equivalents.  (E.g., the power of $q_{\rm opt}$ for discriminating chiralities becomes 0.61, closer to its inclusive value of 0.64.)  The degradations beyond those induced by the phase space bias appear to be driven by a residual population of $\sim$10\% of the events where the softest quark in lab-frame is either over-declustered, heavily contaminated, or fully replaced by an ISR/FSR subjet.}  When $b$-tagging is done through pure kinematics, $q_{\rm opt}$ can be further degraded, and the relative improvement over $q_{\rm soft}$ is reduced to the 10--15\% range.  These conclusions hold independently of whether we work at particle-level versus calorimeter-level, or are considering individual spins versus azimuthal spin correlations.  Indeed, the $b$-tagging is by far the major factor in both absolute and relative performance.  We can also see that the method of Sec.~\ref{sec:untagged}, which in the absence of $b$-tagging uses a superposition of kinematic $W$ reconstructions instead of a binary choice, buys a relative improvement in $q_{\rm opt}$ of about 6\% at calorimeter-level.  For completeness, such a weighted superposition is also applied to the $b$-quark direction, and interestingly shows quite large improvements of up to 50\% relative to the binary choice.

It is clear, then, that aspects of the optimal hadronic polarimeter can survive jet substructure reconstruction and offer substantial gains in spin studies with boosted hadronic top quarks.  The improvements over other polarimeters are largest when $b$-tagging information is available, but persist even when it is not.

To give a sense of numerics for a specific model, consider the KK gluon of~\cite{Agashe:2003zs,Agashe:2006hk,Lillie:2007yh}, with its mass set to 2.5~TeV.  As studied in~\cite{Rehermann:2010vq} with very minimalistic cuts, a 300~fb$^{-1}$ run at 14~TeV could deliver almost 10,000 signal events in the $\mu$+jets channel alone, with $S/B \sim 2$.  This would be further enhanced to $\sim$4 (and the background completely dominated by continuum $t\bar t$) if any $b$-tagging were applied.  Ignoring the background, and just making a rough estimate based on pure signal statistics, the polarization could be measured on the hadronic side to better than 5\% precision.  The vector/axial content $(g_V^2-g_A^2)/(g_V^2+g_A^2)$, which is predicted to be close to zero for this model, could be independently measured to better than 10\% using the azimuthal correlations.  Combining polar decay angle measurements from both hadronic and leptonic sides of the event with azimuthal angle correlations would provide a quite precise picture of the resonance's couplings.

\subsection{Semi-Boosted Tops from Stop Decays}
\label{subsec:stop}

The supersymmetric partner of the top quark, the stop, continues to be a high priority target at the LHC.  A number of dedicated searches for direct QCD production of stop pairs followed by decays $\tilde t \to t \tilde\chi^0_1$ are now complete~\cite{TheATLAScollaboration:2013xha,ATLAS:2013pla,ATLAS:2013cma,Chatrchyan:2013xna,CMS:2013cfa,CMS:2013nia}, and could be indicating that the stop mass is above 700~GeV.  Similar to the heavy resonance examples of the previous subsection, such heavy stops would also produce boosted top quarks in their decays.  Here, we will focus on a stop/neutralino mass point slightly above the experimental limit:  $m_{\tilde t} = 800$~GeV, $m_{\tilde \chi} = 0$.  At LHC14, the cross section for this stop mass is close to 40~fb, implying over 10,000 events produced in a 300~fb$^{-1}$ run, before cuts.  Prospects to measure the stop's effective chirality are then likely very good.  Several recent papers have studied such measurements~\cite{Perelstein:2008zt,Berger:2012an,Bhattacherjee:2012ir,Belanger:2012tm}, including one that exploits hadronic top polarization~\cite{Bhattacherjee:2012ir} in both $l$+jets and all-hadronic channels.\footnote{See also~\cite{Plehn:2012pr,Kaplan:2012gd} for recent phenomenological studies of direct stop pair discovery prospects using hadronic top-jets.}  We will now see whether the optimal hadronic polarimeter can offer any improvements.

For our monte carlo samples, we simulate pure right-handed and left-handed stops in the $l$+jets channel with {\tt MadGraph5} and {\tt PYTHIA6}.  The LSP is chosen to be bino-like, so that stop chirality directly translates to the final top chirality.  The $p_T$'s of tops from this sample peak around 350~GeV.  We can consider tops in this region to be {\it semi-boosted}, since the $\Delta R$ between decay products tends to be larger than normal-sized LHC jets, but the chance of object merging is nontrivial.  Jet substructure methods therefore remain appropriate.   To accommodate the lower boost of the events, we increase the fat-jet radius to 1.5 and decrease the fat-jet $p_T$ threshold to 150~GeV.   We also enforce an absolute minimum $p_T$ of 30~GeV on subjets used after the declustering, as softer subjets could be difficult to separate from pileup noise, and are much more susceptible to measurement uncertainties.  Since $b$-tagging should not be an issue, we demand a $b$-tagged subjet within the top-jet candidate.  Otherwise, the reconstruction is identical to the one used in the previous subsection.

\begin{figure*}[tp!]
\begin{center}
\includegraphics[width=0.44\textwidth]{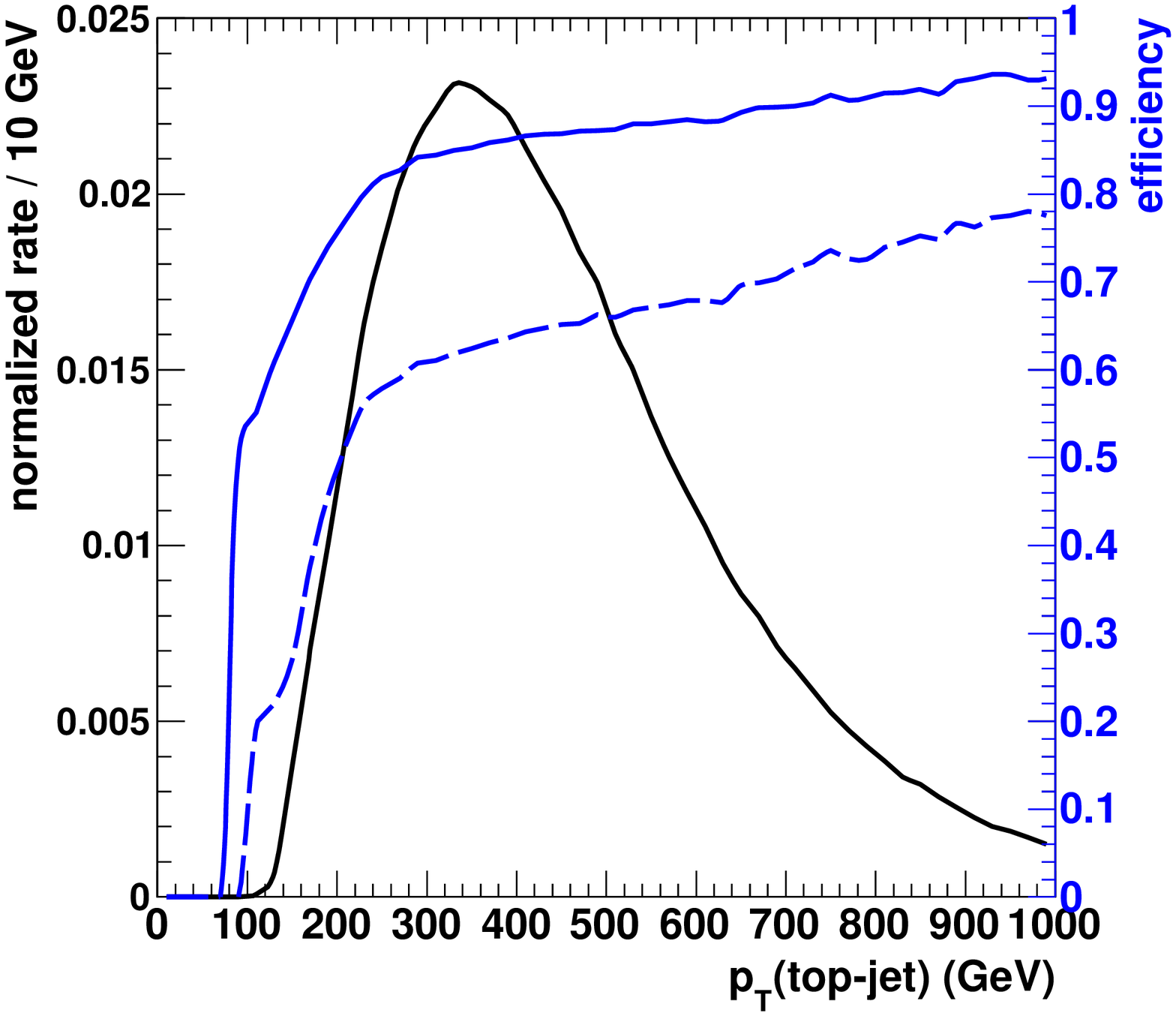} \\
\includegraphics[width=0.44\textwidth]{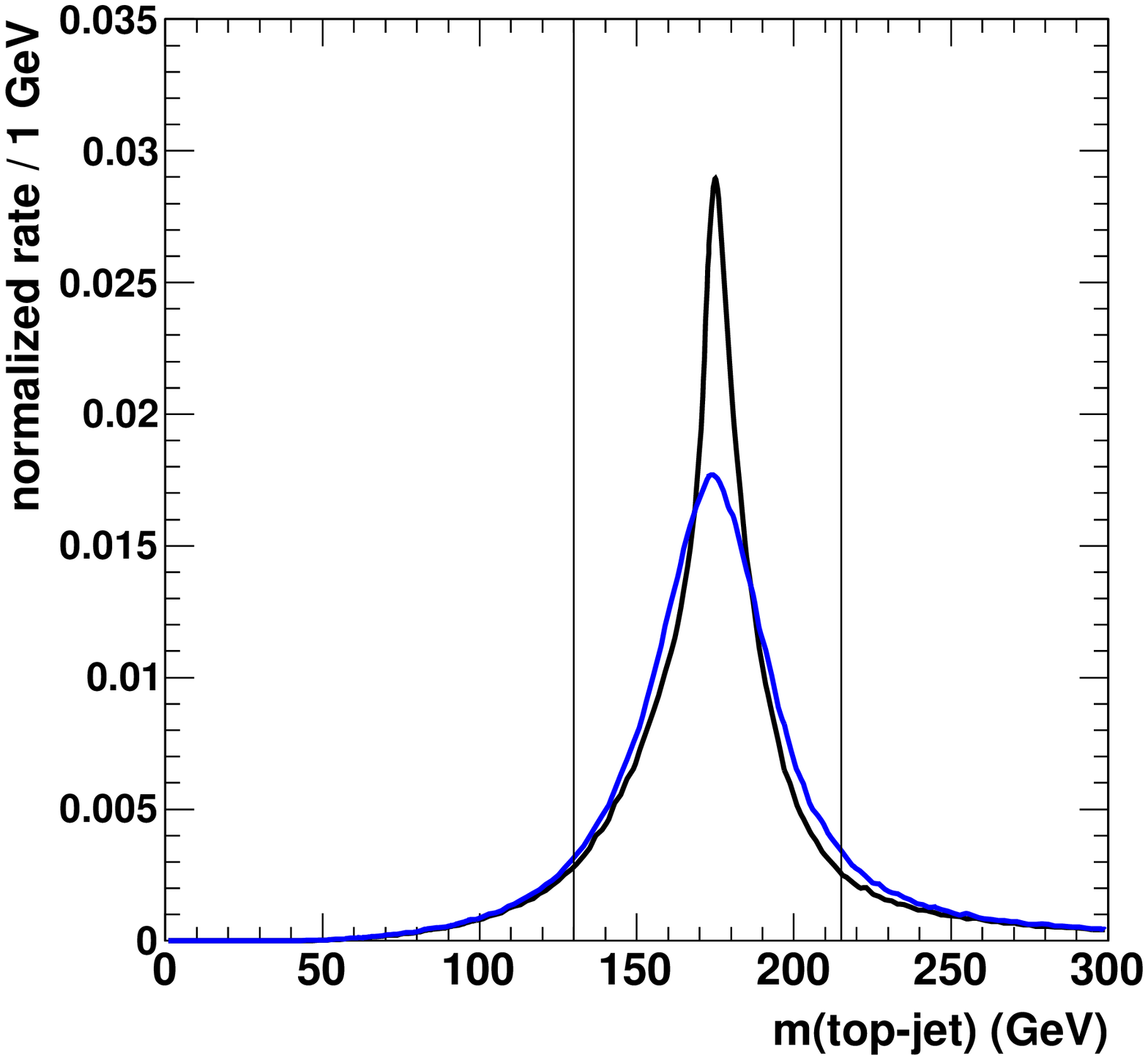}
\includegraphics[width=0.44\textwidth]{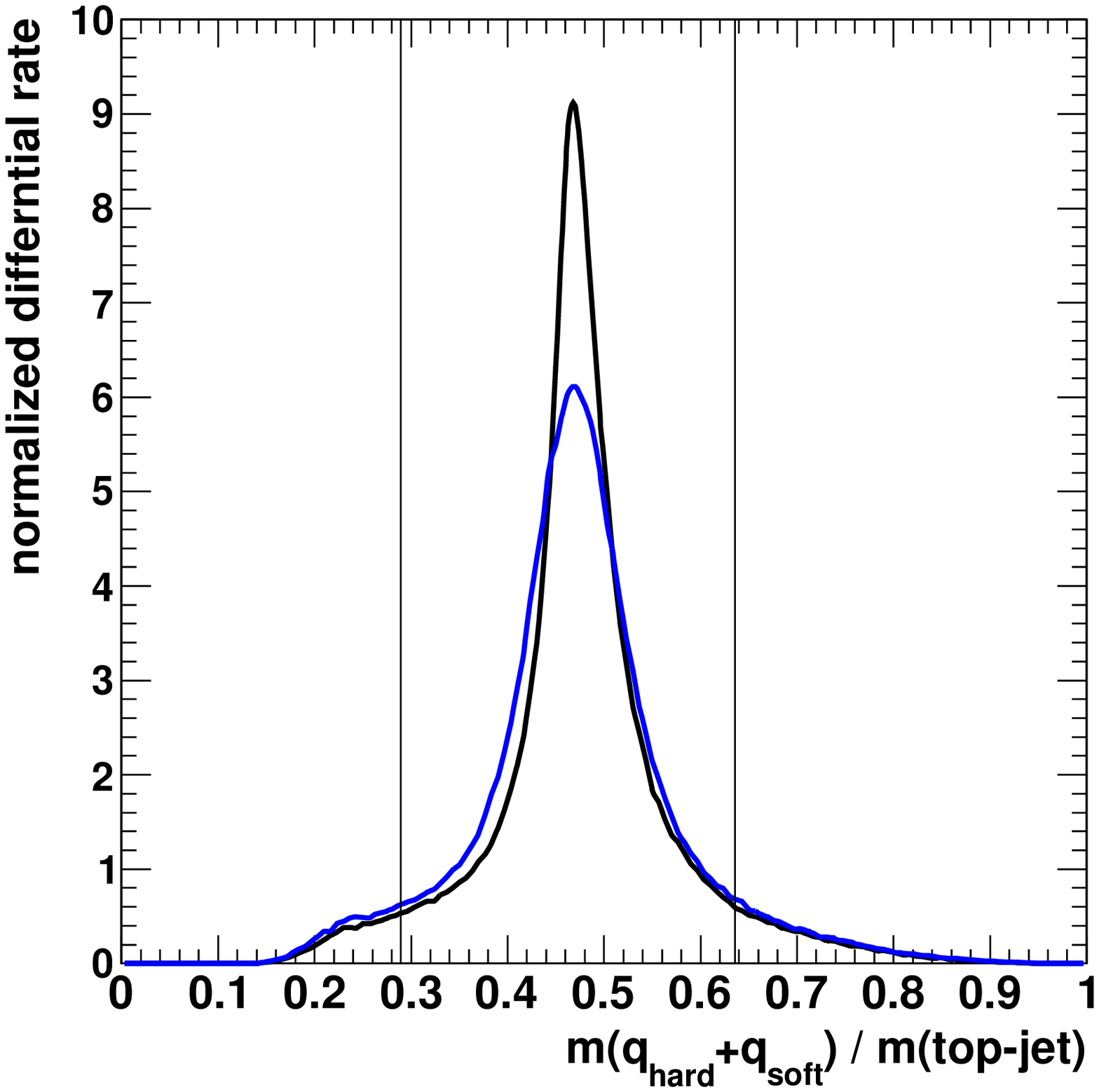}
\caption{Reconstructed distributions for declustered top-jets from stop decay, averaging chiralities:  $p_T$ raw rate and efficiencies at particle-level (top), mass after $b$-tagging (bottom left), and $W$ boson relative mass after $b$-tagging and top mass window (bottom right).  In the $p_T$ plot, the black curve is the raw differential rate before tags or cuts, the solid blue curve is the efficiency for finding a $b$-hadron inside a top-subjet, and the dashed blue curve is further multiplied by the efficiency for passing mass window cuts.  In the mass plots, black is particle-level, and blue is calorimeter-level.  Mass window cuts are indicated by vertical lines.}
\label{fig:stop_distributions}
\end{center}
\end{figure*}

Starting from the inclusive $l$+jets sample, about 40\% of the events pass the most basic reconstruction cuts, such as a mini-isolated lepton and decomposable top-jet with at least three good subjets.  Fig.~\ref{fig:stop_distributions} shows the $p_T$ spectrum and efficiencies of the top-jet candidates as we sequentially demand the $b$-tag and the top/$W$ mass-window cuts.  The figure also shows the mass spectra of the top-jets and $W$ boson candidates.  About 85\% of the top-jets contain a good $b$-subjet, and 75\% of these pass the mass window cuts, leading to a net reconstruction efficiency of $l$+jets $\tilde t \tilde t^*$ events of about 25\%.  The top-jet tagging efficiency exhibits a sharp turn-on at $p_T($top-jet$) \simeq m_t$.  (Though not typically considered for semi-boosted tops, the modified JHU tagger exhibits similar performance.)

\begin{table}
\begin{center}
\begin{tabular}{ l|ccc|ccc }
                        & \multicolumn{3}{c|}{Particle-Level}& \multicolumn{3}{c}{Calorimeter-Level} \\ 
Spin Analyzer \         &\ \ inclusive \ \ &\ $p_T<400$ \ &\ $p_T>400$ \ &\ \ inclusive \ \ &\ $p_T<400$ \ &\ $p_T>400$ \ \\ \hline 
optimal hadronic \      &      0.452       &     0.378    &     0.503    &      0.440       &     0.369    &     0.485       \\ 
soft-jet                &      0.338       &     0.279    &     0.376    &      0.326       &     0.269    &     0.362       \\ 
$b$-jet                 &      0.354       &     0.310    &     0.383    &      0.350       &     0.305    &     0.380       \\ \hline
$\frac{E(l)}{E(l)+E(b)}$&      0.555       &     0.539    &     0.568    &      0.553       &     0.537    &     0.565       \\
\end{tabular}
\end{center}
\caption{Effective analyzing powers of the different spin analyzers in the stop chirality discrimination study, using polar decay angles for the polarization-sensitive variables.  Different columns represent different reconstruction assumptions, and additional cuts to illustrate less-boosted versus more-boosted $p_T$ regions.  The semileptonic top energy-ratio discriminator, with parton-level analyzing power 0.79, is shown to illustrate the degrading effects of analysis cuts and combinatorics.  (Absolute monte carlo statistical errors on inclusive (exclusive) numbers are of order 0.004 (0.006).)}
\label{table:stop_chirality}
\end{table}

The chirality-sensitive $\cos\theta$ distributions (not shown) are qualitatively similar to those in Fig.~\ref{fig:boosted_cosTh}, though more degraded due to the greater combinatoric confusion and kinematic bias.  The effects are especially felt near $\cos\theta \simeq -1$.  Because the stop's rest frame cannot be uniquely reconstructed, we define the polarization axis as the top's direction of flight in lab-frame.  The effective analyzing powers (defined in Sec.~\ref{subsec:resonance}) are reported in Table~\ref{table:stop_chirality}, including for reference the semileptonic top polarimetry variable $E(l)/\big(E(l)+E(b)\big)$~\cite{Shelton:2008nq}, and further breaking down the sample into $p_T($top-jet$) < 400$~GeV and $p_T($top-jet$) > 400$~GeV to compare performances in less-boosted and more-boosted regimes.  Once again, the optimal hadronic polarimeter is always the strongest option for the hadronic top.  Notably, the $q_{\rm soft}$ polarimeter is highly reduced in effectiveness, since soft objects in top-frame are much more likely to be missed in lab-frame.  The main competition here is the $b$-quark, which $q_{\rm opt}$ nonetheless exceeds by 20--30\%.  For the more-boosted tops, both polarimeters become more powerful, approaching the results of Table~\ref{table:boosted_chirality}.  For less-boosted tops, the biases induced by the jet radius and the absolute subjet $p_T$ cutoff become more pronounced, and polarization discrimination uniformly suffers.  These observations are largely unaffected by the presence or absence of the calorimeter model.

With 300~fb$^{-1}$, and scaling by an assumed $b$-tag efficiency of 70\%, the total sample size for this study would be about 500 events.  The absolute statistical error on the polarization $P$ using the optimal polarimeter would be less than 0.2, suggesting very high statistical separation between $P = \pm 1$.  Of course, this simplistic analysis does not take into account backgrounds such as $t\bar t$ ($l$+jets and dileptonic) or $t\bar t+W/Z$, but leaves ample room for additional cuts.  It should also be possible to improve both the acceptance and the analyzing powers for the lower-$p_T$ region by supplementing with more traditional $t\bar t$ reconstructions (i.e., allowing $\Delta R > 1.5$ between hadronic top decay products), or possibly hybrid traditional/substructure methods (analogous to the two-body hybrid method of~\cite{Gouzevitch:2013qca}).  But the substructure-based techniques discussed here will continue to apply for even heavier stops.  The smaller overall cross sections will be somewhat compensated by higher top-jet reconstruction efficiencies and improved quality of polarimetry.

\subsection{Color Dipole Moments}

New physics in $t\bar t$ production need not arise from the on-shell production of new particles, but could appear indirectly in the form of higher-dimension operators.  A large variety of these appear at dimension-six (see, e.g.,~\cite{AguilarSaavedra:2008zc}).  A set particularly relevant for spin correlation and CP studies at the LHC are the chromomagnetic and chromoelectric dipole moment operators (CMDM and CEDM),
\beq
\Delta {\mathcal L}  \,=\, \frac{g_s}{2} \, G_{\mu\nu}^a \, \bar t \left[ T^a \sigma^{\mu\nu} (\mu + i\gamma^5 d) \right] t \, ,
\eeq
with $\sigma^{\mu\nu} \equiv (i/2)[\gamma^\mu,\gamma^\nu]$.  We have implicitly preserved electroweak gauge symmetry with a Higgs VEV insertion that is absorbed into the couplings $\mu$ and $d$, which have the dimensions of length, or inverse mass.  A recent study of the collider phenomenology of these operators~\cite{Baumgart:2012ay}, which we build upon here, noted that a large portion of their effects on $t\bar t$ spin correlations is to induce sine/cosine modulations in the relative azimuthal decay angles of the two tops.  (For additional work on color dipole phenomenology, see references contained therein.)  This variable is analogous in construction to the azimuthal-sum that we studied above in Sec.~\ref{subsec:resonance}, though without the mirror-reflection step.  After boosting to the CM frame, we measure the {\it signed} relative azimuthal angle offset between the hadronic top polarimeter and the semileptonic top's lepton, as measured counterclockwise about the hadronic top's direction of motion.  In~\cite{Baumgart:2012ay}, it was found that comparable sensitivities could be obtained in both $l$+jets and dileptonic channels (using two leptons as polarimeters in the latter case).  Given the introduction of a more powerful hadronic polarimeter in this paper, we can now determine if $l$+jets becomes even better.

For this analysis, we again use $t\bar t$ pairs produced in {\tt MadGraph5} and {\tt PYTHIA6} at LHC14, though now fully inclusively.  The effects of the dipole operators are applied via event-by-event reweightings after generation, keeping only spin correlation effects linear in the new couplings. Unlike previous sections, we apply a fairly traditional reconstruction strategy.  Jets are clustered with anti-$k_T$ $R=0.45$, with thresholds of $p_T(j) > 50$~GeV and $|\eta(j)|<2.5$.  We assume a $b$-tag efficiency of 70\%, as well as mistag rates of 10\% for charms and 2\% for unflavored.  The event must contain at least four jets, at least one of which is tagged.  To keep lepton identification efficiency high, we continue to use mini-isolation instead of traditional isolation.

The global $t\bar t$ system reconstruction closely follows~\cite{Baumgart:2012ay}.  We iterate over all possible partitions of the lepton and jets into a leptonic top $(l\nu j)$ and a hadronic top $(jjj)$, including at least one $b$-jet, and considering both of the possible neutrino solutions.  (Again, the $\met$ magnitude is reduced if no real solutions exist initially.)  In events with at least two $b$-tags, each top-candidate must contain at least one.  The partition that minimizes $\big(m(l\nu j)-m_t\big)^2 + \big(m(jjj)-m_t\big)^2$ defines our semileptonic and hadronic top candidates.  To ensure good quality reconstruction, we apply the same hadronic top-mass and relative $W$-mass cuts as in Sec.~\ref{subsec:resonance}.  We continue to use different cuts for $b$-tagged and untagged hadronic tops, applying a looser $W$-mass window when tagged, and a tighter window to the better $W$ candidate when untagged.  In addition, we require $m(l\nu j) < 215$~GeV.  To construct decay angles when the hadronic top is untagged, we apply the $W$ superposition method of Sec.~\ref{sec:untagged} to help improve the quality of the polarimetry, again assuming dimensionless Breit-Wigner widths of 0.06 (0.12) for particle-level (calorimeter-level).  The total cross section passing all cuts is about 1.6~pb (globally normalizing to NLO), and the reconstructed top $p_T$ is peaked near 220~GeV.  We therefore pick up a large fraction of semi-boosted events simply by virtue of our tight jet $p_T$ cuts, though this in any case works in our favor since the effects of the dipoles are largest at high $p_T$.  Substructure approaches might also offer some improvements here, though we have not explored this, and lower-$p_T$ tops might be folded in with more relaxed cuts.  With the current set of cuts, a complete analysis including backgrounds would yield a final sample consisting of more than 80\% $l$+jets $t\bar t$~\cite{Baumgart:2012ay}.

Correlating the optimal hadronic polarimeter with the lepton, we find induced cosine/sine modulations of strength $(0.66)\mu\times m_t$ and $(0.56)d\times m_t$ for the CMDM and CEDM, respectively.  These results are only mildly affected (at the few-percent level) by the presence or absence of the calorimeter model.  Given 300~fb$^{-1}$ of data, measurements of $\mu\times m_t$ and $d\times m_t$ should be possible with statistical uncertainties of 0.003--0.004.  Once again, the alternative choices for a hadronic polarimeter are less powerful.  For the CMDM, the relative sensitivities of $q_{\rm opt}$/$q_{\rm soft}$/$b$ go as 1/0.71/0.88.  For the CEDM, they go as 1/0.75/0.81.\footnote{Contrary to the results of~\cite{Baumgart:2012ay}, $q_{\rm soft}$ becomes a weaker polarimeter than the $b$-quark.  This may be due to the harder cuts used in the present analysis.}

\section{Conclusions}
\label{sec:conclusions}

This paper has introduced a new and truly optimal hadronic top quark spin analyzer, and verified its improvement relative to other hadronic spin analyzers for a variety of new physics measurements, in a variety of kinematic regimes, and under a variety of reconstruction assumptions.  We envision that this approach can play a role in any polarimetry analysis that utilizes hadronic top decays.  In pursuing high-quality kinematic reconstructions for boosted and semi-boosted tops, we have also explored some novel modifications to existing jet substructure techniques.

The basic logic of the optimal polarimeter construction is to generalize the well-known ``softer light-quark'' choice to a weighted sum of light-quark unit vectors in the top rest frame.  The required weights are just the relative probabilities of the softer or harder light-quark to have come from the down-type quark in the $W$ decay.  The analyzing power of this construction integrates to 0.64 at parton-level at leading order, and is fairly stable as a function of the $W$ helicity angle.

QCD showering and kinematic reconstruction can significantly change the effective analyzing powers.  For example, in several cases we have found the soft-quark choice to underperform the $b$-quark, even though the relationship is reversed at parton-level.  While the optimal hadronic polarimeter also loses some of its power, in all cases we have found it to consistently outperform the alternatives.  The improvement relative to the next-best option is typically 25--30\%.  Simulations of individual top decays with full NLO corrections reveal the same improvement.

We have also studied some further generalizations, including a parton-level likelihood-based polarimeter, and weighted-sum methods to help improve the effective analyzing power when none of the jets/subjets are $b$-tagged.  The former method is technically more powerful than the simpler spin analyzer approach, but yields nearly identical sensitivity.  The latter method appears to offer a small but nontrivial ($\sim$6\%) recuperation in the power of the optimal polarimeter for untagged boosted tops, relative to a simple binary kinematic choice of $b$ and $W$ candidates.

Boosted tops featured prominently in our studies here, as this kinematic regime is growing in importance for new physics searches.  As a potentially useful offshoot, we developed modified versions of both the HEPTopTagger and the JHU top-tagger that exhibit better mapping onto the 3-body parton-level kinematics, as required for polarization measurements.  Both eliminate some declustering parameters (and in some cases introduce new, more targeted ones), and both appear to continue to perform well at semi-boosted $p_T$'s.  While in some ways simpler and more kinematically faithful than the originals, the basic reconstruction requirements and residual radiative pollution still leave their mark as biases in the reconstructable decay angle distributions.  A more systematic study of polarization-sensitive observables under different substructure strategies is warranted.  The more general behavior of our modified top-taggers, such as their ability to reject QCD jets and the impact of further cuts on the polarization sensitivity, would also be interesting to follow-up on.

Moving ahead, we can imagine a couple of other directions for future work.  While we have demonstrated robustness under realistic sets of cuts and reconstructions, the analyses here have only been very coarsely optimized.  Better performance might be achieved with more refined procedures.  Within a given analysis, a full matrix element approach in principle offers the best performance.  This would also fold in the effects of transfer functions within the multidimensional space of top decay angles, and effectively identify more idealized contours for separating out different polarization or correlation hypotheses.  Without committing to such specifics, the gains introduced by our optimal spin analyzer construction are fully portable, and universally raise the baseline level of performance.  However, it remains an open question whether an even more optimal general-purpose polarimetry variable could be constructed, given the additonal kinematic confusions induced by the QCD radiation beyond just the identities of the light quarks.  Certainly any such procedure would be intertwined with the jet algorithm used to reconstruct the decay, and it is likely that some algorithms have better optimal performance than others.


\acknowledgments{We thank Joe Boudreau for a conversation that inspired this work, and for discussions on statistics.  We thank Michael Spannowsky for comments on the draft.  BT was supported by DoE grant No. DE-FG02-95ER40896 and by PITT PACC.}


\appendix

\section{Improving Jet Substructure for Spin Measurements}
\label{sec:appendix}

Optimizing polarimetry with boosted hadronic top quarks requires the application of jet substructure techniques that can accurately assign radiation back to the individual 3-body parton-level decay products.  Many of the procedures on the market (reviewed in~\cite{Abdesselam:2010pt,Altheimer:2012mn,Altheimer:2013yza}) strive to do this anyway, simply to make the best use of the top decay kinematics in discriminating against QCD jet backgrounds.  We have nonetheless identified some simple ways to improve polarimetry performance for both the JHU top-tagger~\cite{Kaplan:2008ie} and the HEPTopTagger~\cite{Plehn:2010st}.  A modified version of the latter serves as our default procedure in this paper, and is described in detail in Sec.~\ref{subsec:resonance}.  We now describe our proposed modifications to the JHU tagger, and briefly illustrate the performance gains in both taggers relative to their default algorithms.

The JHU tagger is based on a two-stage declustering algorithm, where each stage is a variant of the ``mass drop'' method introduced in~\cite{Butterworth:2008iy}.  Starting with the entire fat-jet (clustered with the C/A algorithm), the collection of particles is iteratively declustered, with the softer branch thrown away and the harder branch fed into the next iteration.  The declustering stops when the $p_T$ of each branch relative to the original fat-jet is found to be larger than a threshold $\delta_p$.  An additional parameter $\delta_r$ serves as a collinear cutoff in the declustering.  Typically, $\delta_p = 5$--10\%, and $\delta_r = 0.1$--0.2.  If the initial declustering successfully yields two hard branches according to $\delta_p$ before hitting the cutoff $\delta_r$, these branches individually undergo a second stage of declustering.  If both of these succeed, we have four subjets.  If only one succeeds, we fully reconstitute the failed branch and work with three subjets.  If both fail, we veto the jet.  The sum of these subjets serves as the top-jet candidate.  The original method, which assumes no $b$-tagging, further attempts to kinematically identify the $W$ by finding the subjet-pair that best reconstructs $m_W$, and also places a cut on the reconstructed $|\cW|$ to help reject backgrounds.

To improve the performance of the JHU top-tagger for polarimetry (and likely in more general contexts as well), we recommend the following changes:
\begin{enumerate}
\item Set $\delta_r \to 0$, removing one continuous parameter.\footnote{The omission of this parameter opens the possibility of collinear-unsafety of the procedure.  For top decays, this is not actually a problem given the additional steps and demands on good $m_t$ and $m_W$ reconstructions.  But for QCD background jets viewed at fixed order, a 2-parton final state would never pass the tagger, whereas adding in a hard collinear splitting would give it an opportunity.  To the extent that this could pose a calculational problem, or significantly raise the rate of QCD backgrounds in situations where they are important (such as studies with all-hadronic top pairs), an additional boost-invariant cutoff on the declustering could be applied as a regulator.  For example, a $p_T$-scaled $\delta_r$ cut or veto on declusterings below some mass threshold would each serve such a purpose.  A weak cut on $|\cW|$ might also be adequate, even without an explicit declustering cutoff.  At a bare minimum, the calorimeter segmentation can effectively act to replace $\delta_r$.}
\item In cases with four subjets, reconstitute into exactly three subjets by looking at the two branches found at the first declustering stage, and undoing the second-stage declustering of the lower-mass branch.  We have the option to immediately use these subjets as our quark candidates.
\item Or attempt to further refine by considering alternative top reconstructions.  In cases with four subjets, try to combine the two hardest with either the 3rd-hardest or 4th-hardest.  If the declustering instead yields three subjets, try to break down into four by running another declustering stage on the most massive one, and similarly consider subsets of three.  Amongst these alternative reconstructions and the nominal one in step (2), pick the one that best reconstructs $m_t$.
\item Apply any desired multibody kinematic cuts to the three ``quarks.''  
\end{enumerate}
The first and second steps yield increased efficiency and a tighter top-jet mass peak at $m_t$.  The third (optional) step cures a minor problem with the original tagger, wherein one of the first-stage branches consists of FSR/ISR and the other branch contains the entire top decay.  This comes at the ``price'' of introducing $m_t$ explicitly into the declustering, in a manner that is essentially equivalent to what is done in the HEPTopTagger approach.\footnote{Yet another approach, which also implicitly requires introducing $m_t$, is to shrink the jet cone as $R \propto m_t/p_T$.  This was done in a coarse manner in the original JHU tagger paper, and was explored more systematically in the Snowmass 2013 study~\cite{Calkins:2013ega}.}  The reconstruction rate of events within top and $W$ mass windows further increases by about 10\% when this last step is applied.

We now use the 2.5~TeV $t\bar t$ resonance monte carlo sample to compare four substructure variations:  JHU and HEPTopTagger with/without our modifications.  In all cases we treat the 3-body kinematics as in Sec.~\ref{subsec:resonance}, replacing and in some cases eliminating the original cuts (such as JHU's $|\cW|$ cut and HEPTopTagger's $m_{23}/m_{123}$ cut).  We focus on particle-level events with perfect $b$-tagging.  The original JHU tagger is run with $\delta_p = 0.05$ and $\delta_r = 0.19$, and the modified version uses the same $\delta_p$.  In events where the original JHU tagger yields four subjets, we use a simple recombination prescription to get back three:  keep the recombination that maximizes the velocity of the slowest subjet as viewed in top-frame.  (The exact recombination method is not crucial.)  For the HEPTopTagger, we use the original parameters in~\cite{Plehn:2010st}.

\begin{figure*}[tp!]
\begin{center}
\includegraphics[width=0.44\textwidth]{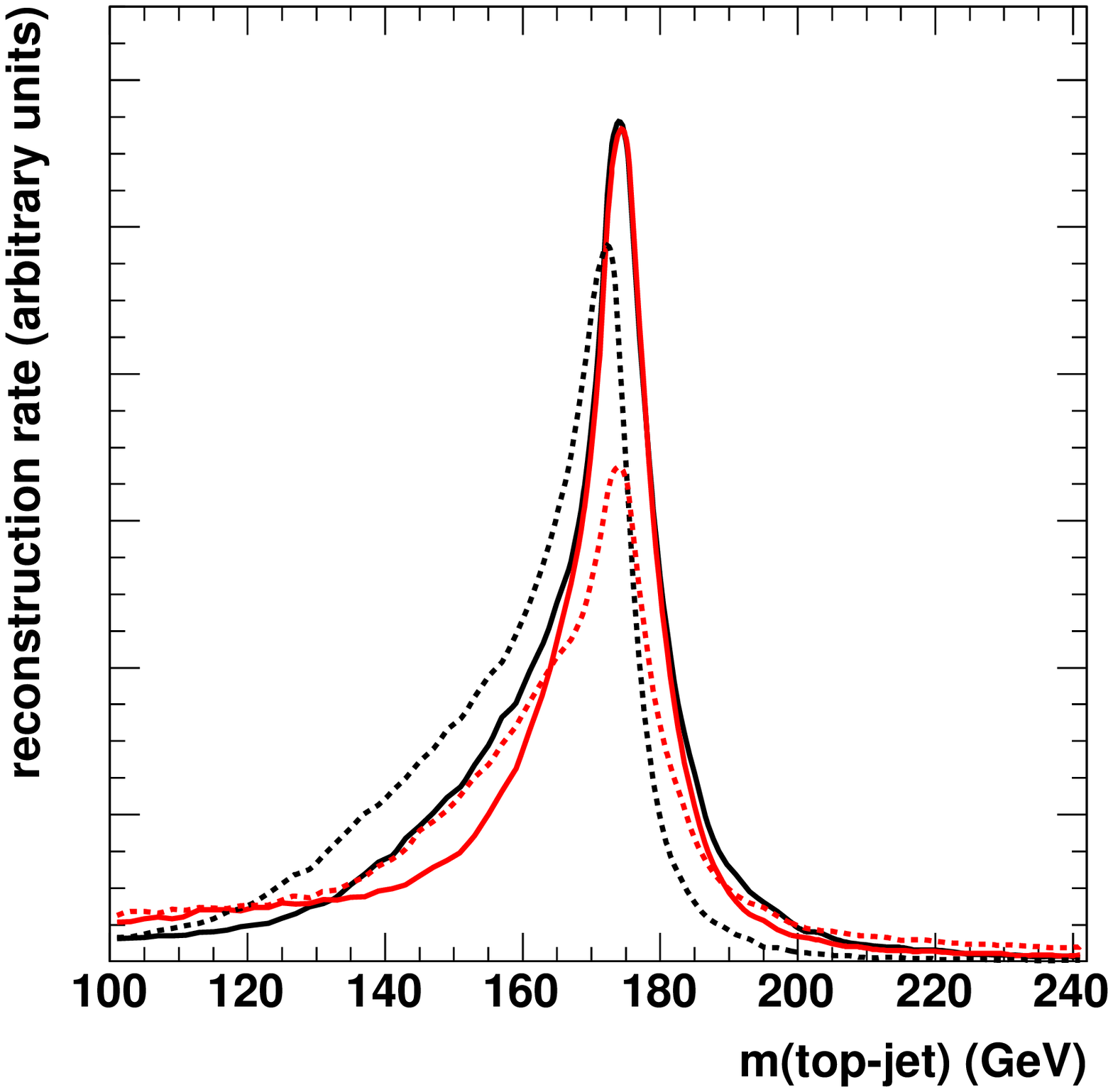}
\includegraphics[width=0.44\textwidth]{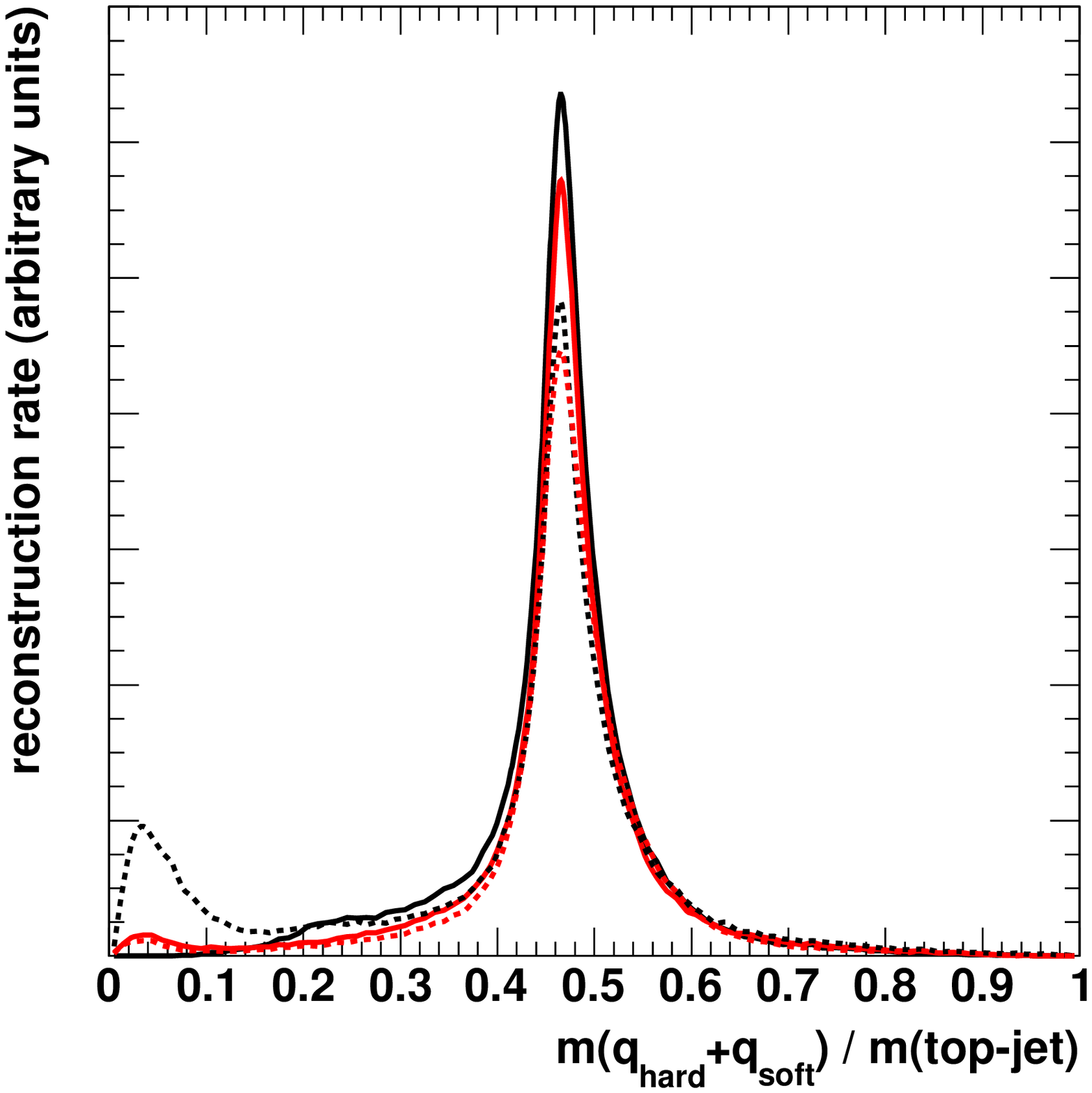}
\caption{Distributions of reconstructed top-jet mass (left) and $W$ boson mass relative to top-jet mass after $b$-tagging and top mass window (right).  Black is HEPTopTagger, and red is JHU top-tagger.  Solid are our modified algorithms, and dotted are the default algorithms.  All simulations are particle-level.}
\label{fig:algorithm_mass_distributions}
\end{center}
\end{figure*}

Fig.~\ref{fig:algorithm_mass_distributions} shows the reconstructed top-jet mass and $W$ candidate relative-mass under these different treatments.  The improvement in the top-jet mass peak for both algorithms is clear.  Notably, while most of the approaches use $m_t$ explicitly and are therefore prone to artificially ``sculpt'' a top peak, the $W$ peak almost always comes out well-reconstructed without further input.  The original HEPTopTagger displays a population of events with a relative $W$ mass near 0, which usually arise when the algorithm accidentally clusters two quarks together and splits one in half.  Our modifications cure this pathology.  The final modified algorithms give very similar distributions, and in particular the core of the top mass peak comes out nearly identical.  However, HEPTopTagger captures more events above and especially below the peak, leading to a 15\% higher total efficiency.  These added events include cases with fairly soft wide-angle subjets missed by the JHU $\delta_p$ criterion.

\begin{figure*}[tp!]
\begin{center}
\includegraphics[width=0.44\textwidth]{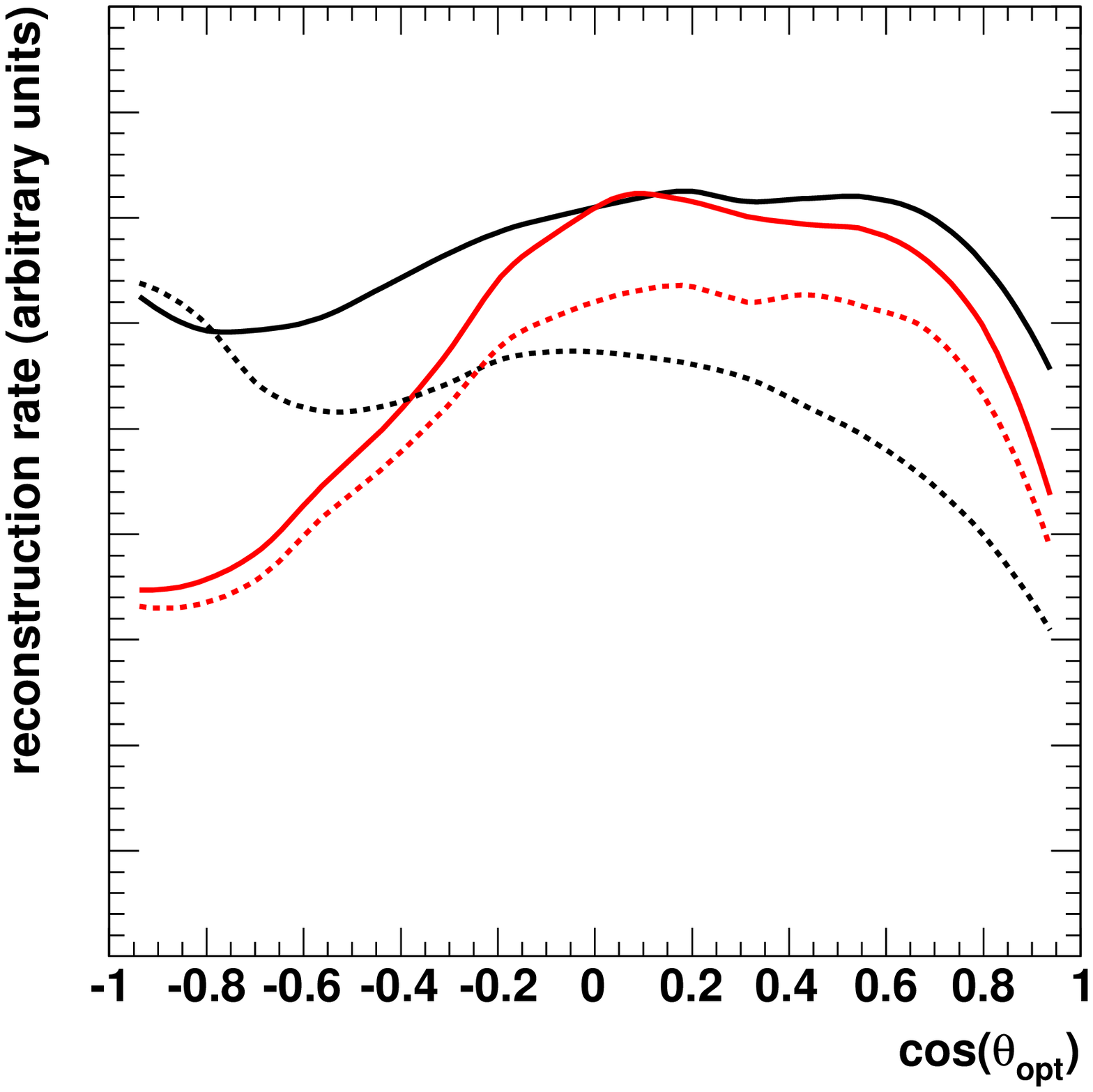}
\includegraphics[width=0.44\textwidth]{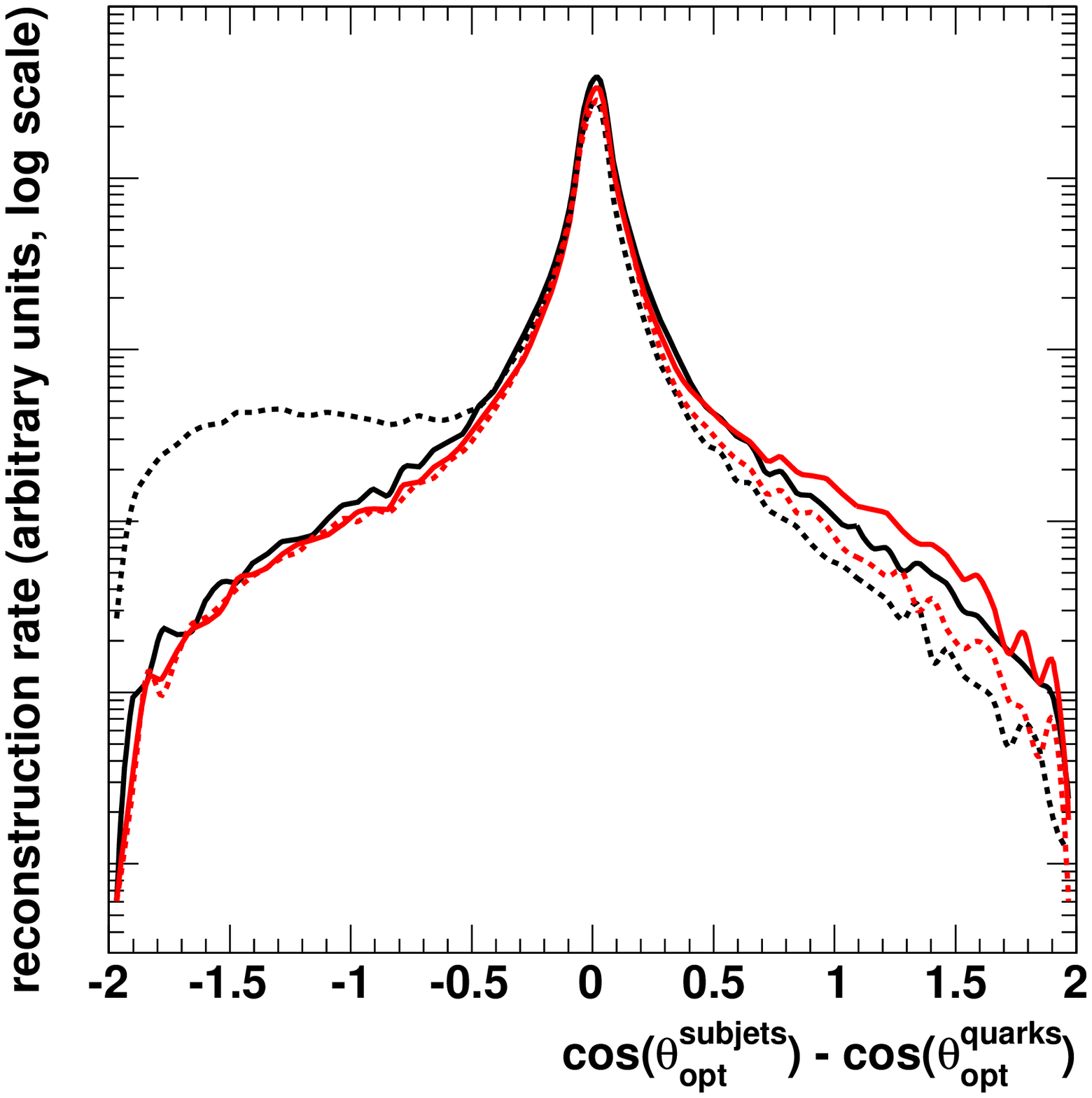}
\caption{Distributions of $\cos\theta_{\rm opt}$ without polarization (left) and the discrepancy between subjets and quarks (right).  Black is HEPTopTagger, and red is JHU top-tagger.  Solid are our modified algorithms, and dotted are the default algorithms.  All simulations are particle-level.}
\label{fig:algorithm_cosTh_distributions}
\end{center}
\end{figure*}

To get a sense for the quality of polarization measurements, we show in Fig.~\ref{fig:algorithm_cosTh_distributions} the distribution of $\cos\theta_{\rm opt}$ in the original unpolarized event sample, as well as the absolute difference between the reconstructed and parton-level values.  For a perfectly unbiased measurement, the rate should be flat over $\cos\theta_{\rm opt}$.  We see that, before modification, the HEPTopTagger displays a spurious peak at $\cos\theta_{\rm opt} \simeq -1$ and a broad tail of misreconstructions, corresponding to the pathological events.  After modification, the distribution is much flatter, and the discrepant tail is removed.  After modification, both algorithms show similar resolution on $\cos\theta_{\rm opt}$, and similar reconstruction rates for $\cos\theta_{\rm opt} > 0$.  However, JHU loses efficiency at $\cos\theta_{\rm opt} < 0$, again where emission against the top's boost tends to cause objects to become too soft to be resolved.

When we apply the event-by-event reweightings for resonances and rerun the sensitivity estimates of Sec.~\ref{subsec:resonance}, the modified JHU tagger yields about 10\% weaker sensitivity to helicity and comparable sensitivity to azimuthal spin correlations.  These numbers do not account for the total efficiency, which gives HEPTopTagger an added advantage of $\sim$8\%.  The relative performances of different hadronic spin analyzer choices remains quite similar, though for JHU $q_{\rm soft}$ weakens and becomes more comparable to the $b$-quark, similar to what we see in semi-boosted studies when arbitrarily soft quarks cannot be reconstructed.

We issue a final word of caution that, although we have rewritten and retuned the two top-taggers (in fact borrowing ideas from one another), the results here have not been systematically optimized.  The inclusion of backgrounds and pileup might also change our conclusions, in particular the ability to use soft subjets.  Other tagger approaches would also be worth exploring, though we have specifically focused on HEPTopTagger and JHU due to their ability to systematically weed out contaminating radiation, even when it is {\it harder} than the softest top decay products.


\bibliography{lit}

\begin{thebibliography}{90}
\expandafter\ifx\csname natexlab\endcsname\relax\def\natexlab#1{#1}\fi
\expandafter\ifx\csname bibnamefont\endcsname\relax
  \def\bibnamefont#1{#1}\fi
\expandafter\ifx\csname bibfnamefont\endcsname\relax
  \def\bibfnamefont#1{#1}\fi
\expandafter\ifx\csname citenamefont\endcsname\relax
  \def\citenamefont#1{#1}\fi
\expandafter\ifx\csname url\endcsname\relax
  \def\url#1{\texttt{#1}}\fi
\expandafter\ifx\csname urlprefix\endcsname\relax\def\urlprefix{URL }\fi
\providecommand{\bibinfo}[2]{#2}
\providecommand{\eprint}[2][]{\url{#2}}

\bibitem[{\citenamefont{Barger et~al.}(1989)\citenamefont{Barger, Ohnemus, and
  Phillips}}]{Barger:1988jj}
\bibinfo{author}{\bibfnamefont{V.~D.} \bibnamefont{Barger}},
  \bibinfo{author}{\bibfnamefont{J.}~\bibnamefont{Ohnemus}}, \bibnamefont{and}
  \bibinfo{author}{\bibfnamefont{R.~J.~N.} \bibnamefont{Phillips}},
  \emph{\bibinfo{title}{{Spin Correlation Effects in the Hadroproduction and
  Decay of Very Heavy Top Quark Pairs}}}, \bibinfo{journal}{Int. J. Mod. Phys.}
  \textbf{\bibinfo{volume}{A4}}, \bibinfo{pages}{617} (\bibinfo{year}{1989}).

\bibitem[{\citenamefont{Kane et~al.}(1992)\citenamefont{Kane, Ladinsky, and
  Yuan}}]{Kane:1991bg}
\bibinfo{author}{\bibfnamefont{G.~L.} \bibnamefont{Kane}},
  \bibinfo{author}{\bibfnamefont{G.}~\bibnamefont{Ladinsky}}, \bibnamefont{and}
  \bibinfo{author}{\bibfnamefont{C.}~\bibnamefont{Yuan}},
  \emph{\bibinfo{title}{{Using the Top Quark for Testing Standard Model
  Polarization and CP Predictions}}}, \bibinfo{journal}{Phys.Rev.}
  \textbf{\bibinfo{volume}{D45}}, \bibinfo{pages}{124} (\bibinfo{year}{1992}).

\bibitem[{\citenamefont{Jezabek}(1994)}]{Jezabek:1994qs}
\bibinfo{author}{\bibfnamefont{M.}~\bibnamefont{Jezabek}},
  \emph{\bibinfo{title}{{Top Quark Physics}}},
  \bibinfo{journal}{Nucl.Phys.Proc.Suppl.} \textbf{\bibinfo{volume}{37B}},
  \bibinfo{pages}{197} (\bibinfo{year}{1994}), \eprint{hep-ph/9406411}.

\bibitem[{\citenamefont{Mahlon and Parke}(1996)}]{Mahlon:1995zn}
\bibinfo{author}{\bibfnamefont{G.}~\bibnamefont{Mahlon}} \bibnamefont{and}
  \bibinfo{author}{\bibfnamefont{S.~J.} \bibnamefont{Parke}},
  \emph{\bibinfo{title}{{Angular Correlations in Top Quark Pair Production and
  Decay at Hadron Colliders}}}, \bibinfo{journal}{Phys.Rev.}
  \textbf{\bibinfo{volume}{D53}}, \bibinfo{pages}{4886} (\bibinfo{year}{1996}),
  \eprint{hep-ph/9512264}.

\bibitem[{\citenamefont{Stelzer and Willenbrock}(1996)}]{Stelzer:1995gc}
\bibinfo{author}{\bibfnamefont{T.}~\bibnamefont{Stelzer}} \bibnamefont{and}
  \bibinfo{author}{\bibfnamefont{S.}~\bibnamefont{Willenbrock}},
  \emph{\bibinfo{title}{{Spin Correlation in Top Quark Production at Hadron
  Colliders}}}, \bibinfo{journal}{Phys.Lett.} \textbf{\bibinfo{volume}{B374}},
  \bibinfo{pages}{169} (\bibinfo{year}{1996}), \eprint{hep-ph/9512292}.

\bibitem[{\citenamefont{Uwer}(2005)}]{Uwer:2004vp}
\bibinfo{author}{\bibfnamefont{P.}~\bibnamefont{Uwer}},
  \emph{\bibinfo{title}{{Maximizing the Spin Correlation of Top Quark Pairs
  Produced at the Large Hadron Collider}}}, \bibinfo{journal}{Phys.Lett.}
  \textbf{\bibinfo{volume}{B609}}, \bibinfo{pages}{271} (\bibinfo{year}{2005}),
  \eprint{hep-ph/0412097}.

\bibitem[{\citenamefont{Mahlon and Parke}(2010)}]{Mahlon:2010gw}
\bibinfo{author}{\bibfnamefont{G.}~\bibnamefont{Mahlon}} \bibnamefont{and}
  \bibinfo{author}{\bibfnamefont{S.~J.} \bibnamefont{Parke}},
  \emph{\bibinfo{title}{{Spin Correlation Effects in Top Quark Pair Production
  at the LHC}}}, \bibinfo{journal}{Phys.Rev.} \textbf{\bibinfo{volume}{D81}},
  \bibinfo{pages}{074024} (\bibinfo{year}{2010}), \eprint{1001.3422}.

\bibitem[{\citenamefont{Baumgart and
  Tweedie}(2013{\natexlab{a}})}]{Baumgart:2012ay}
\bibinfo{author}{\bibfnamefont{M.}~\bibnamefont{Baumgart}} \bibnamefont{and}
  \bibinfo{author}{\bibfnamefont{B.}~\bibnamefont{Tweedie}},
  \emph{\bibinfo{title}{{A New Twist on Top Quark Spin Correlations}}},
  \bibinfo{journal}{JHEP} \textbf{\bibinfo{volume}{1303}}, \bibinfo{pages}{117}
  (\bibinfo{year}{2013}{\natexlab{a}}), \eprint{1212.4888}.

\bibitem[{\citenamefont{Bernreuther and
  Brandenburg}(1994)}]{Bernreuther:1993hq}
\bibinfo{author}{\bibfnamefont{W.}~\bibnamefont{Bernreuther}} \bibnamefont{and}
  \bibinfo{author}{\bibfnamefont{A.}~\bibnamefont{Brandenburg}},
  \emph{\bibinfo{title}{{Tracing CP Violation in the Production of Top Quark
  Pairs by Multiple TeV Proton Proton Collisions}}},
  \bibinfo{journal}{Phys.Rev.} \textbf{\bibinfo{volume}{D49}},
  \bibinfo{pages}{4481} (\bibinfo{year}{1994}), \eprint{hep-ph/9312210}.

\bibitem[{\citenamefont{Beneke et~al.}(2000)\citenamefont{Beneke,
  Efthymiopoulos, Mangano, Womersley, Ahmadov et~al.}}]{Beneke:2000hk}
\bibinfo{author}{\bibfnamefont{M.}~\bibnamefont{Beneke}},
  \bibinfo{author}{\bibfnamefont{I.}~\bibnamefont{Efthymiopoulos}},
  \bibinfo{author}{\bibfnamefont{M.~L.} \bibnamefont{Mangano}},
  \bibinfo{author}{\bibfnamefont{J.}~\bibnamefont{Womersley}},
  \bibinfo{author}{\bibfnamefont{A.}~\bibnamefont{Ahmadov}},
  \bibnamefont{et~al.}, \emph{\bibinfo{title}{{Top Quark Physics}}}
  (\bibinfo{year}{2000}), \eprint{hep-ph/0003033}.

\bibitem[{\citenamefont{Frederix and Maltoni}(2009)}]{Frederix:2007gi}
\bibinfo{author}{\bibfnamefont{R.}~\bibnamefont{Frederix}} \bibnamefont{and}
  \bibinfo{author}{\bibfnamefont{F.}~\bibnamefont{Maltoni}},
  \emph{\bibinfo{title}{{Top Pair Invariant Mass Distribution: A Window on New
  Physics}}}, \bibinfo{journal}{JHEP} \textbf{\bibinfo{volume}{01}},
  \bibinfo{pages}{047} (\bibinfo{year}{2009}), \eprint{0712.2355}.

\bibitem[{\citenamefont{Arai et~al.}(2007)\citenamefont{Arai, Okada, Smolek,
  and Simak}}]{Arai:2007ts}
\bibinfo{author}{\bibfnamefont{M.}~\bibnamefont{Arai}},
  \bibinfo{author}{\bibfnamefont{N.}~\bibnamefont{Okada}},
  \bibinfo{author}{\bibfnamefont{K.}~\bibnamefont{Smolek}}, \bibnamefont{and}
  \bibinfo{author}{\bibfnamefont{V.}~\bibnamefont{Simak}},
  \emph{\bibinfo{title}{{Top Quark Spin Correlations in the Randall-Sundrum
  Model at the CERN Large Hadron Collider}}}, \bibinfo{journal}{Phys.Rev.}
  \textbf{\bibinfo{volume}{D75}}, \bibinfo{pages}{095008}
  (\bibinfo{year}{2007}), \eprint{hep-ph/0701155}.

\bibitem[{\citenamefont{Shelton}(2009)}]{Shelton:2008nq}
\bibinfo{author}{\bibfnamefont{J.}~\bibnamefont{Shelton}},
  \emph{\bibinfo{title}{{Polarized Tops from New Physics: Signals and
  Observables}}}, \bibinfo{journal}{Phys.Rev.} \textbf{\bibinfo{volume}{D79}},
  \bibinfo{pages}{014032} (\bibinfo{year}{2009}), \eprint{0811.0569}.

\bibitem[{\citenamefont{Degrande et~al.}(2011)\citenamefont{Degrande, Gerard,
  Grojean, Maltoni, and Servant}}]{Degrande:2010kt}
\bibinfo{author}{\bibfnamefont{C.}~\bibnamefont{Degrande}},
  \bibinfo{author}{\bibfnamefont{J.-M.} \bibnamefont{Gerard}},
  \bibinfo{author}{\bibfnamefont{C.}~\bibnamefont{Grojean}},
  \bibinfo{author}{\bibfnamefont{F.}~\bibnamefont{Maltoni}}, \bibnamefont{and}
  \bibinfo{author}{\bibfnamefont{G.}~\bibnamefont{Servant}},
  \emph{\bibinfo{title}{{Non-resonant New Physics in Top Pair Production at
  Hadron Colliders}}}, \bibinfo{journal}{JHEP} \textbf{\bibinfo{volume}{1103}},
  \bibinfo{pages}{125} (\bibinfo{year}{2011}), \eprint{1010.6304}.

\bibitem[{\citenamefont{Cao et~al.}(2011)\citenamefont{Cao, Wu, and
  Yang}}]{Cao:2010nw}
\bibinfo{author}{\bibfnamefont{J.}~\bibnamefont{Cao}},
  \bibinfo{author}{\bibfnamefont{L.}~\bibnamefont{Wu}}, \bibnamefont{and}
  \bibinfo{author}{\bibfnamefont{J.~M.} \bibnamefont{Yang}},
  \emph{\bibinfo{title}{{New Physics Effects on Top Quark Spin Correlation and
  Polarization at the LHC: A Comparative Study in Different Models}}},
  \bibinfo{journal}{Phys.Rev.} \textbf{\bibinfo{volume}{D83}},
  \bibinfo{pages}{034024} (\bibinfo{year}{2011}), \eprint{1011.5564}.

\bibitem[{\citenamefont{Baumgart and Tweedie}(2011)}]{Baumgart:2011wk}
\bibinfo{author}{\bibfnamefont{M.}~\bibnamefont{Baumgart}} \bibnamefont{and}
  \bibinfo{author}{\bibfnamefont{B.}~\bibnamefont{Tweedie}},
  \emph{\bibinfo{title}{{Discriminating Top-Antitop Resonances Using Azimuthal
  Decay Correlations}}}, \bibinfo{journal}{JHEP}
  \textbf{\bibinfo{volume}{1109}}, \bibinfo{pages}{049} (\bibinfo{year}{2011}),
  \eprint{1104.2043}.

\bibitem[{\citenamefont{Barger et~al.}(2012)\citenamefont{Barger, Keung, and
  Yencho}}]{Barger:2011pu}
\bibinfo{author}{\bibfnamefont{V.}~\bibnamefont{Barger}},
  \bibinfo{author}{\bibfnamefont{W.-Y.} \bibnamefont{Keung}}, \bibnamefont{and}
  \bibinfo{author}{\bibfnamefont{B.}~\bibnamefont{Yencho}},
  \emph{\bibinfo{title}{{Azimuthal Correlations in Top Pair Decays and the
  Effects of New Heavy Scalars}}}, \bibinfo{journal}{Phys.Rev.}
  \textbf{\bibinfo{volume}{D85}}, \bibinfo{pages}{034016}
  (\bibinfo{year}{2012}), \eprint{1112.5173}.

\bibitem[{\citenamefont{Krohn et~al.}(2011)\citenamefont{Krohn, Liu, Shelton,
  and Wang}}]{Krohn:2011tw}
\bibinfo{author}{\bibfnamefont{D.}~\bibnamefont{Krohn}},
  \bibinfo{author}{\bibfnamefont{T.}~\bibnamefont{Liu}},
  \bibinfo{author}{\bibfnamefont{J.}~\bibnamefont{Shelton}}, \bibnamefont{and}
  \bibinfo{author}{\bibfnamefont{L.-T.} \bibnamefont{Wang}},
  \emph{\bibinfo{title}{{A Polarized View of the Top Asymmetry}}},
  \bibinfo{journal}{Phys.Rev.} \textbf{\bibinfo{volume}{D84}},
  \bibinfo{pages}{074034} (\bibinfo{year}{2011}), \eprint{1105.3743}.

\bibitem[{\citenamefont{Bai and Han}(2012)}]{Bai:2011uk}
\bibinfo{author}{\bibfnamefont{Y.}~\bibnamefont{Bai}} \bibnamefont{and}
  \bibinfo{author}{\bibfnamefont{Z.}~\bibnamefont{Han}},
  \emph{\bibinfo{title}{{Improving the Top Quark Forward-Backward Asymmetry
  Measurement at the LHC}}}, \bibinfo{journal}{JHEP}
  \textbf{\bibinfo{volume}{1202}}, \bibinfo{pages}{135} (\bibinfo{year}{2012}),
  \eprint{1106.5071}.

\bibitem[{\citenamefont{Falkowski
  et~al.}(2013{\natexlab{a}})\citenamefont{Falkowski, Perez, and
  Schmaltz}}]{Falkowski:2011zr}
\bibinfo{author}{\bibfnamefont{A.}~\bibnamefont{Falkowski}},
  \bibinfo{author}{\bibfnamefont{G.}~\bibnamefont{Perez}}, \bibnamefont{and}
  \bibinfo{author}{\bibfnamefont{M.}~\bibnamefont{Schmaltz}},
  \emph{\bibinfo{title}{{Spinning the Top}}}, \bibinfo{journal}{Phys.Rev.}
  \textbf{\bibinfo{volume}{D87}}, \bibinfo{pages}{034041}
  (\bibinfo{year}{2013}{\natexlab{a}}), \eprint{1110.3796}.

\bibitem[{\citenamefont{Han et~al.}(2012)\citenamefont{Han, Katz, Krohn, and
  Reece}}]{Han:2012fw}
\bibinfo{author}{\bibfnamefont{Z.}~\bibnamefont{Han}},
  \bibinfo{author}{\bibfnamefont{A.}~\bibnamefont{Katz}},
  \bibinfo{author}{\bibfnamefont{D.}~\bibnamefont{Krohn}}, \bibnamefont{and}
  \bibinfo{author}{\bibfnamefont{M.}~\bibnamefont{Reece}},
  \emph{\bibinfo{title}{{(Light) Stop Signs}}}, \bibinfo{journal}{JHEP}
  \textbf{\bibinfo{volume}{1208}}, \bibinfo{pages}{083} (\bibinfo{year}{2012}),
  \eprint{1205.5808}.

\bibitem[{\citenamefont{Fajfer et~al.}(2012)\citenamefont{Fajfer, Kamenik, and
  Melic}}]{Fajfer:2012si}
\bibinfo{author}{\bibfnamefont{S.}~\bibnamefont{Fajfer}},
  \bibinfo{author}{\bibfnamefont{J.~F.} \bibnamefont{Kamenik}},
  \bibnamefont{and} \bibinfo{author}{\bibfnamefont{B.}~\bibnamefont{Melic}},
  \emph{\bibinfo{title}{{Discerning New Physics in Top-Antitop Production using
  Top Spin Observables at Hadron Colliders}}}, \bibinfo{journal}{JHEP}
  \textbf{\bibinfo{volume}{1208}}, \bibinfo{pages}{114} (\bibinfo{year}{2012}),
  \eprint{1205.0264}.

\bibitem[{\citenamefont{Yang and Liu}(2013)}]{Yang:2012ib}
\bibinfo{author}{\bibfnamefont{B.}~\bibnamefont{Yang}} \bibnamefont{and}
  \bibinfo{author}{\bibfnamefont{N.}~\bibnamefont{Liu}},
  \emph{\bibinfo{title}{{One-Loop Effects on Top Pair Production in the
  Littlest Higgs Model with T-Parity at the LHC}}},
  \bibinfo{journal}{Eur.Phys.J.} \textbf{\bibinfo{volume}{C73}},
  \bibinfo{pages}{2570} (\bibinfo{year}{2013}), \eprint{1210.5120}.

\bibitem[{\citenamefont{Gabrielli et~al.}(2013)\citenamefont{Gabrielli,
  Racioppi, Raidal, and Veermae}}]{Gabrielli:2012pk}
\bibinfo{author}{\bibfnamefont{E.}~\bibnamefont{Gabrielli}},
  \bibinfo{author}{\bibfnamefont{A.}~\bibnamefont{Racioppi}},
  \bibinfo{author}{\bibfnamefont{M.}~\bibnamefont{Raidal}}, \bibnamefont{and}
  \bibinfo{author}{\bibfnamefont{H.}~\bibnamefont{Veermae}},
  \emph{\bibinfo{title}{{Implications of Effective Axial-Vector Coupling of
  Gluon for $t\bar{t}$ Spin Polarizations at the LHC}}},
  \bibinfo{journal}{Phys.Rev.} \textbf{\bibinfo{volume}{D87}},
  \bibinfo{pages}{054001} (\bibinfo{year}{2013}), \eprint{1212.3272}.

\bibitem[{\citenamefont{Falkowski
  et~al.}(2013{\natexlab{b}})\citenamefont{Falkowski, Mangano, Martin, Perez,
  and Winter}}]{Falkowski:2012cu}
\bibinfo{author}{\bibfnamefont{A.}~\bibnamefont{Falkowski}},
  \bibinfo{author}{\bibfnamefont{M.~L.} \bibnamefont{Mangano}},
  \bibinfo{author}{\bibfnamefont{A.}~\bibnamefont{Martin}},
  \bibinfo{author}{\bibfnamefont{G.}~\bibnamefont{Perez}}, \bibnamefont{and}
  \bibinfo{author}{\bibfnamefont{J.}~\bibnamefont{Winter}},
  \emph{\bibinfo{title}{{Data Driving the Top Quark Forward--Backward Asymmetry
  with a Lepton-Based Handle}}}, \bibinfo{journal}{Phys.Rev.}
  \textbf{\bibinfo{volume}{D87}}, \bibinfo{pages}{034039}
  (\bibinfo{year}{2013}{\natexlab{b}}), \eprint{1212.4003}.

\bibitem[{\citenamefont{Perelstein and Weiler}(2009)}]{Perelstein:2008zt}
\bibinfo{author}{\bibfnamefont{M.}~\bibnamefont{Perelstein}} \bibnamefont{and}
  \bibinfo{author}{\bibfnamefont{A.}~\bibnamefont{Weiler}},
  \emph{\bibinfo{title}{{Polarized Tops from Stop Decays at the LHC}}},
  \bibinfo{journal}{JHEP} \textbf{\bibinfo{volume}{0903}}, \bibinfo{pages}{141}
  (\bibinfo{year}{2009}), \eprint{0811.1024}.

\bibitem[{\citenamefont{Berger et~al.}(2012)\citenamefont{Berger, Cao, Yu, and
  Zhang}}]{Berger:2012an}
\bibinfo{author}{\bibfnamefont{E.~L.} \bibnamefont{Berger}},
  \bibinfo{author}{\bibfnamefont{Q.-H.} \bibnamefont{Cao}},
  \bibinfo{author}{\bibfnamefont{J.-H.} \bibnamefont{Yu}}, \bibnamefont{and}
  \bibinfo{author}{\bibfnamefont{H.}~\bibnamefont{Zhang}},
  \emph{\bibinfo{title}{{Measuring Top Quark Polarization in Top Pair plus
  Missing Energy Events}}}, \bibinfo{journal}{Phys.Rev.Lett.}
  \textbf{\bibinfo{volume}{109}}, \bibinfo{pages}{152004}
  (\bibinfo{year}{2012}), \eprint{1207.1101}.

\bibitem[{\citenamefont{Bhattacherjee et~al.}(2013)\citenamefont{Bhattacherjee,
  Mandal, and Nojiri}}]{Bhattacherjee:2012ir}
\bibinfo{author}{\bibfnamefont{B.}~\bibnamefont{Bhattacherjee}},
  \bibinfo{author}{\bibfnamefont{S.~K.} \bibnamefont{Mandal}},
  \bibnamefont{and} \bibinfo{author}{\bibfnamefont{M.}~\bibnamefont{Nojiri}},
  \emph{\bibinfo{title}{{Top Polarization and Stop Mixing from Boosted Jet
  Substructure}}}, \bibinfo{journal}{JHEP} \textbf{\bibinfo{volume}{1303}},
  \bibinfo{pages}{105} (\bibinfo{year}{2013}), \eprint{1211.7261}.

\bibitem[{\citenamefont{Belanger et~al.}(2013)\citenamefont{Belanger, Godbole,
  Hartgring, and Niessen}}]{Belanger:2012tm}
\bibinfo{author}{\bibfnamefont{G.}~\bibnamefont{Belanger}},
  \bibinfo{author}{\bibfnamefont{R.}~\bibnamefont{Godbole}},
  \bibinfo{author}{\bibfnamefont{L.}~\bibnamefont{Hartgring}},
  \bibnamefont{and} \bibinfo{author}{\bibfnamefont{I.}~\bibnamefont{Niessen}},
  \emph{\bibinfo{title}{{Top Polarization in Stop Production at the LHC}}},
  \bibinfo{journal}{JHEP} \textbf{\bibinfo{volume}{1305}}, \bibinfo{pages}{167}
  (\bibinfo{year}{2013}), \eprint{1212.3526}.

\bibitem[{\citenamefont{Baumgart and
  Tweedie}(2013{\natexlab{b}})}]{Baumgart:2013yra}
\bibinfo{author}{\bibfnamefont{M.}~\bibnamefont{Baumgart}} \bibnamefont{and}
  \bibinfo{author}{\bibfnamefont{B.}~\bibnamefont{Tweedie}},
  \emph{\bibinfo{title}{{Transverse Top Quark Polarization and the $t
  \overline{t}$ Forward-Backward Asymmetry}}}, \bibinfo{journal}{JHEP}
  \textbf{\bibinfo{volume}{1308}}, \bibinfo{pages}{072}
  (\bibinfo{year}{2013}{\natexlab{b}}), \eprint{1303.1200}.

\bibitem[{\citenamefont{Abazov et~al.}(2012)}]{Abazov:2011gi}
\bibinfo{author}{\bibfnamefont{V.~M.} \bibnamefont{Abazov}}
  \bibnamefont{et~al.} (\bibinfo{collaboration}{D0 Collaboration}),
  \emph{\bibinfo{title}{{Evidence for Spin Correlation in $t\bar{t}$
  Production}}}, \bibinfo{journal}{Phys.Rev.Lett.}
  \textbf{\bibinfo{volume}{108}}, \bibinfo{pages}{032004}
  (\bibinfo{year}{2012}), \eprint{1110.4194}.

\bibitem[{\citenamefont{{ATLAS
  collaboration}}(2013)}]{TheATLAScollaboration:2013gja}
\bibinfo{author}{\bibnamefont{{ATLAS collaboration}}},
  \emph{\bibinfo{title}{{Measurements of Spin Correlation in Top-Antitop Quark
  Events from Proton-Proton Collisions at $\sqrt{s} = $7~TeV Using the ATLAS
  Detector}}} (\bibinfo{year}{2013}), \eprint{{ATLAS-CONF-2013-101}}.

\bibitem[{\citenamefont{Chatrchyan
  et~al.}(2013{\natexlab{a}})}]{Chatrchyan:2013wua}
\bibinfo{author}{\bibfnamefont{S.}~\bibnamefont{Chatrchyan}}
  \bibnamefont{et~al.} (\bibinfo{collaboration}{CMS Collaboration}),
  \emph{\bibinfo{title}{{Measurements of $t\bar{t}$ Spin Correlations and
  Top-Quark Polarization Using Dilepton Final States in $pp$ Collisions at
  $\sqrt{s}$ = 7~TeV}}} (\bibinfo{year}{2013}{\natexlab{a}}),
  \eprint{1311.3924}.

\bibitem[{\citenamefont{Chatrchyan
  et~al.}(2013{\natexlab{b}})}]{Chatrchyan:2013xza}
\bibinfo{author}{\bibfnamefont{S.}~\bibnamefont{Chatrchyan}}
  \bibnamefont{et~al.} (\bibinfo{collaboration}{CMS Collaboration}),
  \emph{\bibinfo{title}{{Measurement of the Top-Quark Mass in All-Jets
  $t\bar{t}$ Events in $pp$ Collisions at $\sqrt{s}$=7~TeV}}}
  (\bibinfo{year}{2013}{\natexlab{b}}), \eprint{1307.4617}.

\bibitem[{\citenamefont{Aad et~al.}(2012)}]{ATLAS:2012aj}
\bibinfo{author}{\bibfnamefont{G.}~\bibnamefont{Aad}} \bibnamefont{et~al.}
  (\bibinfo{collaboration}{ATLAS Collaboration}),
  \emph{\bibinfo{title}{{Measurement of the Top Quark Mass with the Template
  Method in the $t \bar{t} \to $ Lepton + Jets Channel Using ATLAS Data}}},
  \bibinfo{journal}{Eur.Phys.J.} \textbf{\bibinfo{volume}{C72}},
  \bibinfo{pages}{2046} (\bibinfo{year}{2012}), \eprint{1203.5755}.

\bibitem[{\citenamefont{Chatrchyan
  et~al.}(2013{\natexlab{c}})}]{Chatrchyan:2013lca}
\bibinfo{author}{\bibfnamefont{S.}~\bibnamefont{Chatrchyan}}
  \bibnamefont{et~al.} (\bibinfo{collaboration}{CMS Collaboration}),
  \emph{\bibinfo{title}{{Searches for Anomalous $t\bar t$ Production in $pp$
  Collisions at $\sqrt(s)=8$~TeV}}}, \bibinfo{journal}{Phys.Rev.Lett.}
  \textbf{\bibinfo{volume}{111}}, \bibinfo{pages}{211804}
  (\bibinfo{year}{2013}{\natexlab{c}}), \eprint{1309.2030}.

\bibitem[{\citenamefont{Aad et~al.}(2013{\natexlab{a}})}]{Aad:2012raa}
\bibinfo{author}{\bibfnamefont{G.}~\bibnamefont{Aad}} \bibnamefont{et~al.}
  (\bibinfo{collaboration}{ATLAS Collaboration}), \emph{\bibinfo{title}{{Search
  for Resonances Decaying into Top-Quark Pairs Using Fully Hadronic Decays in
  $pp$ Collisions with ATLAS at $\sqrt{s}=7$~TeV}}}, \bibinfo{journal}{JHEP}
  \textbf{\bibinfo{volume}{1301}}, \bibinfo{pages}{116}
  (\bibinfo{year}{2013}{\natexlab{a}}), \eprint{1211.2202}.

\bibitem[{\citenamefont{Plehn et~al.}(2010)\citenamefont{Plehn, Spannowsky,
  Takeuchi, and Zerwas}}]{Plehn:2010st}
\bibinfo{author}{\bibfnamefont{T.}~\bibnamefont{Plehn}},
  \bibinfo{author}{\bibfnamefont{M.}~\bibnamefont{Spannowsky}},
  \bibinfo{author}{\bibfnamefont{M.}~\bibnamefont{Takeuchi}}, \bibnamefont{and}
  \bibinfo{author}{\bibfnamefont{D.}~\bibnamefont{Zerwas}},
  \emph{\bibinfo{title}{{Stop Reconstruction with Tagged Tops}}},
  \bibinfo{journal}{JHEP} \textbf{\bibinfo{volume}{1010}}, \bibinfo{pages}{078}
  (\bibinfo{year}{2010}), \eprint{1006.2833}.

\bibitem[{\citenamefont{Kaplan et~al.}(2008)\citenamefont{Kaplan, Rehermann,
  Schwartz, and Tweedie}}]{Kaplan:2008ie}
\bibinfo{author}{\bibfnamefont{D.~E.} \bibnamefont{Kaplan}},
  \bibinfo{author}{\bibfnamefont{K.}~\bibnamefont{Rehermann}},
  \bibinfo{author}{\bibfnamefont{M.~D.} \bibnamefont{Schwartz}},
  \bibnamefont{and} \bibinfo{author}{\bibfnamefont{B.}~\bibnamefont{Tweedie}},
  \emph{\bibinfo{title}{{Top Tagging: A Method for Identifying Boosted
  Hadronically Decaying Top Quarks}}}, \bibinfo{journal}{Phys. Rev. Lett.}
  \textbf{\bibinfo{volume}{101}}, \bibinfo{pages}{142001}
  (\bibinfo{year}{2008}), \eprint{0806.0848}.

\bibitem[{\citenamefont{{CMS Collaboration}}(2013{\natexlab{a}})}]{CMS:2013pfa}
\bibinfo{author}{\bibnamefont{{CMS Collaboration}}},
  \emph{\bibinfo{title}{{Measurement of the $W$-Boson Helicity in Top Decays
  from $t \bar t$ Production in Lepton+Jets Events at the LHC at $\sqrt{s} =
  8$~TeV}}} (\bibinfo{year}{2013}{\natexlab{a}}),
  \eprint{{CMS-PAS-TOP-13-008}}.

\bibitem[{\citenamefont{Chatrchyan
  et~al.}(2013{\natexlab{d}})}]{Chatrchyan:2013jna}
\bibinfo{author}{\bibfnamefont{S.}~\bibnamefont{Chatrchyan}}
  \bibnamefont{et~al.} (\bibinfo{collaboration}{CMS Collaboration}),
  \emph{\bibinfo{title}{{Measurement of the $W$-Boson Helicity in Top-Quark
  Decays from $t\bar{t}$ Production in Lepton+Jets Events in $pp$ Collisions at
  $\sqrt{s} =$ 7~TeV}}}, \bibinfo{journal}{JHEP}
  \textbf{\bibinfo{volume}{1310}}, \bibinfo{pages}{167}
  (\bibinfo{year}{2013}{\natexlab{d}}), \eprint{1308.3879}.

\bibitem[{\citenamefont{{ATLAS
  Collaboration}}(2013{\natexlab{a}})}]{ATLAS:2013tla}
\bibinfo{author}{\bibnamefont{{ATLAS Collaboration}}},
  \emph{\bibinfo{title}{{Combination of the ATLAS and CMS Measurements of the
  $W$-Boson Polarization in Top-Quark Decays}}}
  (\bibinfo{year}{2013}{\natexlab{a}}), \eprint{{ATLAS-CONF-2013-033,
  CMS-PAS-TOP-12-025}}.

\bibitem[{\citenamefont{Krohn et~al.}(2013{\natexlab{a}})\citenamefont{Krohn,
  Schwartz, Lin, and Waalewijn}}]{Krohn:2012fg}
\bibinfo{author}{\bibfnamefont{D.}~\bibnamefont{Krohn}},
  \bibinfo{author}{\bibfnamefont{M.~D.} \bibnamefont{Schwartz}},
  \bibinfo{author}{\bibfnamefont{T.}~\bibnamefont{Lin}}, \bibnamefont{and}
  \bibinfo{author}{\bibfnamefont{W.~J.} \bibnamefont{Waalewijn}},
  \emph{\bibinfo{title}{{Jet Charge at the LHC}}},
  \bibinfo{journal}{Phys.Rev.Lett.} \textbf{\bibinfo{volume}{110}},
  \bibinfo{pages}{212001} (\bibinfo{year}{2013}{\natexlab{a}}),
  \eprint{1209.2421}.

\bibitem[{\citenamefont{{CDF Collaboration}}(2010)}]{CDF10211}
\bibinfo{author}{\bibnamefont{{CDF Collaboration}}},
  \emph{\bibinfo{title}{{Measurement of $t\bar t$ Helicity Fractions and Spin
  Correlation Using Reconstructed Lepton+Jets Events}}} (\bibinfo{year}{2010}),
  \eprint{{CDF NOTE 10211}}.

\bibitem[{\citenamefont{{CMS Collaboration}}(2013{\natexlab{b}})}]{CMS:2013vea}
\bibinfo{author}{\bibnamefont{{CMS Collaboration}}},
  \emph{\bibinfo{title}{{Performance of $b$ Tagging at $\sqrt{s} = 8$~TeV in
  Multijet, $t\bar t$ and Boosted Topology Events}}}
  (\bibinfo{year}{2013}{\natexlab{b}}), \eprint{{CMS-PAS-BTV-13-001}}.

\bibitem[{\citenamefont{{ATLAS Collaboration}}(2009)}]{ATLAS:2009elr}
\bibinfo{author}{\bibnamefont{{ATLAS Collaboration}}},
  \emph{\bibinfo{title}{{ATLAS Sensitivity to the Standard Model Higgs in the
  $HW$ and $HZ$ Channels at High Transverse Momenta}}} (\bibinfo{year}{2009}),
  \eprint{{ATL-PHYS-PUB-2009-088}}.

\bibitem[{\citenamefont{Sjostrand et~al.}(2006)\citenamefont{Sjostrand, Mrenna,
  and Skands}}]{Sjostrand:2006za}
\bibinfo{author}{\bibfnamefont{T.}~\bibnamefont{Sjostrand}},
  \bibinfo{author}{\bibfnamefont{S.}~\bibnamefont{Mrenna}}, \bibnamefont{and}
  \bibinfo{author}{\bibfnamefont{P.}~\bibnamefont{Skands}},
  \emph{\bibinfo{title}{{PYTHIA 6.4 Physics and Manual}}},
  \bibinfo{journal}{JHEP} \textbf{\bibinfo{volume}{05}}, \bibinfo{pages}{026}
  (\bibinfo{year}{2006}), \eprint{hep-ph/0603175}.

\bibitem[{\citenamefont{Campbell and Ellis}(2012)}]{Campbell:2012uf}
\bibinfo{author}{\bibfnamefont{J.~M.} \bibnamefont{Campbell}} \bibnamefont{and}
  \bibinfo{author}{\bibfnamefont{R.~K.} \bibnamefont{Ellis}},
  \emph{\bibinfo{title}{{Top-Quark Processes at NLO in Production and Decay}}}
  (\bibinfo{year}{2012}), \eprint{1204.1513}.

\bibitem[{\citenamefont{Jezabek and Kuhn}(1989)}]{Jezabek:1988iv}
\bibinfo{author}{\bibfnamefont{M.}~\bibnamefont{Jezabek}} \bibnamefont{and}
  \bibinfo{author}{\bibfnamefont{J.~H.} \bibnamefont{Kuhn}},
  \emph{\bibinfo{title}{{QCD Corrections to Semileptonic Decays of Heavy
  Quarks}}}, \bibinfo{journal}{Nucl.Phys.} \textbf{\bibinfo{volume}{B314}},
  \bibinfo{pages}{1} (\bibinfo{year}{1989}).

\bibitem[{\citenamefont{Catani and Seymour}(1997)}]{Catani:1996vz}
\bibinfo{author}{\bibfnamefont{S.}~\bibnamefont{Catani}} \bibnamefont{and}
  \bibinfo{author}{\bibfnamefont{M.}~\bibnamefont{Seymour}},
  \emph{\bibinfo{title}{{A General Algorithm for Calculating Jet Cross-Sections
  in NLO QCD}}}, \bibinfo{journal}{Nucl.Phys.} \textbf{\bibinfo{volume}{B485}},
  \bibinfo{pages}{291} (\bibinfo{year}{1997}), \eprint{hep-ph/9605323}.

\bibitem[{\citenamefont{Corcella and Mitov}(2002)}]{Corcella:2001hz}
\bibinfo{author}{\bibfnamefont{G.}~\bibnamefont{Corcella}} \bibnamefont{and}
  \bibinfo{author}{\bibfnamefont{A.~D.} \bibnamefont{Mitov}},
  \emph{\bibinfo{title}{{Bottom Quark Fragmentation in Top Quark Decay}}},
  \bibinfo{journal}{Nucl.Phys.} \textbf{\bibinfo{volume}{B623}},
  \bibinfo{pages}{247} (\bibinfo{year}{2002}), \eprint{hep-ph/0110319}.

\bibitem[{\citenamefont{Czarnecki et~al.}(1991)\citenamefont{Czarnecki,
  Jezabek, and Kuhn}}]{Czarnecki:1990pe}
\bibinfo{author}{\bibfnamefont{A.}~\bibnamefont{Czarnecki}},
  \bibinfo{author}{\bibfnamefont{M.}~\bibnamefont{Jezabek}}, \bibnamefont{and}
  \bibinfo{author}{\bibfnamefont{J.~H.} \bibnamefont{Kuhn}},
  \emph{\bibinfo{title}{{Lepton Spectra From Decays of Polarized Top Quarks}}},
  \bibinfo{journal}{Nucl.Phys.} \textbf{\bibinfo{volume}{B351}},
  \bibinfo{pages}{70} (\bibinfo{year}{1991}).

\bibitem[{\citenamefont{Czarnecki and Jezabek}(1994)}]{Czarnecki:1994pu}
\bibinfo{author}{\bibfnamefont{A.}~\bibnamefont{Czarnecki}} \bibnamefont{and}
  \bibinfo{author}{\bibfnamefont{M.}~\bibnamefont{Jezabek}},
  \emph{\bibinfo{title}{{Distributions of Leptons in Decays of Polarized Heavy
  Quarks}}}, \bibinfo{journal}{Nucl.Phys.} \textbf{\bibinfo{volume}{B427}},
  \bibinfo{pages}{3} (\bibinfo{year}{1994}), \eprint{hep-ph/9402326}.

\bibitem[{\citenamefont{Fischer et~al.}(2001)\citenamefont{Fischer, Groote,
  Korner, and Mauser}}]{Fischer:2000kx}
\bibinfo{author}{\bibfnamefont{M.}~\bibnamefont{Fischer}},
  \bibinfo{author}{\bibfnamefont{S.}~\bibnamefont{Groote}},
  \bibinfo{author}{\bibfnamefont{J.}~\bibnamefont{Korner}}, \bibnamefont{and}
  \bibinfo{author}{\bibfnamefont{M.}~\bibnamefont{Mauser}},
  \emph{\bibinfo{title}{{Longitudinal, Transverse-Plus and Transverse-Minus
  $W$-Bosons in Unpolarized Top Quark Decays at ${\cal O}(\alpha_s)$}}},
  \bibinfo{journal}{Phys.Rev.} \textbf{\bibinfo{volume}{D63}},
  \bibinfo{pages}{031501} (\bibinfo{year}{2001}), \eprint{hep-ph/0011075}.

\bibitem[{\citenamefont{Groote et~al.}(2013)\citenamefont{Groote, Korner, and
  Tuvike}}]{Groote:2013xt}
\bibinfo{author}{\bibfnamefont{S.}~\bibnamefont{Groote}},
  \bibinfo{author}{\bibfnamefont{J.}~\bibnamefont{Korner}}, \bibnamefont{and}
  \bibinfo{author}{\bibfnamefont{P.}~\bibnamefont{Tuvike}},
  \emph{\bibinfo{title}{{Fully Analytical ${\cal O}(\alpha_s)$ Results for
  On-Shell and Off-Shell Polarized $W$-Boson Decays into Massive Quark
  Pairs}}}, \bibinfo{journal}{Eur.Phys.J.} \textbf{\bibinfo{volume}{C73}},
  \bibinfo{pages}{2454} (\bibinfo{year}{2013}), \eprint{1301.0881}.

\bibitem[{\citenamefont{Brandenburg et~al.}(2002)\citenamefont{Brandenburg, Si,
  and Uwer}}]{Brandenburg:2002xr}
\bibinfo{author}{\bibfnamefont{A.}~\bibnamefont{Brandenburg}},
  \bibinfo{author}{\bibfnamefont{Z.~G.} \bibnamefont{Si}}, \bibnamefont{and}
  \bibinfo{author}{\bibfnamefont{P.}~\bibnamefont{Uwer}},
  \emph{\bibinfo{title}{{QCD-Corrected Spin Analysing Power of Jets in Decays
  of Polarized Top Quarks}}}, \bibinfo{journal}{Phys. Lett.}
  \textbf{\bibinfo{volume}{B539}}, \bibinfo{pages}{235} (\bibinfo{year}{2002}),
  \eprint{hep-ph/0205023}.

\bibitem[{\citenamefont{Alwall et~al.}(2011)\citenamefont{Alwall, Herquet,
  Maltoni, Mattelaer, and Stelzer}}]{Alwall:2011uj}
\bibinfo{author}{\bibfnamefont{J.}~\bibnamefont{Alwall}},
  \bibinfo{author}{\bibfnamefont{M.}~\bibnamefont{Herquet}},
  \bibinfo{author}{\bibfnamefont{F.}~\bibnamefont{Maltoni}},
  \bibinfo{author}{\bibfnamefont{O.}~\bibnamefont{Mattelaer}},
  \bibnamefont{and} \bibinfo{author}{\bibfnamefont{T.}~\bibnamefont{Stelzer}},
  \emph{\bibinfo{title}{{MadGraph 5 : Going Beyond}}}, \bibinfo{journal}{JHEP}
  \textbf{\bibinfo{volume}{1106}}, \bibinfo{pages}{128} (\bibinfo{year}{2011}),
  \eprint{1106.0522}.

\bibitem[{\citenamefont{Norrbin and Sjostrand}(2001)}]{Norrbin:2000uu}
\bibinfo{author}{\bibfnamefont{E.}~\bibnamefont{Norrbin}} \bibnamefont{and}
  \bibinfo{author}{\bibfnamefont{T.}~\bibnamefont{Sjostrand}},
  \emph{\bibinfo{title}{{QCD Radiation off Heavy Particles}}},
  \bibinfo{journal}{Nucl.Phys.} \textbf{\bibinfo{volume}{B603}},
  \bibinfo{pages}{297} (\bibinfo{year}{2001}), \eprint{hep-ph/0010012}.

\bibitem[{\citenamefont{Cacciari et~al.}(2008)\citenamefont{Cacciari, Salam,
  and Soyez}}]{Cacciari:2008gp}
\bibinfo{author}{\bibfnamefont{M.}~\bibnamefont{Cacciari}},
  \bibinfo{author}{\bibfnamefont{G.~P.} \bibnamefont{Salam}}, \bibnamefont{and}
  \bibinfo{author}{\bibfnamefont{G.}~\bibnamefont{Soyez}},
  \emph{\bibinfo{title}{{The Anti-$k_t$ Jet Clustering Algorithm}}},
  \bibinfo{journal}{JHEP} \textbf{\bibinfo{volume}{04}}, \bibinfo{pages}{063}
  (\bibinfo{year}{2008}), \eprint{0802.1189}.

\bibitem[{\citenamefont{Ovyn et~al.}(2009)\citenamefont{Ovyn, Rouby, and
  Lemaitre}}]{Ovyn:2009tx}
\bibinfo{author}{\bibfnamefont{S.}~\bibnamefont{Ovyn}},
  \bibinfo{author}{\bibfnamefont{X.}~\bibnamefont{Rouby}}, \bibnamefont{and}
  \bibinfo{author}{\bibfnamefont{V.}~\bibnamefont{Lemaitre}},
  \emph{\bibinfo{title}{{DELPHES, A Framework for Fast Simulation of a Generic
  Collider Experiment}}} (\bibinfo{year}{2009}), \eprint{0903.2225}.

\bibitem[{\citenamefont{de~Favereau et~al.}(2013)\citenamefont{de~Favereau,
  Delaere, Demin, Giammanco, Lemaître et~al.}}]{deFavereau:2013fsa}
\bibinfo{author}{\bibfnamefont{J.}~\bibnamefont{de~Favereau}},
  \bibinfo{author}{\bibfnamefont{C.}~\bibnamefont{Delaere}},
  \bibinfo{author}{\bibfnamefont{P.}~\bibnamefont{Demin}},
  \bibinfo{author}{\bibfnamefont{A.}~\bibnamefont{Giammanco}},
  \bibinfo{author}{\bibfnamefont{V.}~\bibnamefont{Lemaître}},
  \bibnamefont{et~al.}, \emph{\bibinfo{title}{{DELPHES 3, A Modular Framework
  for Fast Simulation of a Generic Collider Experiment}}}
  (\bibinfo{year}{2013}), \eprint{1307.6346}.

\bibitem[{\citenamefont{Krohn et~al.}(2010)\citenamefont{Krohn, Thaler, and
  Wang}}]{Krohn:2009th}
\bibinfo{author}{\bibfnamefont{D.}~\bibnamefont{Krohn}},
  \bibinfo{author}{\bibfnamefont{J.}~\bibnamefont{Thaler}}, \bibnamefont{and}
  \bibinfo{author}{\bibfnamefont{L.-T.} \bibnamefont{Wang}},
  \emph{\bibinfo{title}{{Jet Trimming}}}, \bibinfo{journal}{JHEP}
  \textbf{\bibinfo{volume}{1002}}, \bibinfo{pages}{084} (\bibinfo{year}{2010}),
  \eprint{0912.1342}.

\bibitem[{\citenamefont{Krohn et~al.}(2013{\natexlab{b}})\citenamefont{Krohn,
  Low, Schwartz, and Wang}}]{Krohn:2013lba}
\bibinfo{author}{\bibfnamefont{D.}~\bibnamefont{Krohn}},
  \bibinfo{author}{\bibfnamefont{M.}~\bibnamefont{Low}},
  \bibinfo{author}{\bibfnamefont{M.~D.} \bibnamefont{Schwartz}},
  \bibnamefont{and} \bibinfo{author}{\bibfnamefont{L.-T.} \bibnamefont{Wang}},
  \emph{\bibinfo{title}{{Jet Cleansing: Pileup Removal at High Luminosity}}}
  (\bibinfo{year}{2013}{\natexlab{b}}), \eprint{1309.4777}.

\bibitem[{\citenamefont{Aad et~al.}(2013{\natexlab{b}})}]{Aad:2013gja}
\bibinfo{author}{\bibfnamefont{G.}~\bibnamefont{Aad}} \bibnamefont{et~al.}
  (\bibinfo{collaboration}{ATLAS Collaboration}),
  \emph{\bibinfo{title}{{Performance of Jet Substructure Techniques for
  Large-$R$ Jets in Proton-Proton Collisions at $\sqrt{s}$ = 7 TeV Using the
  ATLAS Detector}}}, \bibinfo{journal}{JHEP} \textbf{\bibinfo{volume}{1309}},
  \bibinfo{pages}{076} (\bibinfo{year}{2013}{\natexlab{b}}),
  \eprint{1306.4945}.

\bibitem[{\citenamefont{Calkins et~al.}(2013)\citenamefont{Calkins, Chekanov,
  Conway, Dolen, Erbacher et~al.}}]{Calkins:2013ega}
\bibinfo{author}{\bibfnamefont{R.}~\bibnamefont{Calkins}},
  \bibinfo{author}{\bibfnamefont{S.}~\bibnamefont{Chekanov}},
  \bibinfo{author}{\bibfnamefont{J.}~\bibnamefont{Conway}},
  \bibinfo{author}{\bibfnamefont{J.}~\bibnamefont{Dolen}},
  \bibinfo{author}{\bibfnamefont{R.}~\bibnamefont{Erbacher}},
  \bibnamefont{et~al.}, \emph{\bibinfo{title}{{Reconstructing Top Quarks at the
  Upgraded LHC and at Future Accelerators}}} (\bibinfo{year}{2013}),
  \eprint{1307.6908}.

\bibitem[{\citenamefont{Bai et~al.}(2014)\citenamefont{Bai, Katz, and
  Tweedie}}]{Bai:2013xla}
\bibinfo{author}{\bibfnamefont{Y.}~\bibnamefont{Bai}},
  \bibinfo{author}{\bibfnamefont{A.}~\bibnamefont{Katz}}, \bibnamefont{and}
  \bibinfo{author}{\bibfnamefont{B.}~\bibnamefont{Tweedie}},
  \emph{\bibinfo{title}{{Pulling Out All the Stops: Searching for RPV SUSY with
  Stop-Jets}}}, \bibinfo{journal}{JHEP} \textbf{\bibinfo{volume}{1401}},
  \bibinfo{pages}{040} (\bibinfo{year}{2014}), \eprint{1309.6631}.

\bibitem[{\citenamefont{Sjostrand et~al.}(2008)\citenamefont{Sjostrand, Mrenna,
  and Skands}}]{Sjostrand:2007gs}
\bibinfo{author}{\bibfnamefont{T.}~\bibnamefont{Sjostrand}},
  \bibinfo{author}{\bibfnamefont{S.}~\bibnamefont{Mrenna}}, \bibnamefont{and}
  \bibinfo{author}{\bibfnamefont{P.~Z.} \bibnamefont{Skands}},
  \emph{\bibinfo{title}{{A Brief Introduction to PYTHIA 8.1}}},
  \bibinfo{journal}{Comput.Phys.Commun.} \textbf{\bibinfo{volume}{178}},
  \bibinfo{pages}{852} (\bibinfo{year}{2008}), \eprint{0710.3820}.

\bibitem[{\citenamefont{Langacker}(2009)}]{Langacker:2008yv}
\bibinfo{author}{\bibfnamefont{P.}~\bibnamefont{Langacker}},
  \emph{\bibinfo{title}{{The Physics of Heavy $Z^\prime$ Gauge Bosons}}},
  \bibinfo{journal}{Rev.Mod.Phys.} \textbf{\bibinfo{volume}{81}},
  \bibinfo{pages}{1199} (\bibinfo{year}{2009}), \eprint{0801.1345}.

\bibitem[{\citenamefont{Arkani-Hamed et~al.}(2001)\citenamefont{Arkani-Hamed,
  Cohen, and Georgi}}]{ArkaniHamed:2001nc}
\bibinfo{author}{\bibfnamefont{N.}~\bibnamefont{Arkani-Hamed}},
  \bibinfo{author}{\bibfnamefont{A.~G.} \bibnamefont{Cohen}}, \bibnamefont{and}
  \bibinfo{author}{\bibfnamefont{H.}~\bibnamefont{Georgi}},
  \emph{\bibinfo{title}{{Electroweak Symmetry Breaking from Dimensional
  Deconstruction}}}, \bibinfo{journal}{Phys.Lett.}
  \textbf{\bibinfo{volume}{B513}}, \bibinfo{pages}{232} (\bibinfo{year}{2001}),
  \eprint{hep-ph/0105239}.

\bibitem[{\citenamefont{Agashe et~al.}(2003)\citenamefont{Agashe, Delgado, May,
  and Sundrum}}]{Agashe:2003zs}
\bibinfo{author}{\bibfnamefont{K.}~\bibnamefont{Agashe}},
  \bibinfo{author}{\bibfnamefont{A.}~\bibnamefont{Delgado}},
  \bibinfo{author}{\bibfnamefont{M.~J.} \bibnamefont{May}}, \bibnamefont{and}
  \bibinfo{author}{\bibfnamefont{R.}~\bibnamefont{Sundrum}},
  \emph{\bibinfo{title}{{RS1, Custodial Isospin and Precision Tests}}},
  \bibinfo{journal}{JHEP} \textbf{\bibinfo{volume}{0308}}, \bibinfo{pages}{050}
  (\bibinfo{year}{2003}), \eprint{hep-ph/0308036}.

\bibitem[{\citenamefont{{ATLAS
  Collaboration}}(2013{\natexlab{b}})}]{TheATLAScollaboration:2013kha}
\bibinfo{author}{\bibnamefont{{ATLAS Collaboration}}}, \emph{\bibinfo{title}{{A
  Search for $t\bar{t}$ Resonances in the Lepton Plus Jets Final State with
  ATLAS Using 14~fb$^{-1}$ of $pp$ Collisions at $\sqrt{s}=8$~TeV}}}
  (\bibinfo{year}{2013}{\natexlab{b}}), \eprint{{ATLAS-CONF-2013-052}}.

\bibitem[{\citenamefont{Rehermann and Tweedie}(2011)}]{Rehermann:2010vq}
\bibinfo{author}{\bibfnamefont{K.}~\bibnamefont{Rehermann}} \bibnamefont{and}
  \bibinfo{author}{\bibfnamefont{B.}~\bibnamefont{Tweedie}},
  \emph{\bibinfo{title}{{Efficient Identification of Boosted Semileptonic Top
  Quarks at the LHC}}}, \bibinfo{journal}{JHEP}
  \textbf{\bibinfo{volume}{1103}}, \bibinfo{pages}{059} (\bibinfo{year}{2011}),
  \eprint{1007.2221}.

\bibitem[{\citenamefont{Cacciari and Salam}(2006)}]{Cacciari:2005hq}
\bibinfo{author}{\bibfnamefont{M.}~\bibnamefont{Cacciari}} \bibnamefont{and}
  \bibinfo{author}{\bibfnamefont{G.~P.} \bibnamefont{Salam}},
  \emph{\bibinfo{title}{{Dispelling the $N^{3}$ Myth for the $k_t$
  Jet-Finder}}}, \bibinfo{journal}{Phys. Lett.}
  \textbf{\bibinfo{volume}{B641}}, \bibinfo{pages}{57} (\bibinfo{year}{2006}),
  \eprint{hep-ph/0512210}.

\bibitem[{\citenamefont{Dokshitzer et~al.}(1997)\citenamefont{Dokshitzer,
  Leder, Moretti, and Webber}}]{Dokshitzer:1997in}
\bibinfo{author}{\bibfnamefont{Y.~L.} \bibnamefont{Dokshitzer}},
  \bibinfo{author}{\bibfnamefont{G.~D.} \bibnamefont{Leder}},
  \bibinfo{author}{\bibfnamefont{S.}~\bibnamefont{Moretti}}, \bibnamefont{and}
  \bibinfo{author}{\bibfnamefont{B.~R.} \bibnamefont{Webber}},
  \emph{\bibinfo{title}{{Better Jet Clustering Algorithms}}},
  \bibinfo{journal}{JHEP} \textbf{\bibinfo{volume}{08}}, \bibinfo{pages}{001}
  (\bibinfo{year}{1997}), \eprint{hep-ph/9707323}.

\bibitem[{\citenamefont{Abdesselam et~al.}(2011)\citenamefont{Abdesselam,
  Kuutmann, Bitenc, Brooijmans, Butterworth et~al.}}]{Abdesselam:2010pt}
\bibinfo{author}{\bibfnamefont{A.}~\bibnamefont{Abdesselam}},
  \bibinfo{author}{\bibfnamefont{E.~B.} \bibnamefont{Kuutmann}},
  \bibinfo{author}{\bibfnamefont{U.}~\bibnamefont{Bitenc}},
  \bibinfo{author}{\bibfnamefont{G.}~\bibnamefont{Brooijmans}},
  \bibinfo{author}{\bibfnamefont{J.}~\bibnamefont{Butterworth}},
  \bibnamefont{et~al.}, \emph{\bibinfo{title}{{Boosted Objects: A Probe of
  Beyond the Standard Model Physics}}}, \bibinfo{journal}{Eur.Phys.J.}
  \textbf{\bibinfo{volume}{C71}}, \bibinfo{pages}{1661} (\bibinfo{year}{2011}),
  \eprint{1012.5412}.

\bibitem[{\citenamefont{Altheimer et~al.}(2012)\citenamefont{Altheimer, Arora,
  Asquith, Brooijmans, Butterworth et~al.}}]{Altheimer:2012mn}
\bibinfo{author}{\bibfnamefont{A.}~\bibnamefont{Altheimer}},
  \bibinfo{author}{\bibfnamefont{S.}~\bibnamefont{Arora}},
  \bibinfo{author}{\bibfnamefont{L.}~\bibnamefont{Asquith}},
  \bibinfo{author}{\bibfnamefont{G.}~\bibnamefont{Brooijmans}},
  \bibinfo{author}{\bibfnamefont{J.}~\bibnamefont{Butterworth}},
  \bibnamefont{et~al.}, \emph{\bibinfo{title}{{Jet Substructure at the Tevatron
  and LHC: New Results, New Tools, New Benchmarks}}},
  \bibinfo{journal}{J.Phys.G} \textbf{\bibinfo{volume}{G39}},
  \bibinfo{pages}{063001} (\bibinfo{year}{2012}), \eprint{1201.0008}.

\bibitem[{\citenamefont{Altheimer et~al.}(2013)\citenamefont{Altheimer, Arce,
  Asquith, Backus~Mayes, Bergeaas~Kuutmann et~al.}}]{Altheimer:2013yza}
\bibinfo{author}{\bibfnamefont{A.}~\bibnamefont{Altheimer}},
  \bibinfo{author}{\bibfnamefont{A.}~\bibnamefont{Arce}},
  \bibinfo{author}{\bibfnamefont{L.}~\bibnamefont{Asquith}},
  \bibinfo{author}{\bibfnamefont{J.}~\bibnamefont{Backus~Mayes}},
  \bibinfo{author}{\bibfnamefont{E.}~\bibnamefont{Bergeaas~Kuutmann}},
  \bibnamefont{et~al.}, \emph{\bibinfo{title}{{Boosted Objects and Jet
  Substructure at the LHC}}} (\bibinfo{year}{2013}), \eprint{1311.2708}.

\bibitem[{\citenamefont{Butterworth et~al.}(2008)\citenamefont{Butterworth,
  Davison, Rubin, and Salam}}]{Butterworth:2008iy}
\bibinfo{author}{\bibfnamefont{J.~M.} \bibnamefont{Butterworth}},
  \bibinfo{author}{\bibfnamefont{A.~R.} \bibnamefont{Davison}},
  \bibinfo{author}{\bibfnamefont{M.}~\bibnamefont{Rubin}}, \bibnamefont{and}
  \bibinfo{author}{\bibfnamefont{G.~P.} \bibnamefont{Salam}},
  \emph{\bibinfo{title}{{Jet Substructure as a New Higgs Search Channel at the
  LHC}}}, \bibinfo{journal}{Phys.Rev.Lett.} \textbf{\bibinfo{volume}{100}},
  \bibinfo{pages}{242001} (\bibinfo{year}{2008}), \eprint{0802.2470}.

\bibitem[{\citenamefont{Agashe et~al.}(2008)\citenamefont{Agashe, Belyaev,
  Krupovnickas, Perez, and Virzi}}]{Agashe:2006hk}
\bibinfo{author}{\bibfnamefont{K.}~\bibnamefont{Agashe}},
  \bibinfo{author}{\bibfnamefont{A.}~\bibnamefont{Belyaev}},
  \bibinfo{author}{\bibfnamefont{T.}~\bibnamefont{Krupovnickas}},
  \bibinfo{author}{\bibfnamefont{G.}~\bibnamefont{Perez}}, \bibnamefont{and}
  \bibinfo{author}{\bibfnamefont{J.}~\bibnamefont{Virzi}},
  \emph{\bibinfo{title}{{LHC Signals from Warped Extra Dimensions}}},
  \bibinfo{journal}{Phys.Rev.} \textbf{\bibinfo{volume}{D77}},
  \bibinfo{pages}{015003} (\bibinfo{year}{2008}), \eprint{hep-ph/0612015}.

\bibitem[{\citenamefont{Lillie et~al.}(2007)\citenamefont{Lillie, Randall, and
  Wang}}]{Lillie:2007yh}
\bibinfo{author}{\bibfnamefont{B.}~\bibnamefont{Lillie}},
  \bibinfo{author}{\bibfnamefont{L.}~\bibnamefont{Randall}}, \bibnamefont{and}
  \bibinfo{author}{\bibfnamefont{L.-T.} \bibnamefont{Wang}},
  \emph{\bibinfo{title}{{The Bulk RS KK-Gluon at the LHC}}},
  \bibinfo{journal}{JHEP} \textbf{\bibinfo{volume}{0709}}, \bibinfo{pages}{074}
  (\bibinfo{year}{2007}), \eprint{hep-ph/0701166}.

\bibitem[{\citenamefont{{ATLAS
  Collaboration}}(2013{\natexlab{c}})}]{TheATLAScollaboration:2013xha}
\bibinfo{author}{\bibnamefont{{ATLAS Collaboration}}},
  \emph{\bibinfo{title}{{Searches for Direct Scalar Top Pair Production in
  Final States with Two Leptons Using the Stransverse Mass Variable and a
  Multivariate Analysis Technique in $\sqrt{s} = 8$~TeV $pp$ Collisions Using
  20.3~fb$^{-1}$ of ATLAS Data}}} (\bibinfo{year}{2013}{\natexlab{c}}),
  \eprint{{ATLAS-CONF-2013-065}}.

\bibitem[{\citenamefont{{ATLAS
  Collaboration}}(2013{\natexlab{d}})}]{ATLAS:2013pla}
\bibinfo{author}{\bibnamefont{{ATLAS Collaboration}}},
  \emph{\bibinfo{title}{{Search for Direct Top Squark Pair Production in Final
  States with One Isolated Lepton, Jets, and Missing Transverse Momentum in
  $sqrt{s}=8$~TeV $pp$ Collisions using 21~fb$^{-1}$ of ATLAS Data}}}
  (\bibinfo{year}{2013}{\natexlab{d}}), \eprint{{ATLAS-CONF-2013-037}}.

\bibitem[{\citenamefont{{ATLAS
  Collaboration}}(2013{\natexlab{e}})}]{ATLAS:2013cma}
\bibinfo{author}{\bibnamefont{{ATLAS Collaboration}}},
  \emph{\bibinfo{title}{{Search for Direct Production of the Top Squark in the
  All-Hadronic $t\bar t$ + $E_T^{\rm miss}$ Final State in 21~fb$^{-1}$ of
  $p$-$p$ Collisions at $\sqrt{s} = 8$~TeV with the ATLAS Detector}}}
  (\bibinfo{year}{2013}{\natexlab{e}}), \eprint{{ATLAS-CONF-2013-024}}.

\bibitem[{\citenamefont{Chatrchyan
  et~al.}(2013{\natexlab{e}})}]{Chatrchyan:2013xna}
\bibinfo{author}{\bibfnamefont{S.}~\bibnamefont{Chatrchyan}}
  \bibnamefont{et~al.} (\bibinfo{collaboration}{CMS Collaboration}),
  \emph{\bibinfo{title}{{Search for Top-Squark Pair Production in the
  Single-Lepton Final State in $pp$ Collisions at $\sqrt{s} = 8$~TeV}}},
  \bibinfo{journal}{Eur.Phys.J.} \textbf{\bibinfo{volume}{C73}},
  \bibinfo{pages}{2677} (\bibinfo{year}{2013}{\natexlab{e}}),
  \eprint{1308.1586}.

\bibitem[{\citenamefont{{CMS Collaboration}}(2013{\natexlab{c}})}]{CMS:2013cfa}
\bibinfo{author}{\bibnamefont{{CMS Collaboration}}},
  \emph{\bibinfo{title}{{Search for Supersymmetry Using Razor Variables in
  Events with $b$-Jets in $pp$ Collisions at 8~TeV}}}
  (\bibinfo{year}{2013}{\natexlab{c}}), \eprint{{CMS-PAS-SUS-13-004}}.

\bibitem[{\citenamefont{{CMS Collaboration}}(2013{\natexlab{d}})}]{CMS:2013nia}
\bibinfo{author}{\bibnamefont{{CMS Collaboration}}},
  \emph{\bibinfo{title}{{Search for Top Squarks in Multijet Events with Large
  Missing Momentum in Proton-Proton Collisions at 8~TeV}}}
  (\bibinfo{year}{2013}{\natexlab{d}}), \eprint{{CMS-PAS-SUS-13-015}}.

\bibitem[{\citenamefont{Plehn et~al.}(2012)\citenamefont{Plehn, Spannowsky, and
  Takeuchi}}]{Plehn:2012pr}
\bibinfo{author}{\bibfnamefont{T.}~\bibnamefont{Plehn}},
  \bibinfo{author}{\bibfnamefont{M.}~\bibnamefont{Spannowsky}},
  \bibnamefont{and} \bibinfo{author}{\bibfnamefont{M.}~\bibnamefont{Takeuchi}},
  \emph{\bibinfo{title}{{Stop Searches in 2012}}}, \bibinfo{journal}{JHEP}
  \textbf{\bibinfo{volume}{1208}}, \bibinfo{pages}{091} (\bibinfo{year}{2012}),
  \eprint{1205.2696}.

\bibitem[{\citenamefont{Kaplan et~al.}(2012)\citenamefont{Kaplan, Rehermann,
  and Stolarski}}]{Kaplan:2012gd}
\bibinfo{author}{\bibfnamefont{D.~E.} \bibnamefont{Kaplan}},
  \bibinfo{author}{\bibfnamefont{K.}~\bibnamefont{Rehermann}},
  \bibnamefont{and}
  \bibinfo{author}{\bibfnamefont{D.}~\bibnamefont{Stolarski}},
  \emph{\bibinfo{title}{{Searching for Direct Stop Production in Hadronic Top
  Data at the LHC}}}, \bibinfo{journal}{JHEP} \textbf{\bibinfo{volume}{1207}},
  \bibinfo{pages}{119} (\bibinfo{year}{2012}), \eprint{1205.5816}.

\bibitem[{\citenamefont{Gouzevitch et~al.}(2013)\citenamefont{Gouzevitch,
  Oliveira, Rojo, Rosenfeld, Salam et~al.}}]{Gouzevitch:2013qca}
\bibinfo{author}{\bibfnamefont{M.}~\bibnamefont{Gouzevitch}},
  \bibinfo{author}{\bibfnamefont{A.}~\bibnamefont{Oliveira}},
  \bibinfo{author}{\bibfnamefont{J.}~\bibnamefont{Rojo}},
  \bibinfo{author}{\bibfnamefont{R.}~\bibnamefont{Rosenfeld}},
  \bibinfo{author}{\bibfnamefont{G.~P.} \bibnamefont{Salam}},
  \bibnamefont{et~al.}, \emph{\bibinfo{title}{{Scale-Invariant Resonance
  Tagging in Multijet Events and New Physics in Higgs pair Production}}},
  \bibinfo{journal}{JHEP} \textbf{\bibinfo{volume}{1307}}, \bibinfo{pages}{148}
  (\bibinfo{year}{2013}), \eprint{1303.6636}.

\bibitem[{\citenamefont{Aguilar-Saavedra}(2009)}]{AguilarSaavedra:2008zc}
\bibinfo{author}{\bibfnamefont{J.}~\bibnamefont{Aguilar-Saavedra}},
  \emph{\bibinfo{title}{{A Minimal Set of Top Anomalous Couplings}}},
  \bibinfo{journal}{Nucl.Phys.} \textbf{\bibinfo{volume}{B812}},
  \bibinfo{pages}{181} (\bibinfo{year}{2009}), \eprint{0811.3842}.

\end{thebibliography}
\bibliographystyle{apsper}

\end{document}